\newcommand*{\ATLASLATEXPATH}{latex/}
\documentclass[UKenglish,texlive=2011,cernpreprint]{\ATLASLATEXPATH atlasdoc}

\usepackage{\ATLASLATEXPATH atlaspackage}
\usepackage{\ATLASLATEXPATH atlasbiblatex}

\usepackage{\ATLASLATEXPATH atlascontribute}

\usepackage{\ATLASLATEXPATH atlasphysics}
\usepackage{multirow}

\graphicspath{{logos/}{Figures/}}

\usepackage{Photon_performance_paper-defs}
\usepackage{amssymb}

\AtlasTitle{Measurement of the photon identification efficiencies with the ATLAS detector using LHC Run-1 data}

\author{The ATLAS Collaboration}

\AtlasRefCode{PERF-2013-04}

\PreprintIdNumber{CERN-EP-2016-110}

\arXivId{1606.01813}

\AtlasJournalRef{Eur. Phys. J. C 76 (2016) 666}
\AtlasDOI{10.1140/epjc/s10052-016-4507-9}

\AtlasAbstract{%
  The algorithms used by the ATLAS Collaboration to reconstruct
  and identify prompt photons are described.   
  Measurements of the photon identification efficiencies are reported, using
  4.9~\ifb\ of $pp$ collision data collected at the LHC
  at $\sqrt{s} = 7$~\TeV\ and 20.3~\ifb\ at $\sqrt{s} = 8$~\TeV. 
  The efficiencies are measured separately for
  converted and unconverted photons, in four different pseudorapidity
  regions, for transverse momenta between 10~\GeV\ and 1.5~\TeV.  
  The results from the combination of three data-driven techniques are
  compared to the predictions from a simulation of the detector
  response, after correcting the electromagnetic shower momenta in
  the simulation for the average differences observed with respect to
  data. Data-to-simulation efficiency ratios used as correction
  factors in physics measurements are determined to account for the
  small residual efficiency differences. These factors are measured
  with uncertainties between 0.5\% and 10\% in 7~\TeV\ data and between
  0.5\% and 5.6\% in 8~\TeV\ data, depending on the photon
  transverse momentum and pseudorapidity.
}

\hypersetup{pdftitle={ATLAS document},pdfauthor={The ATLAS Collaboration}}

\begin{document}

\maketitle

\tableofcontents

\section{Introduction}
\label{sec:intro}
Several physics processes occurring in proton--proton collisions at the 
Large Hadron Collider (LHC) produce final states with \textit{prompt
  photons}, \textit{i.e.} photons not originating from hadron decays.
The main contributions come from non-resonant production of photons in
association with jets or of photon pairs, with cross sections
respectively of the order of tens of nanobarns or
picobarns~\cite{Aad:2010sp,Aad:2011tw,Aad:2013zba,ATLAS:2012ar,Aad:2011mh,Aad:2012je}. 
The study of such final states, and the measurement of their
production cross sections, are of great interest as they probe
the perturbative regime of QCD and can provide useful information
about the parton
distribution functions of the proton~\cite{dEnterria2012311}. 
Prompt photons are also produced in rarer processes that are key to the
LHC physics programme, such as diphoton decays of the Higgs boson
discovered with a mass near 125~\GeV, produced with a cross section times
branching ratio of about 20 fb at $\sqrt{s}=8$~\TeV~\cite{Aad:2014eha}.  
Finally, some expected signatures of physics beyond the 
Standard Model (SM) are characterised by the presence of prompt
photons in the final state.
These include resonant photon pairs from
graviton decays in models with extra spatial
dimensions~\cite{Aad:2015mna},
pairs of photons accompanied by large missing transverse momentum
produced in the decays of pairs of supersymmetric
particles~\cite{Aad:2015hea}
and events with highly energetic photons and jets from decays of
excited quarks or other exotic scenarios~\cite{Aad:2013cva}.

The identification of prompt photons in hadronic collisions is 
particularly challenging since an overwhelming majority of 
reconstructed photons is due to {\em background photons}.
These are usually real photons originating from hadron decays
in processes with larger cross sections than
prompt-photon production.
An additional smaller component of background photon candidates 
is due to hadrons producing in the detector energy deposits that have
characteristics similar to those of real photons.

Prompt photons are separated from background photons in the ATLAS experiment
by means of selections on quantities describing the shape and
properties of the associated electromagnetic showers and by requiring
them to be isolated from other particles in the event.
An estimate of the efficiency of the photon identification criteria
can be obtained from Monte Carlo (MC) simulation. 
Such an estimate, however, is subject to large, $\mathcal{O}(10\%)$,
systematic uncertainties. 
These uncertainties arise from limited knowledge of the detector
material, from an imperfect description of the shower development and
from the detector response~\cite{Aad:2010sp}.
Ultimately, for high-precision measurements and for accurate
comparisons with the predictions from the SM 
or from theories beyond the SM,
a determination of the photon identification efficiency with 
an uncertainty of $\mathcal{O}(1\%)$ or smaller is needed in a large
energy range from 10~\GeV\ to several \TeV. 
This can only be achieved through the use of data control samples.
However, this can present several difficulties since there is no
single physics process that produces a pure sample of prompt
photons in a large transverse momentum (\ET) range. 

In this document, the reconstruction and identification of photons by
the ATLAS detector are described, as well as the measurements of the
identification efficiency.
This study considers both photons that do (called {\em converted photons} in the following) or
do not convert (called {\em unconverted photons} in the following) to
electron-positron pairs in the detector material upstream of the
ATLAS electromagnetic calorimeter.
The measurements use the full Run-1 $pp$ collision dataset recorded at
centre-of-mass energies of 7 and 8~\TeV. 
The details of the selections and the results are given for the 
data collected in 2012 at $\sqrt{s}=8$ \TeV.
The same algorithms are applied with minor differences 
to the $\sqrt{s}=7$ \TeV\ data collected in 2011.

To overcome the difficulties arising from the absence of a 
single, pure control sample of prompt photons over a large \ET~range, three different data-driven techniques are used. A first method
selects photons from radiative decays of the \Zboson\ boson, i.e. \zllg\ (\RadZ\ method).
A second one extrapolates photon properties from electrons and positrons
from $Z$ boson decays by exploiting the similarity of the photon and
electron interactions with the ATLAS electromagnetic calorimeter
(\EE\ method).
A third approach exploits a technique to determine the fraction of
background present in a sample of isolated photon candidates (\MM). 
Each of these techniques can measure the photon identification
efficiency in complementary but overlapping \ET\ regions with varying
precision. 

This document is organised as follows. 
After an overview of the ATLAS detector in Sect.~\ref{sec:AtlasDet}, 
the photon reconstruction and identification algorithms used in ATLAS
are detailed in Sect.~\ref{sec:reconstructionandidentification}.
Section~\ref{sec:dataset} summarises the data and simulation samples
used and describes the corrections applied to the simulated photon
shower shapes in order to improve agreement with the data.
In Sect.~\ref{sec:datadrivenefficiencymeasurements}
the three data-driven approaches to the measurement of the photon
identification efficiency are described, listing their respective
sources of uncertainty and the precision reached in the
relevant \ET\ ranges.  The results obtained with the $\sqrt{s}=8$
TeV data collected in 2012, their consistency in the overlapping
$\ET$ intervals and the comparison to the MC predictions are presented
in Sect.~\ref{sec:pidresults}.
Results obtained for the identification criteria used during the
2011 data-taking period at $\sqrt{s}=7$ \TeV\ are described in
Sect.~\ref{sec:seventev}.
Finally, Sect.~\ref{sec:pileupdependence} discusses the
impact of multiple inelastic interactions in the same beam
crossing on the photon identification efficiency.

\section{ATLAS detector}
\label{sec:AtlasDet}
The ATLAS experiment~\cite{Aad:2008zzm} is a multi-purpose particle
detector with approximately forward-backward symmetric
cylindrical geometry and nearly 4$\pi$ coverage in solid 
angle.\footnote{ATLAS uses a right-handed coordinate
  system with its origin at the nominal interaction point (IP) in the
  centre of the detector and the $z$-axis along the beam pipe. The
  $x$-axis points from the IP to the centre of the LHC ring, and the
  $y$-axis points upward. Cylindrical coordinates $(r,\phi)$ are used
  in the transverse plane, $\phi$ being the azimuthal angle around the
  beam pipe. The pseudorapidity is defined in terms of the polar angle
  $\theta$ as $\eta=-\ln\tan(\theta/2)$. The photon transverse momentum is
  $\ET=E/\cosh(\eta)$, where $E$ is its energy.} 

The inner tracking detector (ID), surrounded by a thin superconducting
solenoid providing a $2\:\mathrm{T}$ axial magnetic field, provides
precise reconstruction of tracks within a pseudorapidity range $|\eta|
\lesssim 2.5$.
The innermost part of the ID consists of a silicon
pixel detector (50.5~mm~$<r<150$~mm)
providing typically three measurement points for
charged particles originating in the beam-interaction region.
The layer closest to the beam pipe (referred to as the $b$-layer
in this paper)
contributes significantly to precision vertexing and provides
discrimination between prompt tracks and photon conversions.
A semiconductor tracker (SCT) consisting of modules with two layers of
silicon microstrip sensors surrounds the pixel detector, providing
typically eight hits per track at intermediate radii (275~mm~$< r < 560$~mm).
The outermost region of the ID (563~mm~$ < r < 1066$~mm)
is covered by a transition radiation
tracker (TRT) consisting of straw drift tubes filled with a xenon
gas mixture, interleaved with polypropylene/polyethylene transition
radiators. For charged particles with transverse momentum $\pT>0.5$
\GeV\ within its pseudorapidity coverage ($|\eta| \lesssim 2$), the
TRT provides typically 35 hits per track. The distinction
between transition radiation (low-energy photons emitted by
electrons traversing the radiators) and tracking signals is
obtained on a straw-by-straw basis using separate low and high
thresholds in the front-end electronics.
The inner detector allows an accurate reconstruction and
transverse momentum measurement of tracks from the primary
proton--proton collision region. It also identifies tracks from
secondary vertices, permitting the efficient reconstruction of photon
conversions up to a radial distance of about 80~cm from the beamline. 

\begin{figure}[!htbp]
\centering
    {\includegraphics[width=0.7\columnwidth]{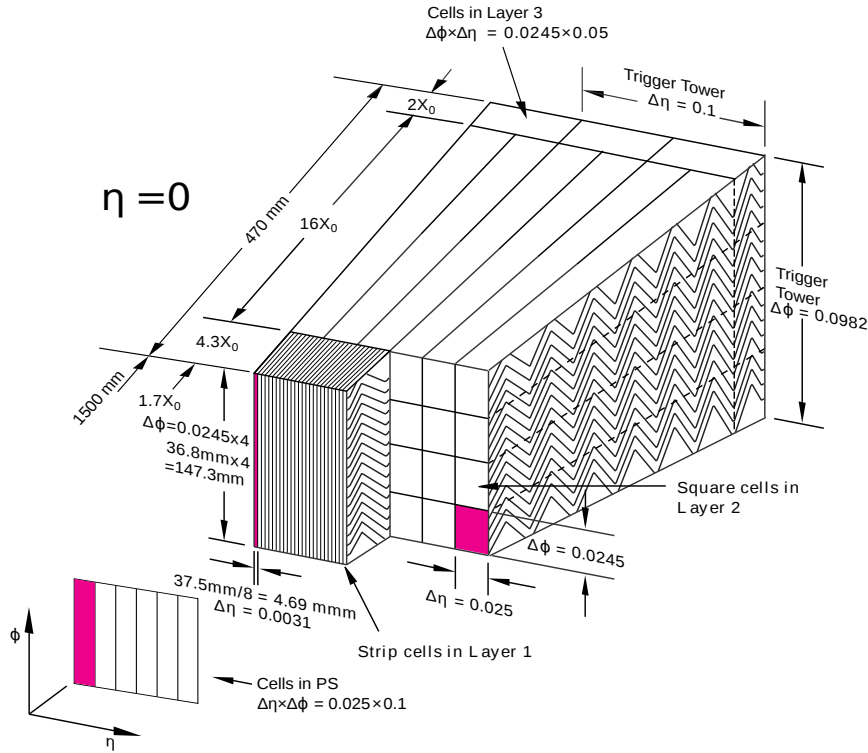}}
    \caption{Sketch of a barrel module (located at $\eta=0$)
      of the ATLAS electromagnetic calorimeter.
      The different longitudinal layers (one presampler, PS, and three 
      layers in the accordion calorimeter) are depicted.
      The granularity in $\eta$ and $\phi$ of the cells of each 
      layer and of the trigger towers is also shown.}
    \label{fig:Detector_EM_Module}
\end{figure}

The solenoid is surrounded by a high-granularity lead/liquid-argon (LAr)
sampling electromagnetic (EM) calorimeter with an accordion geometry.
The EM calorimeter measures the energy and the position of electromagnetic
showers with $\left| \eta \right|<3.2$.
It is divided into a barrel section, covering the
pseudorapidity region $|\eta| < 1.475$, and two end-cap sections,
covering the pseudorapidity regions $1.375 < |\eta| < 3.2$.  
The transition region between the barrel and the end-caps, 
$1.37 < |\eta| < 1.52$, has a large amount of material upstream of the
first active calorimeter layer.
The EM calorimeter is composed, for $|\eta|<2.5$, of three
sampling layers, longitudinal in shower depth.
The first layer has a thickness of about 4.4 radiation lengths 
($X_0$). In the ranges $|\eta|<1.4$ and $1.5< |\eta| <2.4$,
the first layer is segmented into high-granularity strips
in the $\eta$ direction, with a typical cell size of
$0.003\times 0.0982$ in $\Delta\eta \times \Delta \phi$ in the barrel.
For $1.4<|\eta|<1.5$ and $2.4< |\eta| <2.5$ 
the $\eta$-segmentation of the first layer is coarser,
and the cell size is $\Delta \eta \times \Delta \phi = 0.025\times 0.0982$.
The fine $\eta$ granularity of the strips is sufficient to provide,
for transverse momenta up to $\mathcal{O}(100~\mathrm{GeV})$, an
event-by-event discrimination between single photon showers and two
overlapping showers coming from the decays of neutral hadrons, mostly
$\pi^0$ and $\eta$ mesons, in jets in the fiducial pseudorapidity
region $|\eta|<1.37$ or $1.52<|\eta|<2.37$. 
The second layer has a thickness of about
17~$X_0$ and a granularity of $0.025 \times 0.0245$ in
$\Delta\eta \times \Delta \phi$.
It collects most of the energy deposited in the calorimeter by photon
and electron showers.
The third layer has a granularity of $0.05\times 0.0245$ in
$\Delta\eta \times \Delta\phi$ and a depth of about 2~$X_0$. It is 
used to correct for leakage beyond the EM calorimeter of high-energy
showers. In front of the accordion calorimeter, a thin presampler 
layer, covering the pseudorapidity interval $|\eta| < 1.8$, is used to
correct for energy loss upstream of the calorimeter. 
The presampler consists of an active LAr layer with a thickness of
1.1~cm (0.5~cm) in the barrel (end-cap) and has a granularity of 
$\Delta\eta \times \Delta \phi = 0.025 \times 0.0982$.
The material upstream of the presampler has a thickness of about 2~$X_0$
for $|\eta|<0.6$. In the region $0.6<|\eta|<0.8$ this thickness increases
linearly from 2~$X_0$ to 3~$X_0$. For $0.8<|\eta|<1.8$ the material
thickness is about or slightly larger than 3~$X_0$, with the exception
of the transition region between the barrel and the end-caps and the
region near $|\eta|=1.7$, where it reaches 5--6~$X_0$.
A sketch of a barrel module of the electromagnetic calorimeter is shown
in Fig.~\ref{fig:Detector_EM_Module}. 

The hadronic calorimeter surrounds the EM calorimeter.
It consists of an iron--scintillator tile calorimeter in the central region
($|\eta|< 1.7$), and LAr sampling calorimeters with copper and tungsten 
absorbers in the end-cap ($1.5<|\eta|<3.2$) and forward
($3.1<\left| \eta \right|<4.9$) regions.

The muon spectrometer surrounds the calorimeters. It consists of three
large superconducting air-core toroid magnets, each with eight coils,
a system of precision tracking chambers ($\left| \eta \right|<2.7$),
and fast tracking chambers ($\left| \eta \right|<2.4$) for triggering.  

A three-level trigger system selects events to be recorded 
for offline analysis.
A coarser readout granularity (corresponding to the ``towers''
of Fig.~\ref{fig:Detector_EM_Module}) is used by the first-level
trigger, while the full detector granularity is exploited by the
higher-level trigger.
To reduce the data acquisition rate of low-threshold 
triggers, used for collecting various control samples, 
prescale factors ($N$) can be applied to each trigger, 
such that only 1 in $N$ events passing the trigger 
causes an event to be accepted at that trigger level.

\section{Photon reconstruction and identification}
\label{sec:reconstructionandidentification}

\subsection{Photon reconstruction}
\label{ssec:reconstruction}
The electromagnetic shower, originating from an energetic photon's
interaction with the EM calorimeter, deposits a significant amount of
energy in a small number of neighbouring calorimeter cells.
As photons and electrons have very similar signatures in the EM
calorimeter, their reconstruction proceeds in parallel.
The electron reconstruction, including a dedicated, cluster-seeded
track-finding algorithm to increase the efficiency for the reconstruction of
low-momentum electron tracks, is described in Ref.~\cite{ATLAS-CONF-2014-032}.
The reconstruction of unconverted and converted photons proceeds in the
following way:
\begin{itemize}
\item seed clusters of EM calorimeter cells are searched for;
\item tracks reconstructed in the inner detector are loosely matched
  to seed clusters;
\item tracks consistent with originating from a photon conversion are used to
  create conversion vertex candidates;
\item conversion vertex candidates are matched to seed clusters;
\item a final algorithm decides whether a seed cluster corresponds to an
  unconverted photon, a converted photon or a single electron based on
  the matching to conversion vertices or tracks and on the cluster and track(s)
  four-momenta.
\end{itemize}
In the following the various steps of the reconstruction algorithms are
described in more detail.

The reconstruction of photon candidates in the region $|\eta|<2.5$
begins with the creation of a preliminary set of seed clusters of
EM calorimeter cells.
Seed clusters of size $\Delta\eta \times \Delta\phi = 0.075
\times 0.123$ with transverse momentum above 2.5~\GeV\ are formed by a
sliding-window algorithm~\cite{Lampl:1099735}.
After an energy comparison, duplicate clusters of lower energy
are removed from nearby seed clusters.
From MC simulations, the efficiency of the initial 
cluster reconstruction is estimated to be greater than 99\% for 
photons with $\ET > 20$~\GeV.

Once seed clusters are reconstructed, a search is performed for inner detector
tracks~\cite{Cornelissen:1020106,Cornelissen:2008zzc} that are loosely matched to the clusters, in order to identify 
and reconstruct electrons and photon conversions.
Tracks are loosely matched to a cluster if the angular distance between
the cluster barycentre and the extrapolated track's intersection point with
the second sampling layer of the calorimeter is smaller than
0.05 (0.2) along $\phi$ in the direction of (opposite to) the bending  
of the tracks in the magnetic field of the ATLAS solenoid, and smaller
than 0.05 along $\eta$ for tracks with hits in the silicon detectors,
 {\em i.e.} the pixel and SCT detectors.
Tracks with hits in the silicon detectors are extrapolated from the point
of closest approach to the primary vertex, while tracks without hits in
the silicon detectors are extrapolated from the last measured point.
The track is extrapolated to the position corresponding to the 
expected maximum energy deposit for EM showers.
To efficiently select low-momentum tracks that may have
suffered significant bremsstrahlung losses before reaching the
calorimeter, a similar matching procedure is applied after rescaling
the track momentum to the measured cluster energy. The previous matching
requirements are applied except that the $\phi$ difference in the direction of
bending should be smaller than 0.1.
Tracks that are loosely matched to a cluster and 
with hits in the silicon detectors are refitted with a Gaussian-sum-filter
technique~\cite{Fruhwirth2003131, ATLAS-CONF-2012-047},
to improve the track parameter resolution, 
and are retained for the reconstruction of 
electrons and converted photons.

``Double-track'' conversion vertex candidates are reconstructed from
pairs of oppositely charged tracks in the ID that are likely to be
electrons.
For each track the likelihood to be an electron, based on high-threshold 
hits and time-over-threshold of low-threshold hits in the TRT, is required to
be at least 10\% (80\%) for tracks with (without) hits in the silicon 
detectors.
Since the tracks of a photon conversion are parallel at the place of 
conversion,
geometric requirements are used to select the track pairs.
Track pairs are classified into three categories, whether 
both tracks (Si--Si), none (TRT--TRT) or only one of them (Si--TRT) have
hits in the silicon detectors.
Track pairs fulfilling the following requirements are retained:
\begin{itemize}
\item $\Delta\cot\theta$ between the two tracks (taken at the tracks'
  points of closest approach to the primary vertex) is less than 0.3 for Si--Si 
  track pairs and 0.5 for track pairs with at least one track without 
  hits in the silicon detectors.
  This requirement is not applied for TRT--TRT track pairs with both tracks within
  $|\eta|<0.6$.
\item The distance of closest approach between the two tracks is less
  than 10~mm for Si--Si track pairs and 50~mm for track pairs with at 
  least one track without hits in the silicon detectors.
\item The difference between the sum of the radii of the helices 
  that can be constructed from the electron and positron tracks and
  the distance between the centres of the two helices is 
  between $-5$ and $5$~mm, between $-50$ and 10~mm, or between $-25$ and 10~mm,
  for Si--Si, TRT--TRT and Si--TRT track pairs, respectively.
\item $\Delta\phi$ between the two tracks (taken at the estimated vertex
  position before the conversion vertex fit) is less than 0.05 for Si--Si
  track pairs and 0.2 for tracks pairs with at least one track without
  hits in the silicon detectors.
\end{itemize}

A constrained conversion vertex fit with three degrees of freedom is performed
using the five measured helix parameters of each of the two participating
tracks with the constraint that the tracks are parallel at the vertex. 
Only the vertices satisfying the following requirements are retained:
\begin{itemize}
\item The $\chi^2$ of the conversion vertex fit is less than 50. 
  This loose requirement suppresses fake candidates from random 
  combination of tracks while being highly efficient for true
  photon conversions.
\item The radius of the conversion vertex, defined as the distance 
  from the vertex to the beamline in the transverse plane, 
  is greater than 20~mm, 70~mm or 250~mm for vertices from Si--Si, 
  Si--TRT and TRT--TRT track pairs, respectively.
\item
  The difference in $\phi$ between the vertex position and the direction of
  the reconstructed conversion is less than 0.2.
\end{itemize}

The efficiency to reconstruct photon conversions as double-track
vertex candidates decreases significantly for conversions taking place in the 
outermost layers of the ID. This effect is due to photon conversions
in which one of the two produced electron tracks is not reconstructed either
because it is very soft (asymmetric conversions
where one of the two tracks has $\pT < 0.5$~\GeV), or because the two tracks
are very close to each other and cannot be adequately separated.
For this reason, tracks without hits in the $b$-layer 
that either have 
an electron likelihood greater than 95\%, or have no hits in the TRT, 
are considered as ``single-track'' conversion vertex candidates.
In this case, since a conversion vertex fit cannot be performed, the conversion
vertex is defined to be the location of the first measurement of the track.
Tracks which pass through a passive region of the $b$-layer 
are not considered as single-track conversions
unless they are missing a hit in the second pixel layer.

As in the loose track matching, the matching of the conversion
vertices to the clusters relies on an extrapolation of
the conversion candidates to the second sampling layer of the calorimeter, 
and the comparison of the extrapolated $\eta$ and $\phi$ coordinates
to the $\eta$ and $\phi$ coordinates of the cluster centre. The details
of the extrapolation depend on the type of the conversion vertex
candidate.
\begin{itemize}
\item For double-track conversion vertex candidates for which the track
  transverse momenta differ by less than a factor of four from each other, 
  each track is extrapolated to the second sampling layer of the calorimeter
  and is required to be matched to the cluster.
\item For double-track conversion vertex candidates for which the track
  transverse momenta differ by more than a factor of four from each other, the 
  photon direction is reconstructed from the electron and positron
  directions determined by the conversion vertex fit, and used to perform a
  straight-line extrapolation to the second sampling layer of the calorimeter,
  as expected for a neutral particle.
\item For single-track conversion vertex candidates, the track is 
  extrapolated from its last measurement.
\end{itemize}
Conversion vertex candidates built from tracks with hits in
the silicon detectors are considered matched to a cluster if the
angular distance between the extrapolated conversion vertex candidate and
the cluster centre is smaller than 0.05 in both $\eta$ and $\phi$.
If the extrapolation is performed for single-track
conversions, the window in $\phi$ is increased to 0.1 in the direction
of the bending. 
For tracks without hits in the silicon detectors, the matching
requirements are tighter:
\begin{itemize}
\item
  The distance in $\phi$ between the extrapolated track(s) and the
  cluster is less than 0.02 (0.03) in the direction of (opposite to) 
  the bending.
  In the case where the conversion vertex candidate is extrapolated as a
  neutral particle, the distance is required to be less than 0.03 on
  both sides.
\item 
  The distance in $\eta$ between the extrapolated track(s) and the
  cluster is less than 0.35 and 0.2 in the barrel and end-cap
  sections of the TRT, respectively. 
  The criteria are significantly looser than in the $\phi$ direction
  since the TRT does not provide a measurement of the pseudorapidity
  in its barrel section.
  In the case that the conversion vertex candidate is extrapolated as a
  neutral particle, the distance is required to be less than 0.35. 
\end{itemize}
In the case of multiple conversion vertex candidates matched to the same 
cluster, the final conversion vertex candidate is chosen as follows:
\begin{itemize}
\item preference is given to double-track candidates over
  single-track candidates;
\item if both conversion vertex candidates are formed from the 
  same number of tracks, preference is given to the candidate 
  with more tracks with hits in the silicon detectors; 
\item if the conversion vertex candidates are formed from the same number of
  tracks with hits in the silicon detectors, preference is given to
  the vertex candidate with smaller radius.
\end{itemize}

The final arbitration between the unconverted photon, converted photon
and electron hypotheses for the reconstructed EM clusters is performed
in the following way~\cite{ATL-PHYS-PUB-2011-007}:
\begin{itemize}
\item
Clusters to which neither a conversion vertex candidate
nor any track has been matched during the electron reconstruction
are considered unconverted photon candidates.
\item
Electromagnetic clusters matched to a conversion vertex candidate
are considered converted photon candidates.
For converted photon candidates that are also reconstructed as electrons,
the electron track is evaluated against the track(s) originating from the 
conversion vertex candidate matched to the same cluster:
\begin{enumerate}
\item If the track coincides with a track coming from the conversion vertex,
the converted photon candidate is retained.
\item The only exception to the previous rule is the case of a
  double-track conversion vertex candidate where the coinciding track
  has a hit in the $b$-layer, while the other track lacks one (for
  this purpose, a missing hit in a disabled $b$-layer module is
  counted as a hit
\footnote{About 6.3\% of the $b$-layer modules were disabled at the end
  of Run 1 due to individual module failures like low-voltage or
  high-voltage powering faults or data transmission faults. During the
  shutdown following the end of Run 1, repairs reduced
  the $b$-layer fault fraction to 1.4\%}).  
\item If the track does not coincide with any of the tracks assigned to the
conversion vertex candidate, the converted photon candidate is removed, unless
the track $\pT$ is smaller than the $\pT$ of the converted photon candidate.  
\end{enumerate}
\item
Single-track converted photon candidates are recovered from objects
that are only reconstructed as electron candidates with $\pT > 2$~\GeV\ and
$E/p < 10$ ($E$ being the cluster energy and $p$ the track momentum),
if the track has no hits in the silicon detectors.
\item
Unconverted photon candidates are recovered from reconstructed electron 
candidates if the electron candidate has a corresponding track 
without hits in the silicon detectors and with $\pT < 2$~\GeV,
or if the electron candidate is not considered as single-track
converted photon and its matched track has a transverse momentum
lower than 2~\GeV\ or $E/p$ greater than $10$. The corresponding
electron candidate is then removed from the event.
Using this procedure around 85\% of the unconverted photons erroneously
categorised as electrons are recovered.
\end{itemize}

From MC simulations, 96\% of prompt photons with $\ET>25$~\GeV\ are expected to 
be reconstructed as photon candidates, while the remaining 4\% are
incorrectly reconstructed as electrons but not as photons.
The reconstruction efficiencies of photons with transverse momenta of
a few tens of~\GeV\ (relevant for the search for Higgs boson decays to
two photons) are checked in data with a technique described in
Ref.~\cite{Aad:2014nim}.
The results point to inefficiencies and fake rates that exceed by
up to a few percent the predictions from MC simulation. 
The efficiency to reconstruct photon conversions decreases at high $\ET$ 
($>150$~GeV), where it becomes more difficult to separate the 
two tracks from the conversions. Such conversions
with very close-by tracks are often not recovered as single-track conversions
because of the tighter selections, including the
transition radiation requirement,
applied to single-track conversion candidates. 
The overall photon reconstruction efficiency is thus reduced to about 90\% for 
$\ET$ around 1~\TeV.

The final photon energy measurement is performed using information from the
calorimeter, with a cluster size that depends on the photon
classification.\footnote{For converted photon candidates, the
  energy calibration procedure uses the following as additional inputs:
  (i) $\pT/\ET$ and the momentum balance of the two conversion tracks
  if both tracks are reconstructed by the silicon detectors,
  and (ii) the conversion radius for photon candidates with
  transverse momentum above 3~\GeV.}
In the barrel, a cluster of size 
$\Delta\eta \times \Delta\phi =0.075\times 0.123$ is used for unconverted
photon candidates, while a cluster of size $0.075 \times 0.172$ is
used for converted photon candidates to compensate for the opening
between the conversion products in the $\phi$ direction due to the
magnetic field of the ATLAS solenoid. 
In the end-cap, a cluster size of $0.125\times 0.123$ is used for all
candidates. 
The photon energy calibration, which accounts for upstream energy loss and both
lateral and longitudinal leakage, is based on the same procedure that is
used for electrons~\cite{Aad:2011mk, Aad:2014nim}
but with different calibration factors for converted and unconverted
photon candidates.
In the following the photon transverse momentum $\ET$
is computed from the photon cluster's calibrated energy $E$ and the
pseudorapidity $\eta_2$ of the barycentre of the cluster in the
second layer of the EM calorimeter as $\ET=E/\cosh(\eta_2)$.

\subsection{Photon identification}
\label{ssec:identification}
To distinguish prompt photons from background photons, photon 
identification with high signal efficiency and high
background rejection is required for transverse momenta
from 10~\GeV\ to the \TeV\ scale.
Photon identification in ATLAS is based on a set of cuts
on several discriminating variables. Such variables,
listed in Table~\ref{tab:IsEM-DV} and described in 
Appendix~\ref{app:IsEM-DV}, characterise the lateral and
longitudinal shower development in the electromagnetic calorimeter
and the shower leakage fraction in the hadronic calorimeter. 
Prompt photons typically produce narrower energy deposits in the 
electromagnetic calorimeter and have smaller leakage to the hadronic
one compared to background photons from jets, due to the presence of
additional hadrons near the photon candidate in the latter case. In
addition, background candidates from isolated $\pi^0\to\gamma\gamma$ decays
-- unlike prompt photons -- are often characterised by 
two separate local energy maxima in the finely segmented strips of the 
first layer, due to the small separation between the two photons.
The distributions of the discriminating variables for both the prompt
and background photons are affected by
additional soft $pp$ interactions that may accompany the
hard-scattering collision, referred to as in-time pile-up, 
as well as by out-of-time pile-up arising from bunches before or 
after the bunch where the event of interest was triggered. 
Pile-up results in the presence of low-\et\ activity in the detector, 
including energy deposition in the electromagnetic calorimeter. This
effect tends to broaden the distributions of the discriminating variables
and thus to reduce the separation between prompt and background photon 
candidates.
 
Two reference selections, a \textit{loose} one and a \textit{tight} one, 
are defined. The \textit{loose} selection is based only on shower shapes in
the second layer of the electromagnetic calorimeter and on the energy
deposited in the hadronic calorimeter, and is used by the photon triggers.
The loose requirements are designed to provide
a high prompt-photon identification efficiency with respect to reconstruction.
Their efficiency rises from 97\% at $\ET^\gamma = 20~\GeV$ to above
99\% for $\ET^\gamma > 40~\GeV$ for both the converted and unconverted photons,
and the corresponding background rejection factor is about
1000~\cite{ATL-PHYS-PUB-2011-007}.
The rejection factor is defined as the ratio of the 
number of initial jets with $\pT>40$~\GeV\ 
in the acceptance of the calorimeter to the number of
reconstructed background photon candidates satisfying the
identification criteria.
The \textit{tight} selection adds information from the finely segmented strip
layer of the calorimeter, which provides good rejection of hadronic
jets where a neutral meson carries most of the jet energy.
The \textit{tight} criteria are separately optimised for unconverted and
converted photons to provide a photon identification efficiency of
about 85\% for photon candidates with transverse energy $\ET > 40~\GeV$
and a corresponding background rejection factor of about
5000~\cite{ATL-PHYS-PUB-2011-007}. 

The selection criteria
are different in seven intervals of the reconstructed photon
pseudorapidity ($0.0$--$0.6$, $0.6$--$0.8$, $0.8$--$1.15$,
$1.15$--$1.37$, $1.52$--$1.81$, $1.81$--$2.01$, $2.01$--$2.37$) to account for the
calorimeter geometry and for different 
effects on the shower shapes from the material upstream of the 
calorimeter, which is highly non-uniform as a function of $|\eta|$. 
 
The photon identification criteria were first optimised prior to 
the start of the data-taking in 2010, on simulated samples of prompt photons
from $\gamma+$jet, diphoton and $H\to\gamma\gamma$ events and samples of 
background photons in QCD multi-jet
events~\cite{ATL-PHYS-PUB-2011-007}.
Before the 2011 data-taking, the \textit{loose} and the \textit{tight}
selections 
were loosened, without further re-optimisation, in order to reduce
the systematic effects associated to the differences between the 
calorimetric variables measured from data and their description 
by the ATLAS simulation.
Prior to the 8~\TeV\ run in 2012, the identification criteria were
reoptimised based on improved simulations in which the values of the
shower shape variables are slightly shifted to improve the agreement
with the data shower shapes, as described in the next section.
To cope with the higher pile-up expected during the 2012 data taking,
the criteria on the shower shapes more sensitive to pile-up
were relaxed while the others were tightened.

The discriminating variables
that are most sensitive to pile-up are found to be the energy
leakage in the hadronic calorimeter and the shower width in
the second sampling layer of the EM calorimeter.

\begin{table*}[tp]
  \centering
  \def\arraystretch{1.5}
  \begin{tabular}{lp{0.51\textwidth}l|cc}
  \hline \hline
  Category & Description & Name & \textit{loose} & \textit{tight} \\ 
  \hline \hline
  Acceptance & $|\eta|<2.37$, with $1.37<|\eta|<1.52$ excluded  & -- & $\checkmark$ & $\checkmark$ \\
  Hadronic leakage & Ratio of $\et$ in the first sampling layer of the hadronic calorimeter to $\et$ of the 
                     EM cluster (used over the range $|\eta| < 0.8$ or $|\eta| > 1.37$)  & $\Rhadone$ & $\checkmark$ & $\checkmark$ \\ 
                   & Ratio of $\et$ in the hadronic calorimeter to $\et$ of the EM cluster 
                     (used over the range $0.8 < |\eta| < 1.37$)  & $\Rhad$ & $\checkmark$ & $\checkmark$ \\
  EM Middle layer  & Ratio of 3 $\times$ 7 $\eta\times\phi$ to 7 $\times$ 7 cell energies & $\Reta$ & $\checkmark$ & $\checkmark$ \\
                   & Lateral width of the shower  & $\wetatwo$ & $\checkmark$ & $\checkmark$ \\
                   & Ratio of 3$\times$3 $\eta\times\phi$ to 3$\times$7 cell energies & $\Rphi$  & & $\checkmark$\\
  EM Strip layer   & Shower width calculated from three strips around the strip with maximum energy deposit  & $\wthree$ & & $\checkmark$ \\
                   & Total lateral shower width  & $\wtot$  & & $\checkmark$ \\
                   & Energy outside the core of the three central strips but within seven strips divided by energy within the three central strips   & $\Fside$  & & $\checkmark$ \\ 
                   & Difference between the energy associated with the second maximum in the strip layer and the energy reconstructed in the strip with
  the minimum value found between the first and second maxima  & $\DeltaE$ & & $\checkmark$ \\
                   & Ratio of the energy difference associated with the largest and second largest energy deposits to the sum of these
  energies & $\Eratio$  & & $\checkmark$\\
      \hline \hline
  \end{tabular}
  \caption{Discriminating variables used for \textit{loose} and \textit{tight} photon identification.}
  \label{tab:IsEM-DV}
\end{table*}

\subsection{Photon isolation}
\label{ssec:isolation}

The identification efficiencies presented in this article are measured
for photon candidates passing an isolation requirement, similar to those
applied to reduce hadronic background in cross-section measurements
or searches for exotic processes with photons~\cite{Aad:2010sp,Aad:2011tw,Aad:2013zba,ATLAS:2012ar,Aad:2011mh,Aad:2012je,Aad:2014eha,Aad:2015mna,ATLAS-CONF-2014-001,Aad:2013cva}.
In the data taken at $\sqrt{s}=8$~\TeV, the calorimeter isolation 
transverse energy \Etiso~\cite{ATLAS_Higgs} is required to be lower than 4~\GeV.
This quantity is computed from positive-energy three-dimensional 
topological clusters of calorimeter cells~\cite{Lampl:1099735} reconstructed 
in a cone of size $\Delta R = \sqrt{(\Delta\eta)^2+(\Delta\phi)^2}=0.4$ 
around the photon candidate.

The contributions to \Etiso\ from the photon itself and 
from the underlying event and pile-up are subtracted.
The correction for the photon energy outside the cluster is computed
as the product of the photon transverse energy and a coefficient
determined from separate simulations of converted and unconverted
photons.
The underlying event and pile-up energy correction is computed on
an event-by-event basis using the method described in
Refs.~\cite{Cacciari:area} and~\cite{Cacciari:UE}.
A $k_\mathrm{T}$ jet-finding algorithm~\cite{Ellis_kt,Catani_kt} of
size parameter $R =0.5$ is used to reconstruct all jets without any
explicit transverse momentum threshold, starting from the
three-dimensional topological clusters reconstructed in the
calorimeter. Each jet is assigned an area $A_\mathrm{jet}$ via a Voronoi
tessellation~\cite{Voronoi} of the $\eta$--$\phi$ space. According to this
algorithm, every point within a jet's assigned 
area is closer to the axis of that jet than to the axis of any other jet. 
The ambient transverse energy density $\rho_\mathrm{UE}(\eta)$ from
pile-up and the underlying 
event is taken to be the median of the transverse energy
densities $\pT^\mathrm{jet}/A_\mathrm{jet}$ of jets
with pseudorapidity $|\eta|<1.5$ or $1.5<|\eta|<3.0$.
The area of the photon isolation cone is then multiplied by $\rho_\mathrm{UE}$
to compute the correction to \Etiso. 
The estimated ambient transverse energy fluctuates significantly
event-by-event, reflecting the fluctuations in the underlying event
and pile-up activity in the data. The typical size of the correction is
2~\GeV\ in the central region and 1.5~\GeV\ in the forward region. 

A slight dependence of the identification efficiency on the isolation
requirement is observed, as discussed in
Section~\ref{sec:comparisonwithsimulation}. 

\section{Data and Monte Carlo samples}
\label{sec:dataset}
The data used in this study consist of the 7~\TeV\ and
8~\TeV\ proton--proton collisions recorded by the \mbox{ATLAS detector} during
2011 and 2012 in LHC Run 1. They correspond respectively to 4.9~\ifb\ and
20.3~\ifb\ of integrated luminosity after requiring good data
quality. The mean number of interactions per bunch crossing, $\mu$, was
9 and 21 on average in the $\sqrt{s}=7$ and 8~\TeV\ datasets, respectively.

The $Z$ boson radiative decay and the electron extrapolation methods use 
data collected with the lowest-threshold lepton triggers with
prescale factors equal to one and thus exploit the full luminosity of Run 1.
For the data collected in 2012 at $\sqrt{s}=8$~\TeV, the transverse momentum
thresholds for single-lepton triggers are 25 (24) \GeV\ for
$\ell=e~(\mu)$, while those for dilepton triggers are 12 (13) \GeV.
For the data collected in 2011 at $\sqrt{s}=7$~\TeV, the transverse momentum
thresholds for single-lepton triggers are 20 (18) \GeV\ for
$\ell=e~(\mu)$, while those for dilepton triggers are 12 (10) \GeV.
The matrix method uses events collected with single-photon 
triggers with loose identification requirements 
and large prescale factors, and thus exploits only a 
fraction of the total luminosity.
Photons reconstructed near regions of the calorimeter affected by
read-out or high-voltage failures~\cite{1748-0221-9-07-P07024} are
rejected. 

Monte Carlo samples are processed through a full simulation
of the ATLAS detector response~\cite{ATLASsimulation} using
\textsc{Geant4}~\cite{dataset:GEANT4} 4.9.4-patch04.
Pile-up $pp$ interactions in the same and nearby bunch crossings
are included in the simulation. 
The MC samples are reweighted to reproduce the distribution
of $\mu$ and the length of the luminous region observed in data
(approximately 54~cm and 48~cm in the data taken at $\sqrt{s}=7$ and 8~\TeV,
respectively).
Samples of prompt photons are generated with
PYTHIA8~\cite{Sjostrand:2007gs,Sjostrand:2006za}. Such samples include the
leading-order $\gamma$ + jet events from $q g \rightarrow q \gamma$ and $q \bar{q} \rightarrow g
\gamma$ hard scattering, as well as 
prompt photons from quark fragmentation in QCD dijet events. About $10^7$
events are generated, covering the whole transverse momentum spectrum
under study.
Samples of background photons in jets are produced by generating with 
PYTHIA8 all tree-level 2$\rightarrow$2 QCD processes, removing
$\gamma$ + jet events from quark fragmentation.
Between $1.2\times 10^6$ and $5\times 10^6$
$Z\to \ell\ell\gamma$ ($\ell=e,\mu$) events 
are generated with SHERPA~\cite{sherpa2} or with 
POWHEG~\cite{Frixione:2007vw,Alioli:2010xd} interfaced to
PHOTOS~\cite{Golonka:2005pn} for the modelling of QED final-state
radiation and to PYTHIA8 
for showering, hadronisation and modelling of the underlying event.
About $10^7$ $Z(\to\ell\ell)$+jet events are generated 
for both $\ell=e$ and $\ell=\mu$ with each of the following three event 
generators: POWHEG interfaced to PYTHIA8; ALPGEN~\cite{Mangano:2002ea}
interfaced to HERWIG~\cite{herwig} and JIMMY~\cite{jimmy} for
showering, hadronisation and modelling of the underlying event; and SHERPA.
A sample of MC $\Hboson \rightarrow \Zboson\gamma$
events~\cite{ZgammaPaper} is also used to compute the efficiency in
the simulation for photons with transverse momentum between 10 and
15~\GeV, since the $Z\to \ell\ell\gamma$ samples have a
generator-level requirement on the minimum true photon transverse 
momentum of 10~\GeV\ which biases the reconstructed transverse momentum 
near the threshold. A two-dimensional reweighting of the pseudorapidity 
and transverse momentum spectra of the photons is applied to match the
distributions of those reconstructed in $Z\to\ell\ell\gamma$ events.
For the analysis of $\sqrt{s}=7$~TeV data, all simulated samples 
(photon+jet, QCD multi-jet, $Z(\to \ell\ell)+$jet and
$Z\to\ell\ell\gamma$) are generated with PYTHIA6.

For the analysis of 8~\TeV\ data, the events are simulated and
reconstructed using the model of the ATLAS detector described
in Ref.~\cite{Aad:2014nim}, based on
an improved {\em in situ} determination of the passive material
upstream of the electromagnetic calorimeter.
This model is characterised by the presence of additional material 
(up to 0.6 radiation lengths) in the end-cap and a $50\%$ 
smaller uncertainty in the material budget with respect to the 
previous model, which is used for the study of 7~\TeV\ data.

The distributions of the photon transverse shower shapes
in the ATLAS MC simulation do not perfectly match the ones observed in data.
While the shapes of these distributions
in the simulation are rather similar to those found in the data,
small systematic differences in their average values are observed.
On the other hand, the longitudinal electromagnetic shower profiles
are well described by the simulation.
The differences between the data and MC distributions are 
parameterised as simple shifts and applied to the MC simulated values
to match the distributions in data. 
These shifts are calculated by minimising the $\chi^2$ between the
data and the shifted MC distributions of photon candidates satisfying the
\textit{tight} identification criteria and the calorimeter isolation
requirement described in the previous section.
The shifts are computed in intervals of the reconstructed photon
pseudorapidity and transverse momentum. The pseudorapidity intervals
are the same as those used to define the photon selection criteria.
The \ET\ bin boundaries are 0, 15, 20, 25, 30, 40, 50, 60, 80, 100 and
1000~\GeV. 
The typical size of the correction factors is 10\% of the RMS of the
distribution of the corresponding variable in data.
For the variable $\Reta$, for which the level of
agreement between the data and the simulation is worst, the size of the 
correction factors is 50\% of the RMS of the distribution.
The corresponding correction to the prompt-photon efficiency
predicted by the simulation varies with pseudorapidity between
$-10\%$ and $-5\%$ for photon transverse momenta close to 10~GeV, and
approaches zero for transverse momenta above 50~GeV.

Two examples of the simulated discriminating variable distributions
before and after corrections, for converted photon candidates originating
from $Z$ boson radiative decays, are shown in Fig.~\ref{fig:FF_examples}.
For comparison, the distributions observed in data for candidates passing
the $Z$ boson radiative decay selection illustrated in Sect.~\ref{sec:photonsfromZradiativedecays},
are also shown. Better agreement between the shower shape distributions
in data and in the simulation after applying such corrections is clearly
visible.

\begin{figure}[!ht]
   \centering
   \subfloat[]{\label{fig:FF_examples_a}\includegraphics[width=0.49\textwidth]{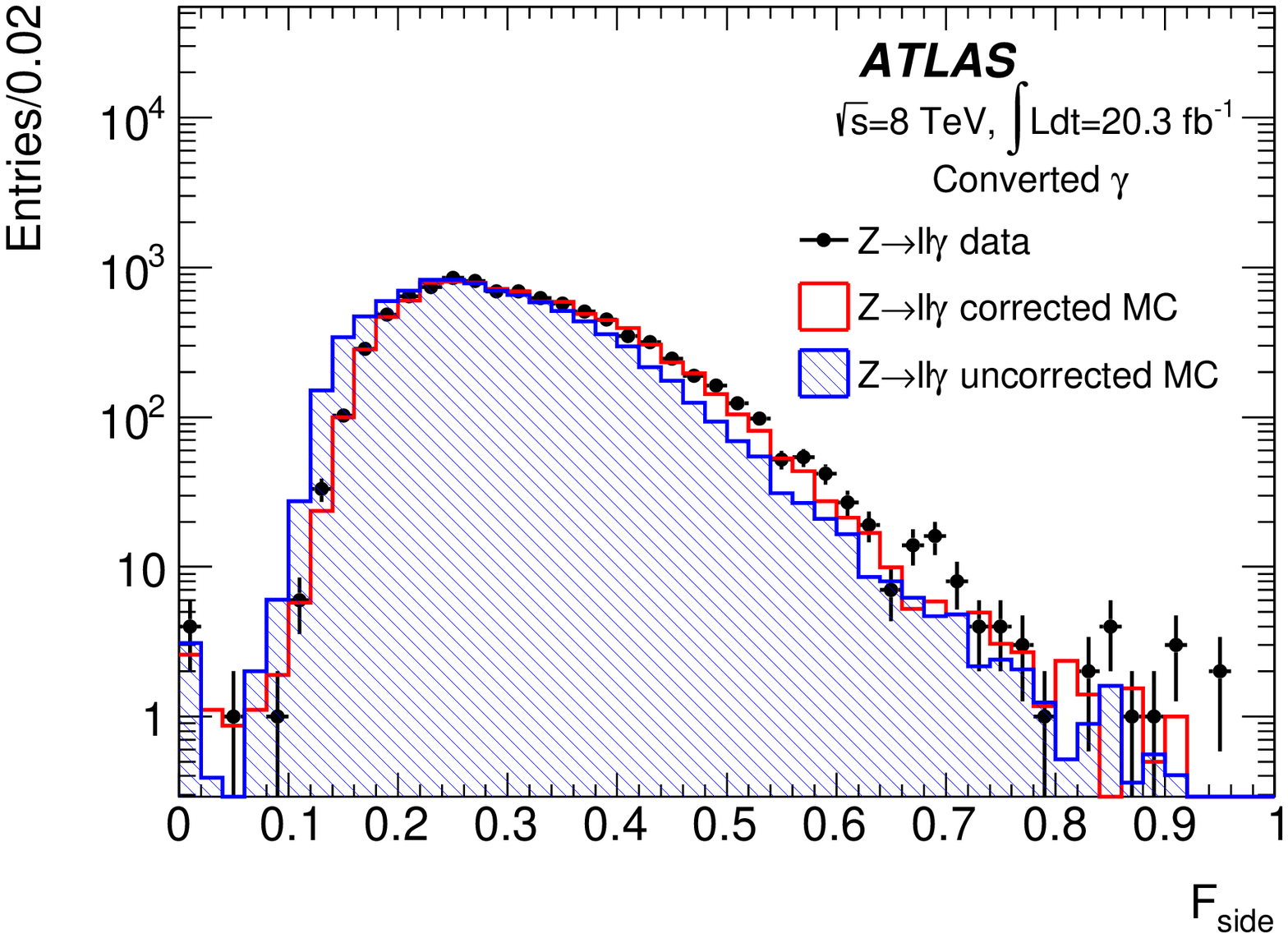}}
   \subfloat[]{\label{fig:FF_examples_b}\includegraphics[width=0.49\textwidth]{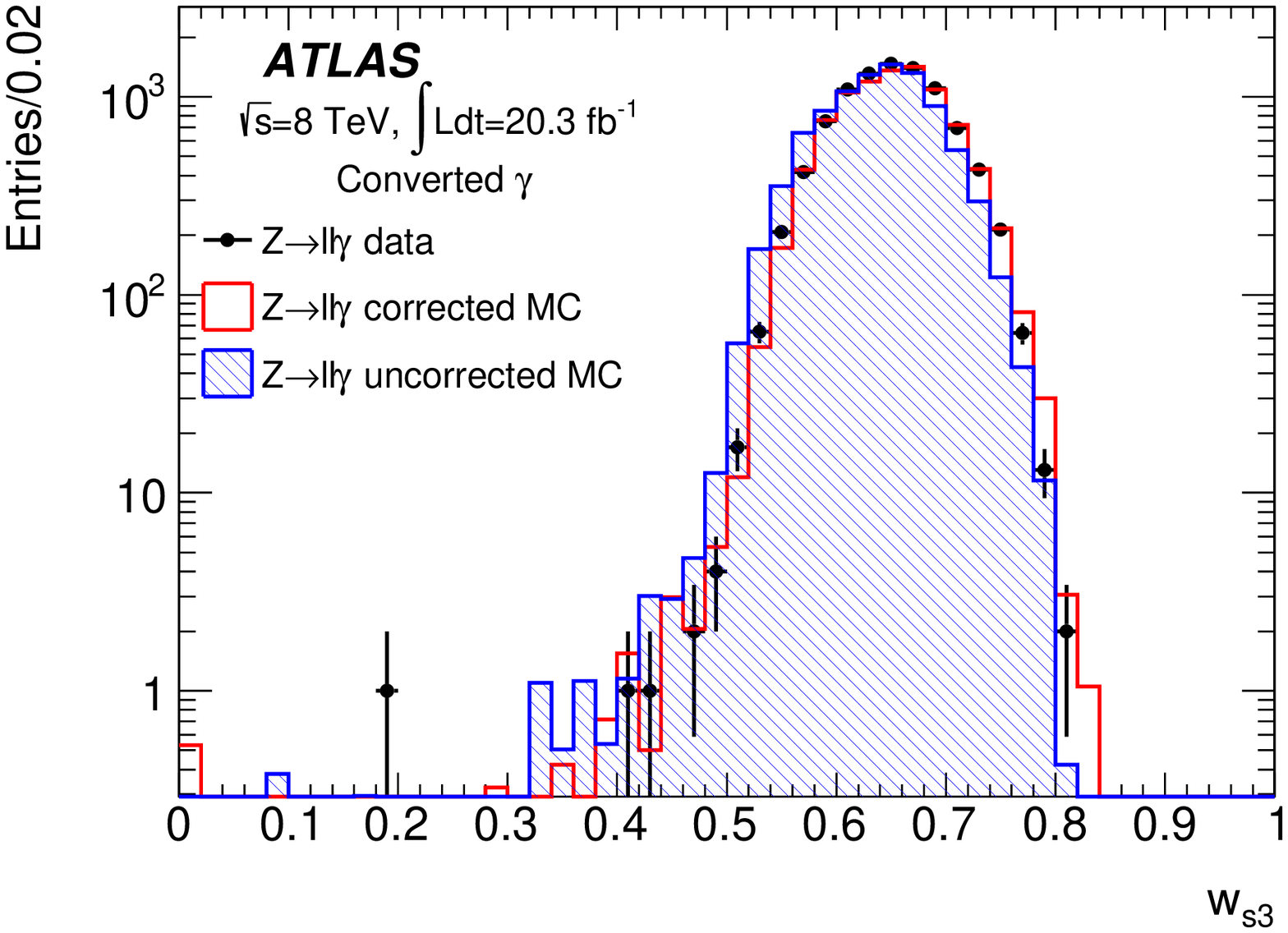}}
   \caption{Distributions of the calorimetric discriminating variables
     (a) $\Fside$ and (b) $\wthree$ for converted photon
     candidates with $\ET > 20~\GeV$ and $|\eta| <2.37$ (excluding
     $1.37<|\eta|<1.52$) selected from $Z\rightarrow\ell\ell\gamma$
     events obtained from the 2012 data sample (dots). The
     distributions for true photons from simulated
     $Z\rightarrow\ell\ell\gamma$ events (blue hatched and red
     hollow histograms) are
     also shown, after reweighting their two-dimensional \ET\ vs $\eta$
     distributions to match that of the data candidates. The blue hatched
     histogram corresponds to the uncorrected simulation and the red hollow
     one to the simulation corrected by the average shift between data
     and simulation distributions determined from the inclusive sample
     of isolated photon candidates passing the \textit{tight} selection per bin
     of ($\eta$, \ET) and for converted and unconverted photons
     separately.
     The photon candidates must be isolated 
     but no shower-shape criteria are applied. The photon
     purity of the data sample, {\em i.e.} the fraction of prompt photons,
     is estimated to be approximately 99\%.}  
   \label{fig:FF_examples} 
\end{figure}

\section{Techniques to measure the photon identification efficiency}
\label{sec:datadrivenefficiencymeasurements}

The photon identification efficiency, \effID, is defined as the ratio
of the number of isolated photons passing the \textit{tight} identification
selection to the total number of isolated photons.
Three data-driven techniques are developed in order to measure 
this efficiency for reconstructed photons over a wide transverse
momentum range. 

The \RadZ\ method uses a clean sample of prompt, isolated photons from
radiative leptonic decays of the \Zboson\ boson, $Z\rightarrow\ell\ell\gamma$,
in which a photon produced from the final-state radiation of one of
the two leptons is selected without imposing any criteria on the photon
discriminating variables.
Given the luminosity of the data collected in Run 1, this method
allows the measurement of the photon identification efficiency only
for $10~\GeV \lesssim \ET \lesssim 80$~\GeV.

In the \EE\ method, a large and pure sample of electrons selected from
$Z\rightarrow ee$ decays with a tag-and-probe technique is used to
deduce the distributions of the discriminating variables for photons 
by exploiting the similarity between the electron and the photon EM showers.
Given the typical \ET\ distribution of electrons from \Zboson\ boson
decays and the Run-1 luminosity, this method provides precise results
for $30~\GeV \lesssim \ET \lesssim 100~\GeV$.

The \MM\ uses the discrimination between prompt photons 
and background photons provided by their
isolation from tracks in the ID to extract the sample
purity before and after applying the \textit{tight} identification requirements.
This method provides results for transverse momenta from 20~\GeV\
to 1.5~\TeV. 

The three measurements are performed for photons with pseudorapidity
in the fiducial region of the EM calorimeter in which the first layer 
is finely segmented along $\eta$: $|\eta|<1.37$ or $1.52<|\eta|<2.37$.
The identification efficiency is measured as a function of $\ET$ in
four pseudorapidity intervals: $|\eta|<0.6$,
$0.6<|\eta|<1.37$, $1.52<|\eta|<1.81$ and $1.81<|\eta|<2.37$.
Since there are not many data events with high-$\ET$ photons, the highest
$\ET$ bin in which the measurement with the matrix method is
performed corresponds to the large interval 250~GeV$<\ET<1500$~GeV
(the upper limit corresponding to the transverse energy 
of the highest-$\ET$ selected photon candidate).
In this range a majority of the photon candidates have transverse
momenta below about 400~GeV (the $\ET$ distribution of the selected photon
candidates in this interval has an average value of
300~GeV and an RMS value of 70~GeV). However, from the simulation
the photon identification efficiency is expected
to be constant at the few per-mil level in this $\ET$ range.

\subsection{Photons from \Zboson\ boson radiative decays}
\label{sec:photonsfromZradiativedecays}
Radiative $\Zboson\to\ell\ell\gamma$ decays are selected by placing kinematic
requirements on the dilepton pair, the invariant mass of the three
particles in the final state and quality requirements on the two leptons.
The reconstructed photon candidates are required to be isolated in the
calorimeter but no selection is applied to their discriminating variables.

Events are collected using the lowest-threshold unprescaled
single-lepton or dilepton triggers. 

Muon candidates are formed from tracks reconstructed both in the
ID and in the muon
spectrometer~\cite{Aad:2014zya}, with transverse momentum
larger than 15~\GeV\ and pseudorapidity $|\eta|<2.4$.
The muon tracks 
are required to have at least one hit in the innermost pixel layer, 
one hit in the other two pixel layers, five hits in the SCT,
and at most two missing hits in the two silicon detectors.
The muon track isolation, defined as the sum of the transverse momenta of
the tracks inside a cone of size $\Delta R=\sqrt{(\Delta\eta)^2 +
 (\Delta\phi)^2}=0.2$  around the muon, excluding the muon track,
is required to be smaller than 10\% of the muon $\pT$.

Electron candidates are required to have $\ET > 15$~\GeV, 
and $|\eta|<1.37$ or $1.52<|\eta|<2.47$.
Electrons are required to satisfy \textit{medium} identification
criteria~\cite{Aad:2014fxa} based on tracking and transition 
radiation information from the ID, shower shape variables 
computed from the lateral and longitudinal profiles of the energy
deposited in the EM calorimeter, and track--cluster matching quantities.

For both the electron and muon candidates, the longitudinal ($z_{0}$) and
transverse ($d_0$) impact parameters of the reconstructed tracks
with respect to the primary vertex with at least three associated tracks
and with the largest $\sum p_\mathrm{T}^2$ of the associated tracks
are required to satisfy $|z_0|<10$ mm and $|d_{0}|/\sigma_{d_{0}}<10$,
respectively, where $\sigma_{d_{0}}$ is the estimated $d_0$ uncertainty.

The $\Zboson\rightarrow \ell\ell\gamma$ candidates are selected by requiring
two opposite-sign charged leptons of the same flavour satisfying the previous
criteria and one isolated photon candidate with $\ET>10$~\GeV\ and 
$|\eta|<1.37$ or $1.52<|\eta|<2.37$. 
An angular separation $\Delta R>0.2$ (0.4) between the photon and each 
of the two muons (electrons) is required so that the energy
deposited by the leptons in the calorimeter does not bias the photon discriminating
variables.
In the selected events, the triggering leptons are required to match
one (or in the case of dilepton triggered events, both) of the $Z$ 
candidate's leptons.

The two-dimensional distribution of the dilepton invariant mass, 
$m_{\ell\ell}$, versus the invariant mass of the three final-state 
particles, $m_{\ell\ell\gamma}$, in events selected in $\sqrt{s}=8$
\TeV\ data is shown in Fig.~\ref{fig:2D_Zllgamma_invaraint_mass}.
\begin{figure}[!htpb]
  \centering
      {\includegraphics[width=0.7\columnwidth]{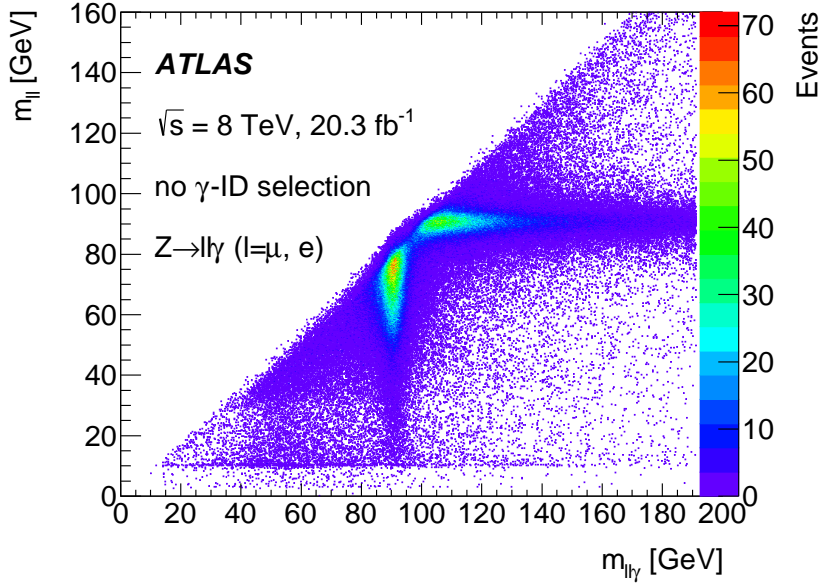}}
      \caption{Two-dimensional distribution of $m_{\ell\ell\gamma}$ and
        $m_{\ell\ell}$ for all reconstructed
        $\Zboson\rightarrow\ell\ell\gamma$ candidates after loosening
        the selection applied to $m_{\ell\ell\gamma}$ and
        $m_{\ell\ell}$. 
        No photon identification requirements are applied.
        Events from initial-state ($m_{\ell\ell}\approx m_Z$) 
        and final-state ($m_{\ell\ell\gamma}\approx m_Z$)
        radiation are clearly visible.}
      \label{fig:2D_Zllgamma_invaraint_mass}
\end{figure}
The selected sample is dominated by \Zboson+jet background events
in which one jet is misreconstructed as a photon.
These events, which have a cross section about three orders of
magnitudes higher than $\ell\ell\gamma$ events,
have $m_{\ell\ell}\approx m_Z$ and
$m_{\ell\ell\gamma}\gtrsim m_Z$, while final-state radiation
$\Zboson\to\ell\ell\gamma$ events have $m_{\ell\ell}\lesssim m_Z$
and $m_{\ell\ell\gamma}\approx m_Z$, where $m_Z$ is the \Zboson\ boson 
pole mass. To significantly reduce the \Zboson+jet background,
the requirements of $40~\GeV <m_{\ell\ell}<83~\GeV$ and $80~\GeV <m_{\ell\ell\gamma}<96~\GeV$ are thus applied.

After the selection, about 54000 unconverted and about 19000
converted isolated photon candidates are selected in the
$\Zboson\rightarrow\mu\mu\gamma$ channel, and 32000 unconverted and 
12000 converted isolated photon candidates are selected 
in the $\Zboson\rightarrow ee\gamma$ channel.  
The residual background contamination from $Z$+jet events is
estimated through a maximum-likelihood fit
(called ``template fit'' in the following)
to the $m_{\ell\ell\gamma}$ distribution of selected events after
discarding the $80~\GeV <m_{\ell\ell\gamma}<96~\GeV$ requirement. 
The data are fit to a sum of the photon and background contributions.
The photon and background $m_{\ell\ell\gamma}$ distributions (``templates'')
are extracted from the $\Zboson\to\ell\ell\gamma$ and \Zboson+jet simulations,
corrected to take into account known data--MC differences in the
photon and lepton energy scales and resolution and in the lepton
efficiencies.
The signal and background yields are determined from the data by maximising
the likelihood.
Due to the small number of selected events in data and simulation,
these fits are performed only for two photon transverse momentum
intervals, $10~\GeV <\ET<15~\GeV$ and $\ET>15$~\GeV, and integrated
over the photon pseudorapidity, since the signal purity is found to
be similar in the four photon $|\eta|$ intervals within statistical
uncertainties.

Figure~\ref{fig:template_fit_ET_10_15_mu} shows the result of the 
fit for unconverted photon candidates with transverse momenta between
10~\GeV\ and 15~\GeV. 
\begin{figure}[!htbp]
    \centering
    {\includegraphics[width=0.7\columnwidth]{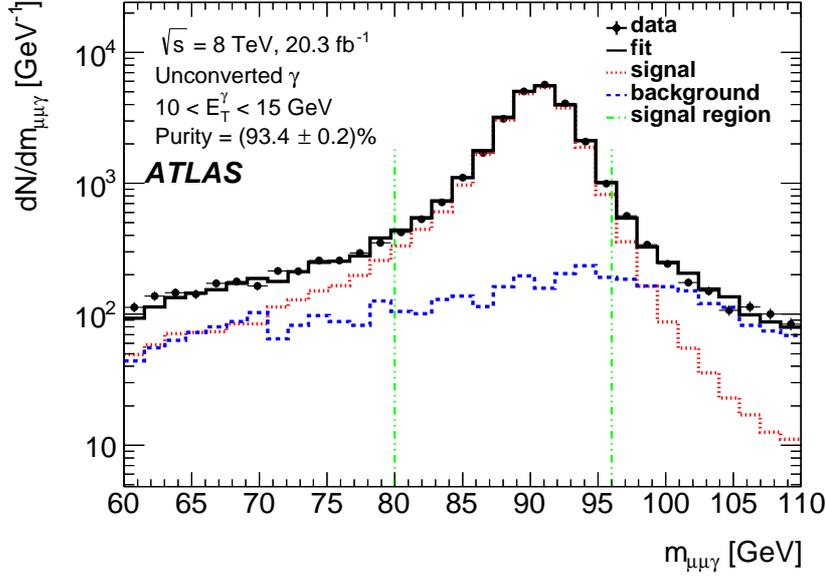}}
    \caption{Invariant mass ($m_{\mu\mu\gamma}$) distribution of events
      in which the unconverted photon has $10~\GeV<\ET<15$~\GeV,
      selected in data at $\sqrt{s}= 8$~\TeV\ after applying all the $\Zboson\to\mu\mu\gamma$ selection
      criteria except that on $m_{\mu\mu\gamma}$ (black dots).
      No photon identification requirements are applied.
      The solid black line represents the result of fitting the data
      distribution to a sum of the signal (red dashed line) and
      background (blue dotted line) invariant mass distributions obtained
      from simulations.}
    \label{fig:template_fit_ET_10_15_mu}
\end{figure}
The fraction of residual background in the region 
$80~\GeV<m_{\ell\ell\gamma}<96$~\GeV\ decreases rapidly with the reconstructed 
photon transverse momentum, from $\approx$ 10\% for
$10~\GeV<\ET<15$~\GeV\ to $\le$ 2\% for higher-$\ET$ regions. 
A similar fit is also performed for the subsample in which the photon
candidates are required to satisfy the \textit{tight} identification criteria. 

The identification efficiency as a function of $\ET$ is estimated 
as the fraction of all the selected probes in a certain $\ET$ interval 
passing the \textit{tight} identification requirements. 
For $10~\GeV<\ET<15$~\GeV, both the numerator and denominator are corrected 
for the average background fraction determined from the template fit.
For $\ET>15$~\GeV, the background is neglected in the nominal result,
and a systematic uncertainty is assigned as the difference between 
the nominal result and the efficiency that would be obtained taking
into account the background fraction determined from the template fit
in this $\ET$ region.
Additional systematic uncertainties related to the signal and background 
$m_{\ell\ell\gamma}$ distributions are estimated by repeating the previous 
fits with templates extracted from alternative MC event generators
(POWHEG interfaced to PHOTOS and PYTHIA8 for $Z\to\ell\ell\gamma$ and
ALPGEN for $Z$+jet, $Z\to\ell\ell$). 
The total relative uncertainty in the efficiency, dominated by the 
statistical component, increases from 1.5--3\% (depending on $\eta$ and 
whether the photon was reconstructed as a converted or an unconverted
candidate) for $10~\GeV<\ET<15$~\GeV\ to 5--20\% for $\ET>40$~\GeV.

\subsection{Electron extrapolation}
\label{sec:electronextrapolation}
The similarity between the electromagnetic showers induced by
isolated electrons and photons in the EM calorimeter is exploited to
extrapolate the expected photon distributions of the discriminating
variables. The photon identification efficiency is thus estimated from the
distributions of the same variables in a pure and large
sample of electrons with $\ET$ between 30~\GeV\ and 100~\GeV\ obtained
from $\Zboson\rightarrow ee$ decays using a tag-and-probe
method~\cite{Aad:2014fxa}.
Events collected with single-electron triggers are selected if
they contain two opposite-sign electrons with $\ET>25$~\GeV,
$|\eta|<1.37$ or $1.52<|\eta|<2.47$,
at least one hit in the pixel detector and seven hits in the silicon
detectors, $\Etiso<4$~\GeV\ and invariant mass $80~\GeV<m_{ee}<120$~\GeV.
The tag electron is required to match the trigger object and
to pass the \textit{tight} electron identification requirements.
A sample of about $9\times 10^6$ electron probes is collected.
Its purity is determined from the $m_{ee}$ spectrum of the selected events
by estimating the background, whose normalisation
is extracted using events with $m_{ee}>120$~\GeV\  
and whose shape is obtained from events in which the  
probe electron candidate fails both the isolation and identification
requirements.
The purity varies slightly with $\ET$ and $|\eta|$, but is always
above 99\%.

The differences between the photon and electron distributions of
the discriminating variables are studied using simulated samples of
prompt photons and electrons from $\Zboson\to ee$ decays, separately
for converted and unconverted photons. 
The shifts of the photon discriminating variables described in
Sect.~\ref{sec:dataset} are not applied, since it is observed that the
photon and electron distributions are biased in a similar way in the
simulation.

Photon conversions produce electron--positron pairs which
are usually sufficiently collimated to produce overlapping showers in
the calorimeter, giving rise to single clusters with distributions of the
discriminating variables similar to those of an isolated electron. The
largest differences between electrons and converted photons are found
in the $\Rphi$ distribution, due to the bending of electrons and
positrons in opposite directions in the 
$r$--$\phi$ plane, which leads to a broader $\Rphi$ distribution for
converted photons. However, the $\Rphi$ requirement used for the
identification of converted photons is relatively loose, and a
test on MC simulated samples shows that, by directly applying the
converted photon identification criteria to an electron sample, the
\effID\ obtained from electrons overestimates the efficiency
for converted photons by at most 3\%.

The showers induced by unconverted photons are more likely to begin 
later than those induced by electrons, and thus to be narrower in the first
layer of the EM calorimeter.  Additionally, the lack of
photon-trajectory bending in the $\phi$
plane makes the $\Rphi$ distribution particularly different from that
of electrons. Therefore, if the unconverted-photon selection criteria
are directly applied to an electron sample, the \effID\ obtained from
these electrons is about 20--30\% smaller than the efficiency for
unconverted photons with the same pseudorapidity and transverse momentum.

To reduce such effects a mapping technique based on a Smirnov
transformation~\cite{RandomGeneration} is used for both
the unconverted and converted photons.
For each discriminating variable $x$, the cumulative distribution
functions (CDF) of simulated electrons and photons,
$\mathrm{CDF}_e(x)$ and $\mathrm{CDF}_\gamma(x)$, are calculated.  
The transform $f(x)$ is thus defined by $\mathrm{CDF}_e(x) =
\mathrm{CDF}_\gamma(f(x))$.
The discriminating variable of the electron probes selected in data
can then be corrected on an event-by-event basis by applying the
transform $f(x)$ to obtain the expected one for photons in data.
Figure~\ref{fig:smirnov-example} illustrates the
process for one shower shape ($R_{\phi}$).
These Smirnov transformations are invariant under systematic shifts which
are fully correlated between the electron and photon distributions.
Due to the differences in the $|\eta|$ and \ET\ distributions
of the source and target samples, the dependence of the shower shapes
on $|\eta|$, \ET, and whether the photon was reconstructed as
a converted or an  unconverted candidate, this process is applied separately
for converted and unconverted photons, and in various regions of
\ET\ and $|\eta|$. 
The efficiency of the identification criteria is determined
from the extrapolated photon distributions of the discriminating
variables.
\begin{figure}[!htbp] 
  \centering
  \subfloat[]{\label{fig:smirnov-example_a}\includegraphics[width=.5\columnwidth]{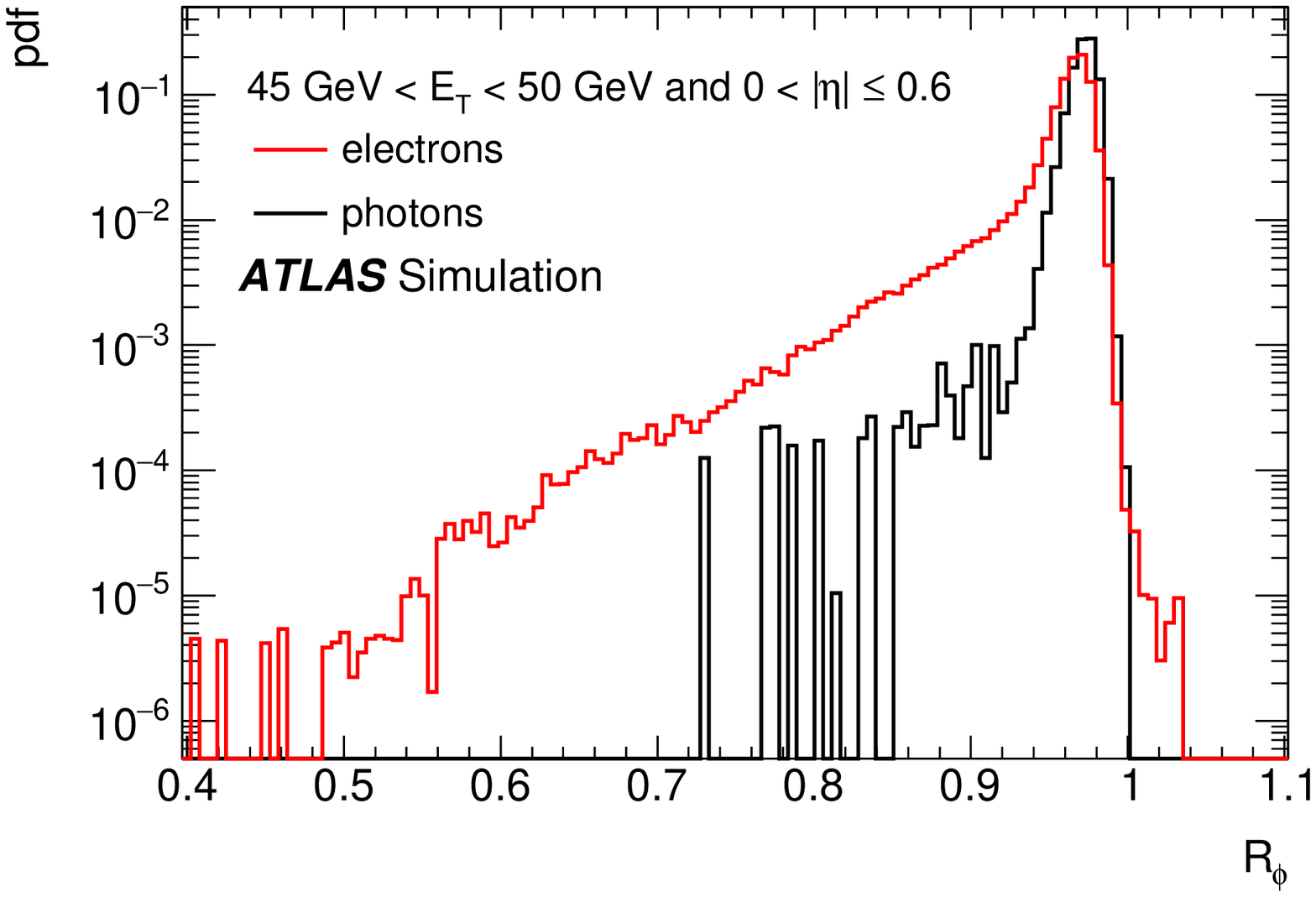}}
  \subfloat[]{\label{fig:smirnov-example_b}\includegraphics[width=.5\columnwidth]{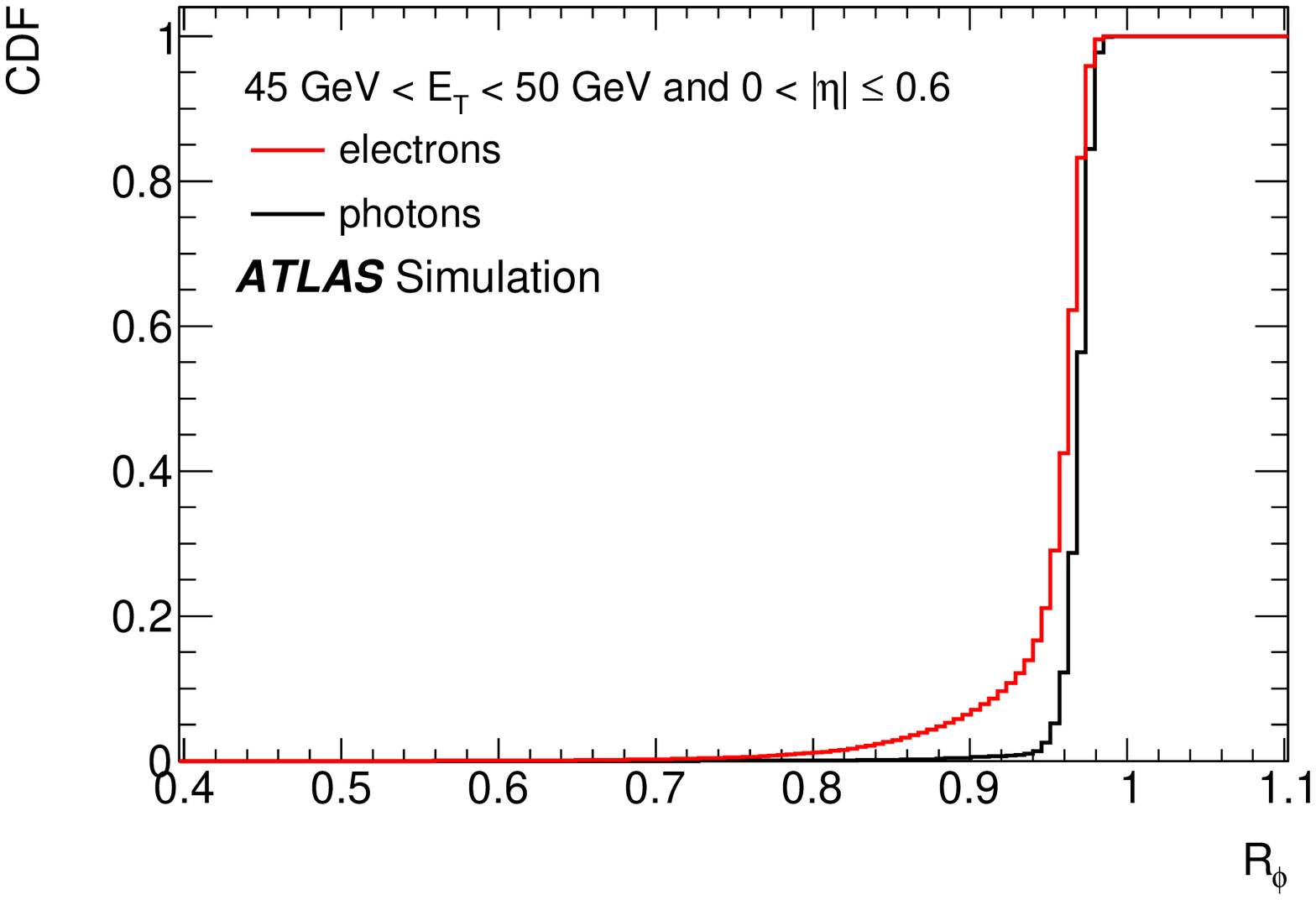}}
  
  \subfloat[]{\label{fig:smirnov-example_c}\includegraphics[width=.5\columnwidth]{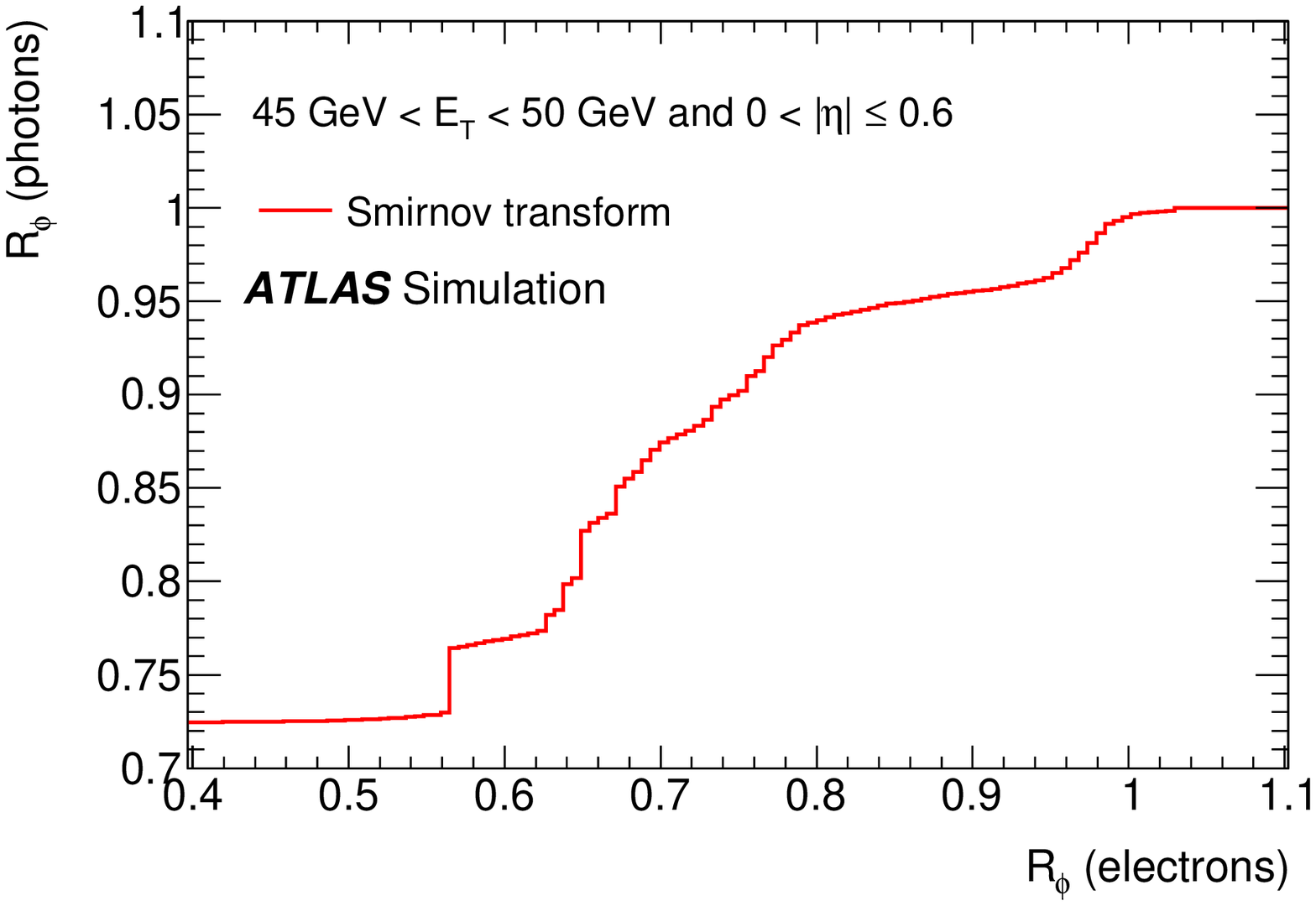}}
  \subfloat[]{\label{fig:smirnov-example_d}\includegraphics[width=.5\columnwidth]{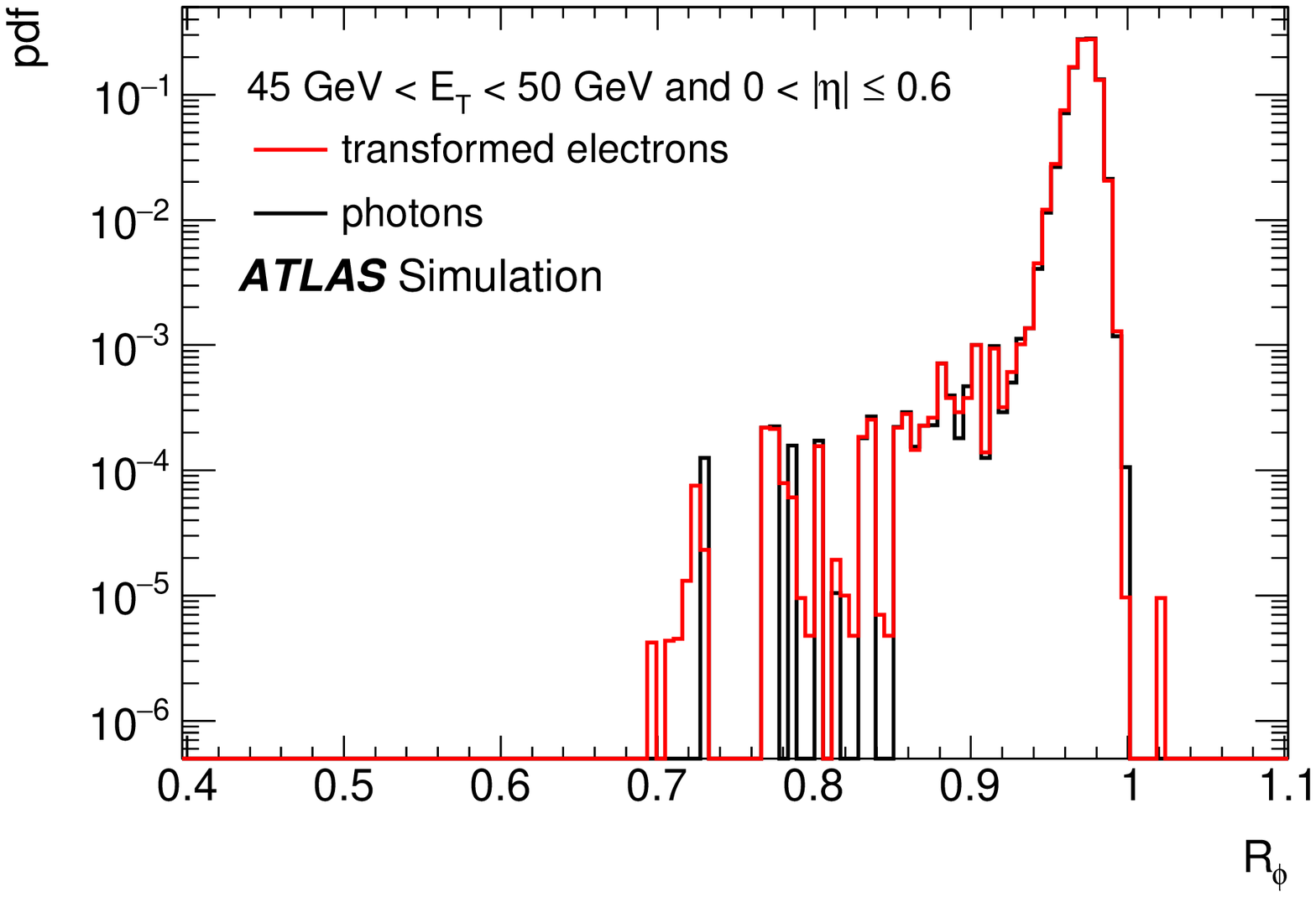}}
  \caption{Diagram illustrating the process of Smirnov
    transformation. $R_{\phi}$ is chosen as an example discriminating
    variable whose distribution is particularly different between
    electrons and (unconverted) photons. The $R_{\phi}$ probability
    density function (pdf) in each sample (a) is used to 
    calculate the respective CDF (b). From the 
    two CDFs, a Smirnov transformation can be derived (c).
    Applying the transformation leads to an $R_{\phi}$ 
    distribution of the transformed electrons which closely
    resembles the photon distribution (d).} 
  \label{fig:smirnov-example}
\end{figure}

The following three sources of systematic uncertainty are considered
for this analysis:
\begin{itemize}
\item
  As the Smirnov transformations are obtained independently for each shower
  shape, the estimated photon identification efficiency can be biased
  if the correlations among the discriminating variables are
  significantly different between electrons and photons.
  Non-closure tests are performed on the simulation, comparing the
  identification efficiency of true prompt photons with the efficiency
  extrapolated from electron probes selected with the same requirement
  as in data and applying the extrapolation procedure. The differences
  between the true and extrapolated efficiencies are at the level
  of 1\% or less, with a few exceptions for unconverted photons, for
  which maximum differences of 2\% are found.
\item
  The results are also affected by the uncertainties in the modelling of
  the shower shape distributions and correlations in the photon and
  electron simulations used to extract the mappings. The largest
  uncertainties in the distributions of the discriminating variables
  originate from limited knowledge of the material upstream of the
  calorimeter. The extraction 
  of the mappings is repeated using alternative MC samples based on a
  detector simulation with a conservative estimate of additional
  material in front of the calorimeter~\cite{Aad:2011mk}. This detector
  simulation is considered as conservative enough to cover any
  mismodelling of the distributions of the discriminating variables.
  The extracted $\effID$ differs from the nominal one by typically less
  than 1\% for converted photons and 2\% for unconverted ones, with
  maximum deviations of 2\% and 3.5\% in the worst cases, respectively. 
\item
  Finally, the effect of a possible background contamination in the
  selected electron probes in data is found to be smaller than 0.5\% in
  all $\ET$, $|\eta|$ intervals for both the converted and unconverted
  photons.
\end{itemize}
The total uncertainty is dominated by its systematic component and
ranges from 1.5\% in the central region to 7.5\% in the highest
\ET\ bin in the endcap region, with typical values of 2.5\%.

\subsection{Matrix method}
\label{sec:matrixmethod}
An inclusive sample of about $7\times 10^6$ isolated photon candidates is
selected using single-photon triggers by requiring at least one photon
candidate with transverse momentum 20~\GeV~$<\et<1500$~\GeV\ and
isolation energy $\Etiso<4$~\GeV, matched to the photon trigger
object passing the \textit{loose} identification requirements.

The distribution of the track isolation of selected candidates in data is
used to discriminate between prompt and background photon candidates,
before and after applying the \textit{tight} identification criteria.
The track isolation variable used for the measurement of the efficiency of
unconverted photon candidates, $p_\mathrm{T}^\mathrm{iso}$, 
is defined as the scalar sum of the transverse momenta of the tracks,
with transverse momentum above 0.5~\GeV\ and distance of closest
approach to the primary vertex along $z$ less than 0.5 mm, within
a hollow cone of $0.1<\Delta R<0.3$ around the photon direction.
For the measurement of the efficiency of the converted photon candidates, the track
isolation variable $\nu_\mathrm{trk}^\mathrm{iso}$ is defined as the number
of tracks, passing the previous requirements, within a hollow cone of
$0.1<\Delta R<0.4$ around the photon direction.
Unconverted photon candidates with $p_\mathrm{T}^\mathrm{iso}<1.2$~\GeV\
and converted photon candidates with $\nu_\mathrm{trk}^\mathrm{iso}=0$
are considered to be isolated from tracks.
The track isolation variables and requirements were chosen to minimise
the total uncertainty in the identification efficiency after including
both the statistical and systematic components.

The yields of prompt and background photons in the selected sample (``ALL'' sample),
$N^\mathrm{S}_\mathrm{all}$ and $N^\mathrm{B}_\mathrm{all}$, and in the sample of candidates satisfying
the {\em tight} identification criteria (``PASS'' sample),
$N^\mathrm{S}_\mathrm{pass}$ and  $N^\mathrm{B}_\mathrm{pass}$,
are obtained by solving a system of four equations:
\begin{eqnarray}
N^\mathrm{T}_\mathrm{all} &=& N^\mathrm{S}_\mathrm{all} + N^\mathrm{B}_\mathrm{all} \label{eq:NT_all} \nonumber ,\\
N^\mathrm{T}_\mathrm{pass} &=& N^\mathrm{S}_\mathrm{pass} + N^\mathrm{B}_\mathrm{pass} \label{eq:NT_pass} \nonumber , \\
N^\mathrm{T,iso}_\mathrm{all} &=& \varepsilon^\mathrm{S}_\mathrm{all}\times N^\mathrm{S}_\mathrm{all} + \varepsilon^\mathrm{B}_\mathrm{all}\times N^\mathrm{B}_\mathrm{all} \label{eq:NI_all} \nonumber ,\\
N^\mathrm{T,iso}_\mathrm{pass} &=& \varepsilon^\mathrm{S}_\mathrm{pass}\times N^\mathrm{S}_\mathrm{pass} + \varepsilon^\mathrm{B}_\mathrm{pass}\times N^\mathrm{B}_\mathrm{pass} \label{eq:NI_pass} .
\end{eqnarray}
Here $N^\mathrm{T}_\mathrm{all}$ and $N^\mathrm{T}_\mathrm{pass}$ are the total numbers of
candidates in the ALL and PASS samples respectively, while
$N^\mathrm{T,iso}_\mathrm{all}$ and $N^\mathrm{T,iso}_\mathrm{pass}$ are
the numbers of candidates in the ALL and PASS samples
that pass the track isolation requirement. The quantities
$\varepsilon_\mathrm{all}^\mathrm{S(B)}$ and
$\varepsilon_\mathrm{pass}^\mathrm{S(B)}$ are the efficiencies of the track isolation
requirements for prompt (background) photons in the ALL and PASS samples.

Equation~(\ref{eq:NI_pass}) implies that the fractions $f_\mathrm{pass}$ and
$f_\mathrm{all}$ of prompt photons in the ALL and in the PASS samples can be written as:
\begin{eqnarray}
  f_\mathrm{pass} &=& \frac{\varepsilon_\mathrm{pass}-\varepsilon_\mathrm{pass}^\mathrm{B}}{\varepsilon_\mathrm{pass}^\mathrm{S}-\varepsilon_\mathrm{pass}^\mathrm{B}}\nonumber\\
  f_\mathrm{all} &=& \frac{\varepsilon_\mathrm{all}-\varepsilon_\mathrm{all}^\mathrm{B}}{\varepsilon_\mathrm{all}^\mathrm{S}-\varepsilon_\mathrm{all}^\mathrm{B}}
 \label{eq:effID_matrix_purities}
\end{eqnarray}
where $\varepsilon_\mathrm{pass(all)} = N^\mathrm{T,iso}_\mathrm{pass(all)}/N^\mathrm{T}_\mathrm{pass(all)}$ is the fraction of \textit{tight} (all) photon candidates in data that satisfy the track isolation criteria. 

The identification efficiency $\effID = N^\mathrm{S}_\mathrm{pass}/N^\mathrm{S}_\mathrm{all}$ is thus:
\begin{equation}
  \effID = \frac{N^\mathrm{T}_\mathrm{pass}}{N^\mathrm{T}_\mathrm{all}}\left(\frac{\varepsilon_\mathrm{pass}-\varepsilon_\mathrm{pass}^\mathrm{B}}{\varepsilon_\mathrm{pass}^\mathrm{S}-\varepsilon_\mathrm{pass}^\mathrm{B}}\right)\left(\frac{\varepsilon_\mathrm{all}-\varepsilon_\mathrm{all}^\mathrm{B}}{\varepsilon_\mathrm{all}^\mathrm{S}-\varepsilon_\mathrm{all}^\mathrm{B}}\right)^{-1}.
 \label{eq:effID_matrix_solution}
\end{equation}

The prompt-photon track isolation efficiencies, 
$\varepsilon_\mathrm{all}^\mathrm{S}$ and $\varepsilon_\mathrm{pass}^\mathrm{S}$, are estimated from
simulated prompt-photon events. The difference between
the track isolation efficiency for electrons collected in data and
simulation with a tag-and-probe $\Zboson\to ee$ selection is taken as a
systematic uncertainty. An additional systematic uncertainty in 
the prompt-photon track isolation efficiencies is estimated by
conservatively varying the fraction of fragmentation photons in the
simulation by $\pm 100\%$. The overall uncertainties in
$\varepsilon_\mathrm{all}^\mathrm{S}$ and $\varepsilon_\mathrm{pass}^\mathrm{S}$ are below 1\%. 

The background-photon track isolation efficiencies,
$\varepsilon^\mathrm{B}_\mathrm{all}$ and $\varepsilon^\mathrm{B}_\mathrm{pass}$, are
estimated from data samples enriched in background 
photons. For the measurement of $\varepsilon^\mathrm{B}_\mathrm{all}$,
the control sample of all photon candidates not meeting at least one
of the \textit{tight} identification criteria is used.
In order to obtain $\varepsilon_\mathrm{pass}^\mathrm{B}$, a relaxed version
of the \textit{tight} identification criteria is defined. 
The \textit{relaxed tight} selection consists of those candidates
which fail at least one of the requirements on four discriminating
variables computed from the energy in the cells of the first EM
calorimeter layer ($F_\mathrm{side}$, $w_{s3}$, $\Delta E$,
$E_\mathrm{ratio}$), but satisfy the remaining \textit{tight}
identification criteria. 
The four variables which are removed from the \textit{tight} selection to
define the \textit{relaxed tight} one are computed from the energy
deposited in a few strips of the first compartment of the LAr EM
calorimeter near the one with the largest deposit and are chosen
since they have negligible correlations with the photon isolation. 
Due to the very small correlation (few \%) between the track isolation
and these discriminating variables, the background-photon track
isolation efficiency is similar for photons satisfying \textit{tight}
or \textit{relaxed tight} criteria.
The differences between the track isolation efficiencies for
background photons satisfying \textit{tight} or \textit{relaxed tight}
criteria are included in the systematic uncertainties.
The contamination from prompt photons in the background enriched samples
is accounted for in this procedure by using as an additional input the fraction 
of signal events passing or failing the \textit{relaxed tight} requirements, as
determined from the prompt-photon simulation. 
The fraction of prompt photons in the
background control samples decreases from about 20\% 
to 1\%, with increasing photon transverse momentum.
The whole procedure is tested with a simulated sample of $\gamma+$jet
and dijet events, and the difference between the true track isolation
efficiency for background photons and the one estimated with this 
procedure is taken as a systematic uncertainty.
An additional systematic uncertainty, due to the use of the
prompt-photon simulation to estimate the fraction of signal photons in the
background control regions, is estimated by re-calculating these
fractions using alternative MC samples based on a detector simulation
with a conservative estimate of additional material in front of the
calorimeter.
The typical total relative uncertainty in the
background-photon track isolation efficiency is 2--4\%.

As an example, Fig.~\ref{fig:mm-trackisoeff} shows the track isolation
efficiencies as a function of $\ET$ for prompt and background
unconverted photon candidates with $|\eta|<0.6$ in the ALL and 
PASS samples, as well as the fractions of all or \textit{tight} photon
candidates in data that satisfy the track isolation criteria. 
From these measurements the photon identification efficiency is derived,
according to Eq.~(\ref{eq:effID_matrix_solution}).
The track isolation efficiency for prompt-photon candidates is 
essentially independent of the photon transverse momentum.
For background candidates, the track isolation efficiency initially
decreases with $\ET$, since candidates with larger $\ET$ are produced
from more energetic jets, which are therefore characterised by a larger
number of tracks near the photon candidate. At higher transverse
energies, typically above 200~\GeV, the boost of such tracks causes
some of them to fall within the inner cone $(\Delta R<0.1)$ of the
isolation cone around the photon and the isolation efficiency
for background candidates therefore increases.

The total systematic uncertainty decreases with the transverse energy.
It reaches 6\% below 40~\GeV, and amounts to 0.5--1\% at higher \ET,
where the contribution of this method is the most important. 

\begin{figure}[!htpb]
  \centering
  \subfloat[]{\label{fig:mm-trackisoeff_a}\includegraphics[width=.5\textwidth]{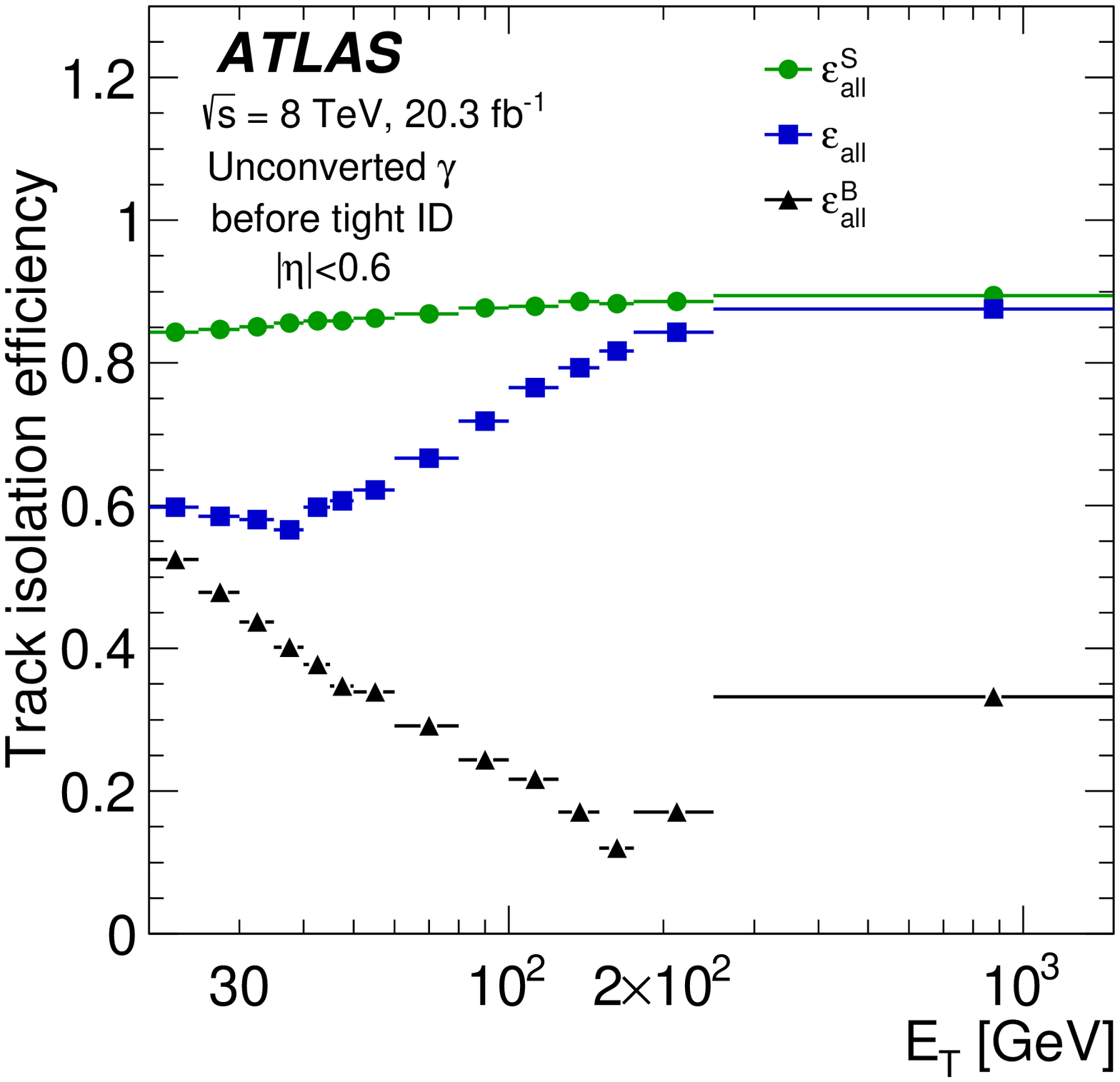}}
  \subfloat[]{\label{fig:mm-trackisoeff_b}\includegraphics[width=.5\textwidth]{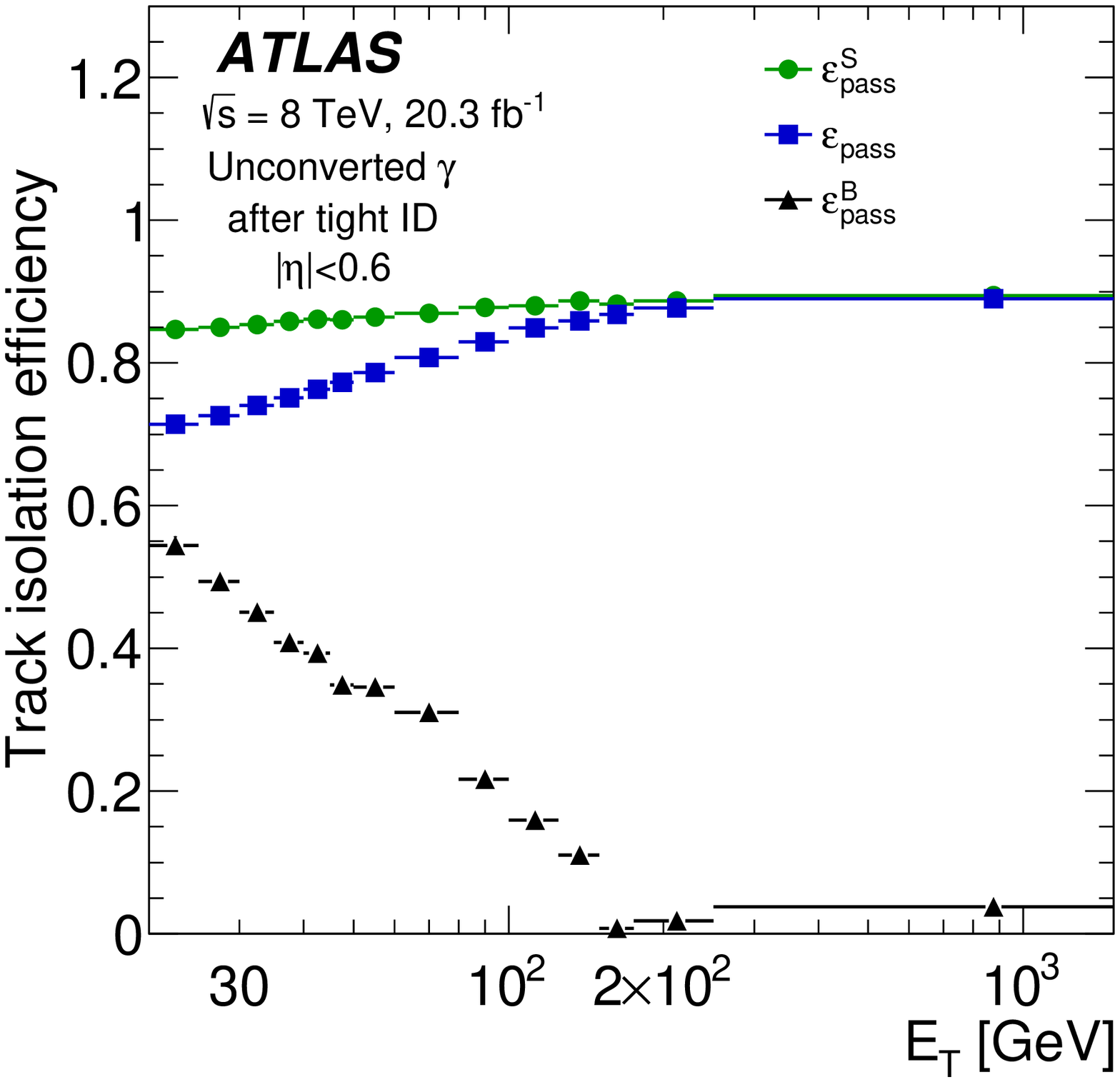}}
  \caption{Track isolation efficiencies as a function of $\ET$
    for unconverted prompt
    (green circles) and background (black triangles) photon candidates
    within $|\eta|< 0.6$ in (a) the inclusive sample or (b) passing
    \textit{tight} identification requirements.
    The efficiencies are estimated combining the simulation and
    data control samples.
    The blue square markers show the track isolation efficiency
    for candidates selected in data.} 
  \label{fig:mm-trackisoeff}
\end{figure}

The final result is obtained by multiplying the measured efficiency
by a correction factor which takes into account the preselection of
the sample using photon triggers, which already apply some  
loose requirements to the photon discriminating variables.
The correction factor, equal to the ratio of the \textit{tight}
identification efficiency for all reconstructed photons to that for
photons matching the trigger object that triggers the event, is
obtained from a corrected simulation of photon+jet events.
This correction is slightly lower than unity, by less
than 1\% for $\ET>50$ \GeV\ and by 2--3\% for \ET\ = 20~\GeV.
The systematic uncertainty from this correction is negligible
compared to the other sources of uncertainty.

\section{Photon identification efficiency results at $\sqrt{s}=8$~\TeV}
\label{sec:pidresults}

\subsection{Efficiencies measured in data}
\label{segc:datadrivenefficiencycomparison}

The identification efficiency measurements for $\sqrt{s}=8$~\TeV\ obtained
from the three data-driven methods discussed in the previous section
are compared in Figs.~\ref{fig:dd_eff_unconv_barrel_and_endcap}
and~\ref{fig:dd_eff_conv_barrel_and_endcap}. 
The $Z\to ee\gamma$ and $Z\to \mu\mu\gamma$ results agree within
uncertainties and are thus combined, following a procedure described
in the next section, and only the combined values are shown
in the figures.
In a few \ET\ bins in which the central values of
the $Z\to ee\gamma$ and the $Z\to \mu\mu\gamma$ results differ by more than
the combined uncertainty, the latter is increased to cover
the full difference between the two results.

\begin{figure}[!ht]
   \centering
   \subfloat[]{\label{fig:dd_eff_unconv_barrel_a}\includegraphics[width=.5\textwidth]{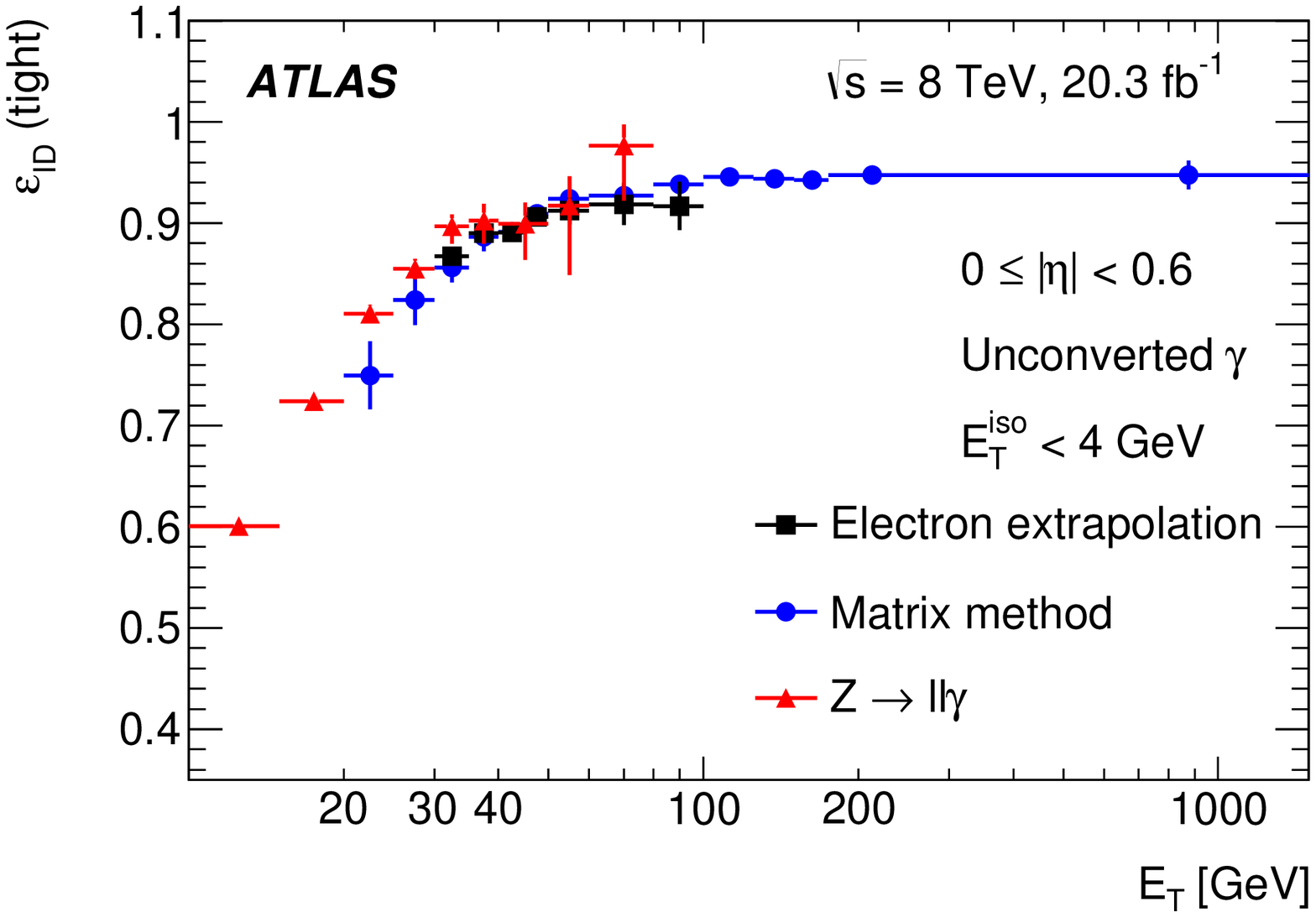}}
   \subfloat[]{\label{fig:dd_eff_unconv_barrel_b}\includegraphics[width=.5\textwidth]{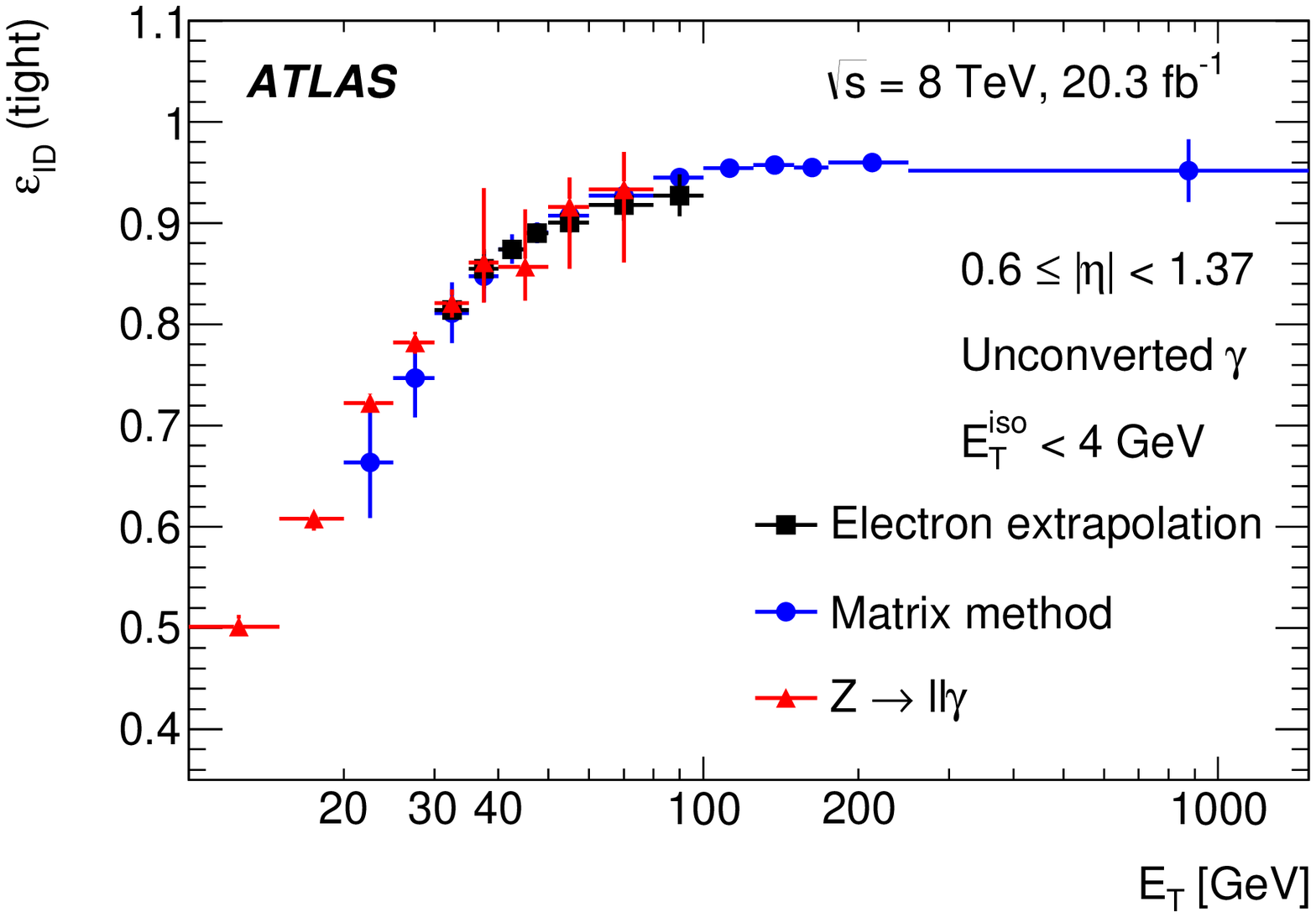}}\\
   \subfloat[]{\label{fig:dd_eff_unconv_endcap_a}\includegraphics[width=.5\textwidth]{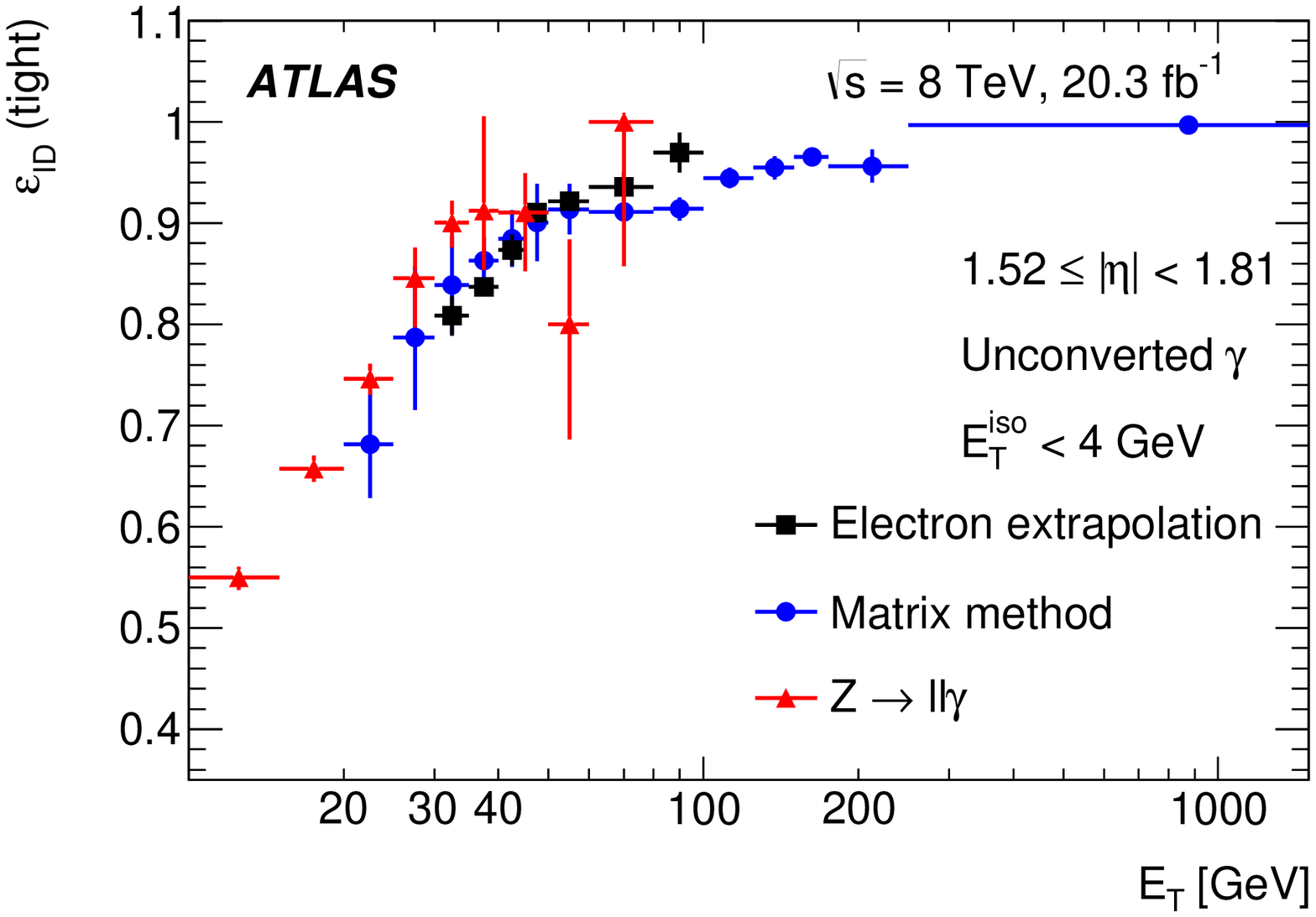}}
   \subfloat[]{\label{fig:dd_eff_unconv_endcap_b}\includegraphics[width=.5\textwidth]{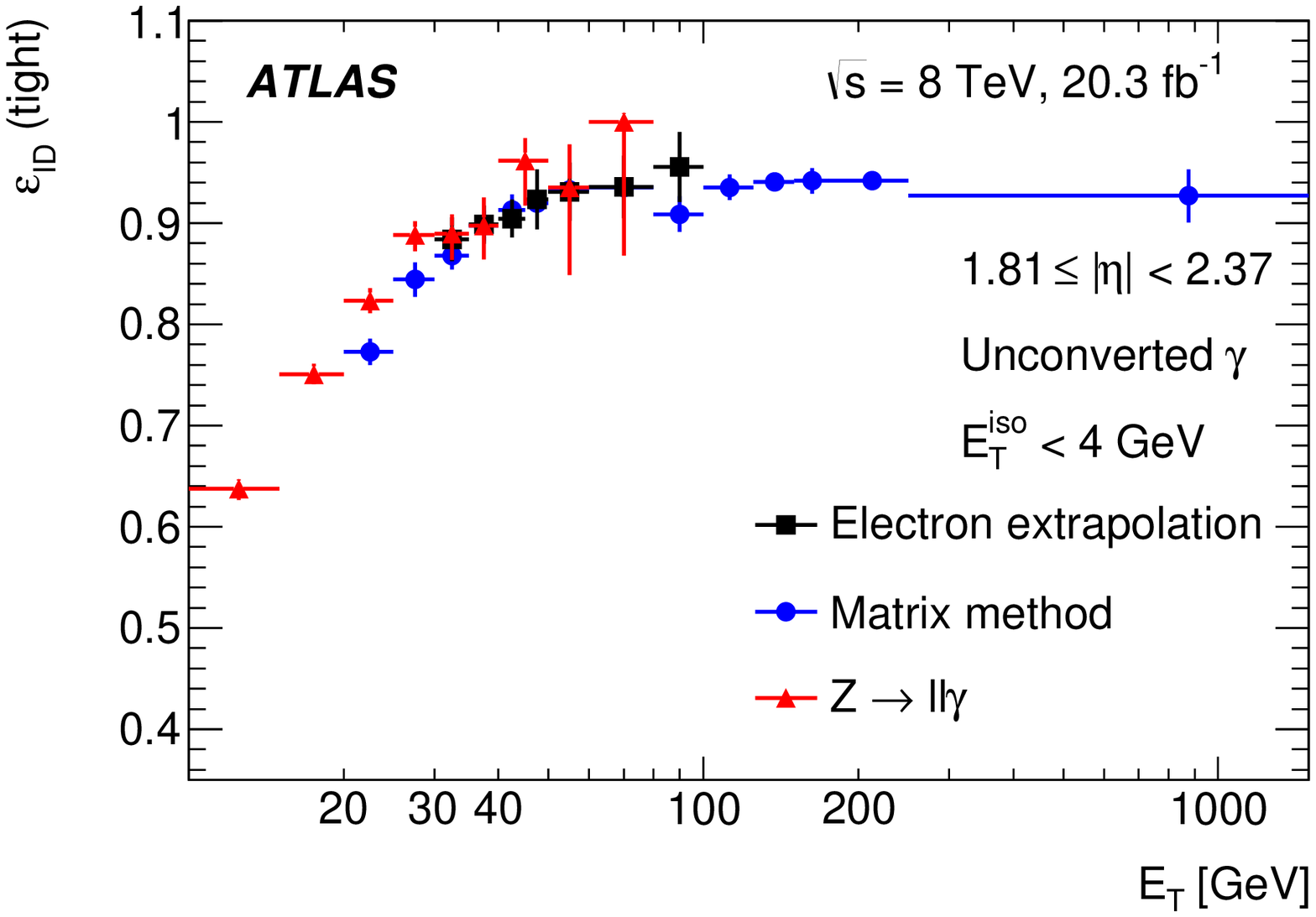}}
   \caption{Comparison of the data-driven measurements of the
     identification efficiency for unconverted photons as a function of $\ET$
     in the region $10~\GeV < \ET < 1500~\GeV$, for the four
     pseudorapidity intervals (a) $|\eta|<0.6$, (b) $0.6\leq|\eta|<1.37$,
     (c) $1.52\leq|\eta|<1.81$, and (d) $1.81\leq|\eta|<2.37$.
     The error bars represent the sum in quadrature
     of the statistical and systematic uncertainties estimated in each
     method.
} 
   \label{fig:dd_eff_unconv_barrel_and_endcap} 
\end{figure}

\begin{figure}[!ht] 
   \centering
   \subfloat[]{\label{fig:dd_eff_conv_barrel_a}\includegraphics[width=.5\textwidth]{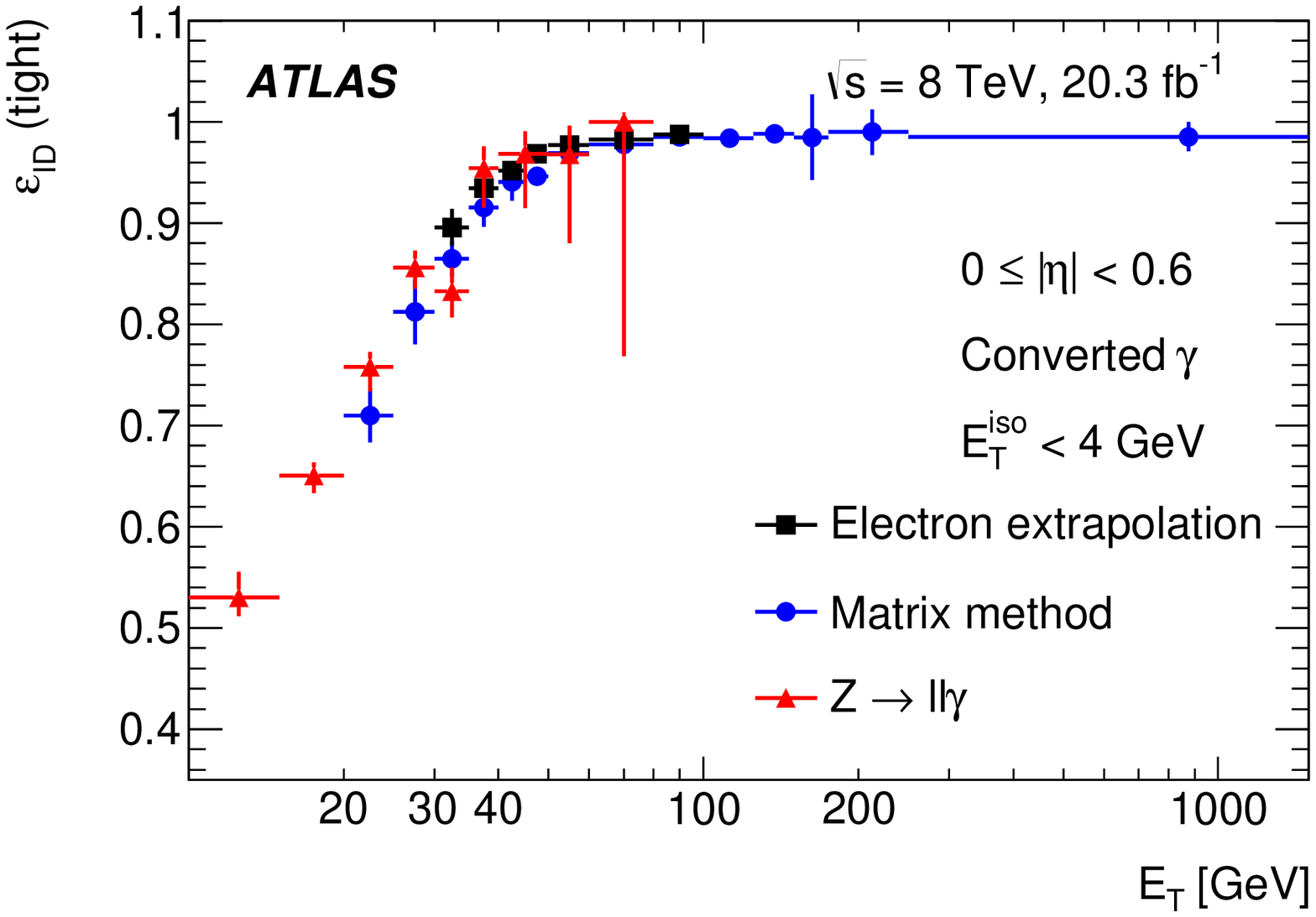}}
   \subfloat[]{\label{fig:dd_eff_conv_barrel_b}\includegraphics[width=.5\textwidth]{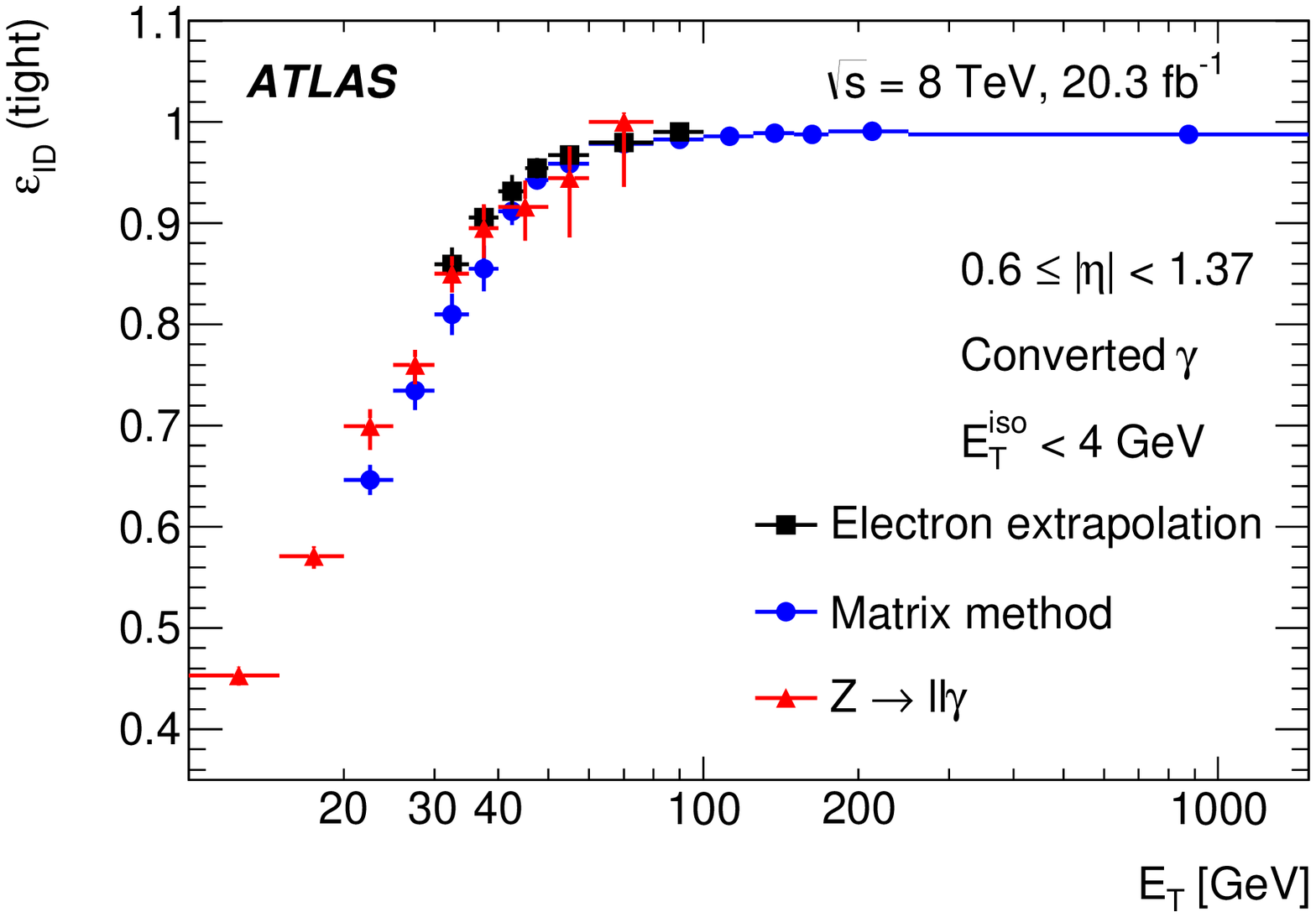}} \\
   \subfloat[]{\label{fig:dd_eff_conv_endcap_a}\includegraphics[width=.5\textwidth]{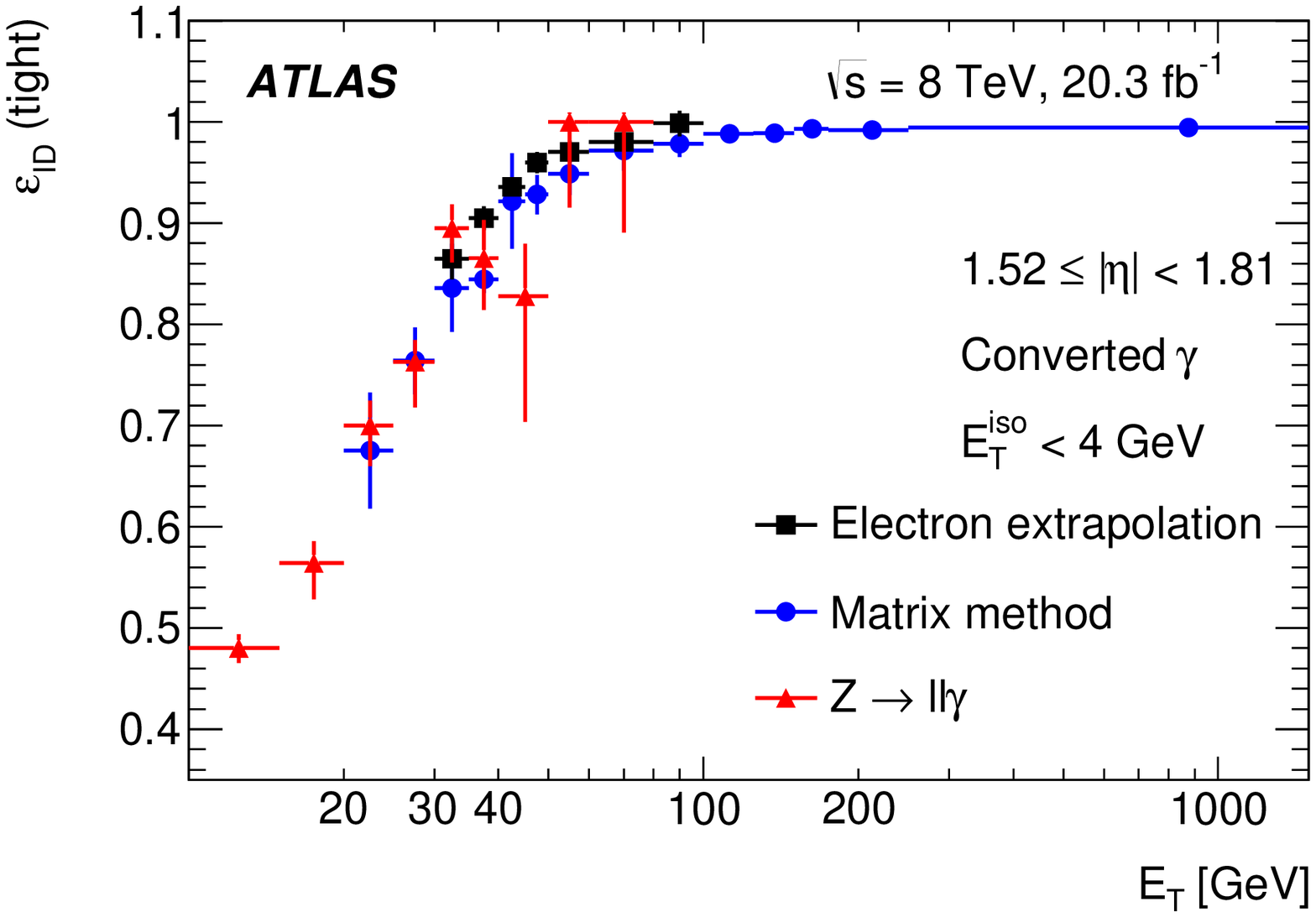}}
   \subfloat[]{\label{fig:dd_eff_conv_endcap_b}\includegraphics[width=.5\textwidth]{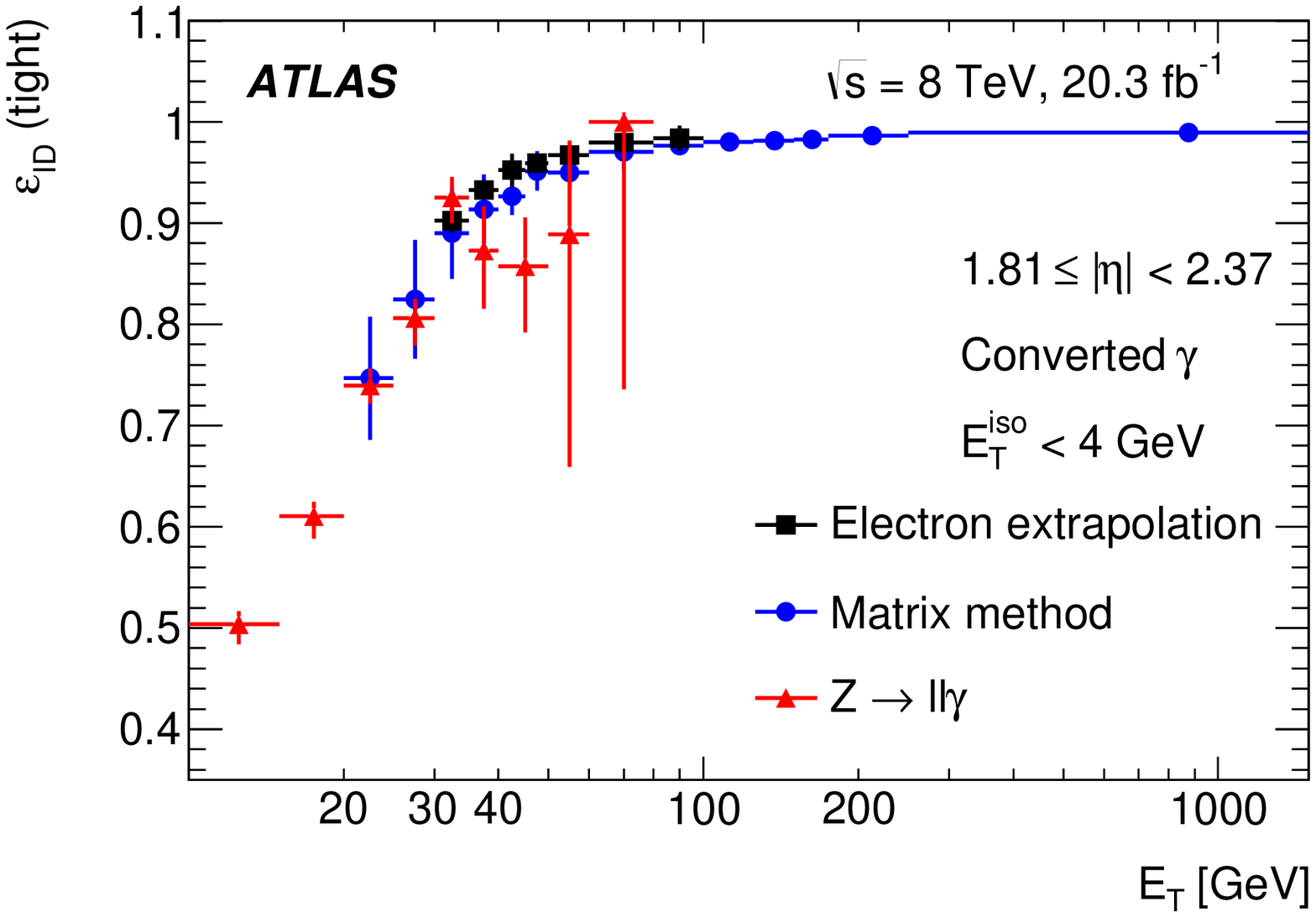}}
   \caption{Comparison of the data-driven measurements of the
     identification efficiency for converted photons as a function of $\ET$
     in the region $10~\GeV < \ET < 1500~\GeV$, for the four
     pseudorapidity intervals (a) $|\eta|<0.6$, (b) $0.6\leq|\eta|<1.37$,
     (c) $1.52\leq|\eta|<1.81$, and (d) $1.81\leq|\eta|<2.37$.
     The error bars represent the quadratic sum
     of the statistical and systematic uncertainties estimated in each
     method.
   } 
   \label{fig:dd_eff_conv_barrel_and_endcap} 
\end{figure} 

In the photon transverse momentum regions in which the different 
measurements overlap, the results from each method are consistent with each
other within the uncertainties.
Relatively large fluctuations of the radiative \Zboson\ decay
measurements are seen, due to their large statistical uncertainties. 

The photon identification efficiency increases from 50--65\%
(45--55\%) for unconverted (converted) photons at $\ET\approx 10$~\GeV\ to
94--100\% at $\ET\gtrsim 100$~\GeV, and is larger than about 90\% for
$\ET>40$~\GeV. 
The absolute uncertainty in the measured efficiency is around 1\%
(1.5\%) for unconverted (converted) photons for $\ET<30$~\GeV\ and
around 0.4--0.5\% for both types of photons above 30~\GeV\ for the 
most precise method in a given bin.

\subsection{Comparison with the simulation}
\label{sec:comparisonwithsimulation}

In this section the results of the data-driven efficiency
measurements are compared to the identification
efficiencies predicted in the simulation.
The comparison is performed both before and after applying
the shower shape corrections.

Prompt photons produced in photon+jet events have different
kinematic distributions than photons originating in radiative $Z$
boson decays. Moreover, some of the photons in $\gamma$+jet events --
unlike those from $Z$ boson decays -- originate in parton fragmentation.
Such photons have lower identification efficiency than the photons
produced directly in the hard-scattering process, due to the energy
deposited in the calorimeter by the hadrons produced almost
collinearly with the photon in the fragmentation. 
After applying an isolation requirement, however, the fragmentation
photons usually represent a small fraction of the selected sample,
typically below 10\% for low transverse momenta and rapidly decreasing
to a few \%  with increasing $\ET$.
The difference in identification efficiency between photons from
radiative $Z$ boson decays and from $\gamma$+jet events is thus
expected to be small. To account for such a difference,
the efficiency measured in data with the radiative \Zboson\ boson decay
method is compared to the prediction from simulated $Z\to\ell\ell\gamma$ events
(Figs.~\ref{fig:radz_eff_unconv_vs_mc_barrel_and_endcap} and
\ref{fig:radz_eff_conv_vs_mc_barrel_and_endcap}), while the efficiency
measured in data with the electron extrapolation and matrix
methods is compared to the prediction from simulated photon+jet events
(Figs.~\ref{fig:eemm_eff_unconv_vs_mc_barrel_and_endcap} and
\ref{fig:eemm_eff_conv_vs_mc_barrel_and_endcap}).  
 
\begin{figure}[!htbp]
   \centering
   \subfloat[]{\label{fig:radz_eff_unconv_barrel_a}\includegraphics[width=.5\textwidth]{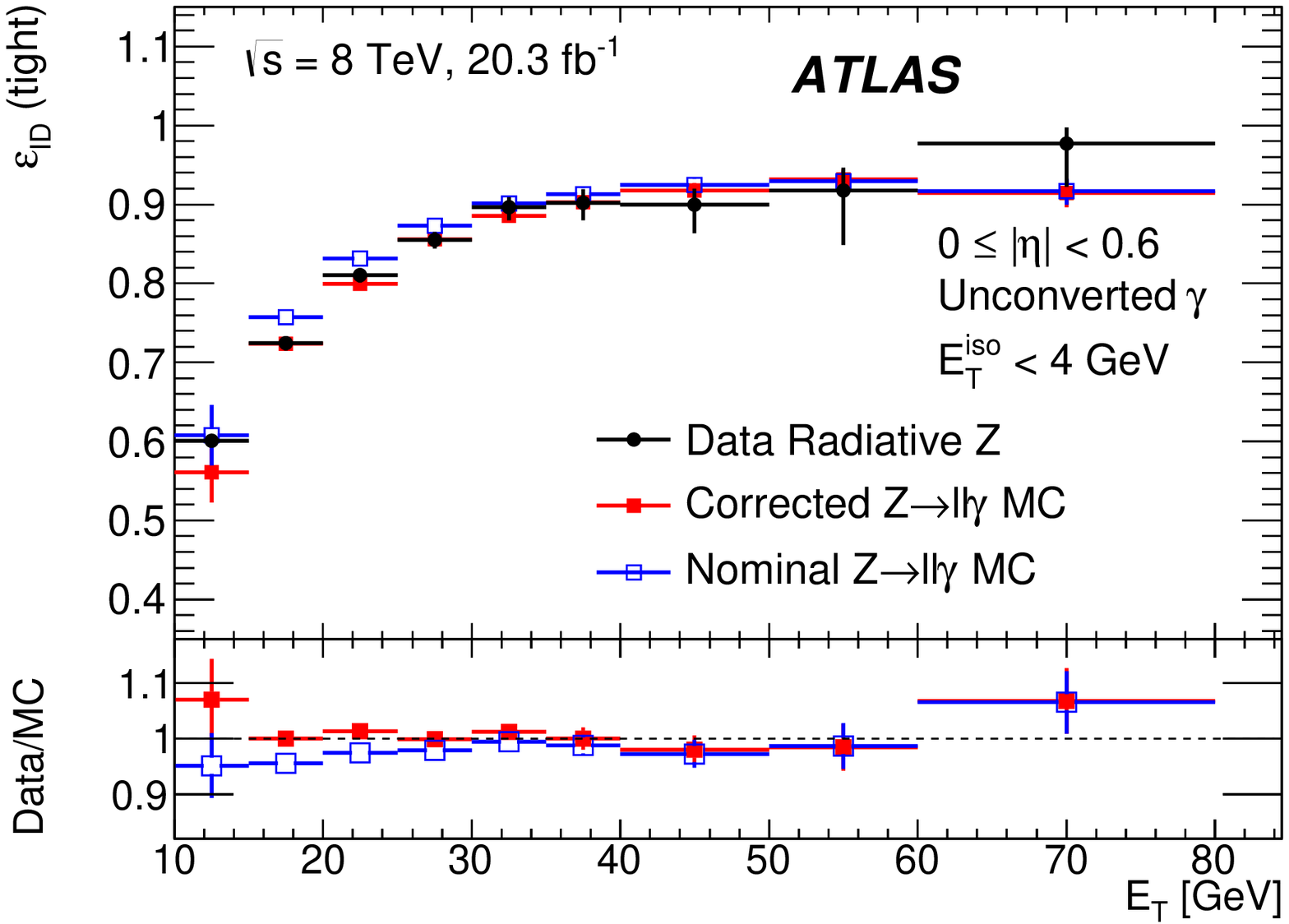}}
   \subfloat[]{\label{fig:radz_eff_unconv_barrel_b}\includegraphics[width=.5\textwidth]{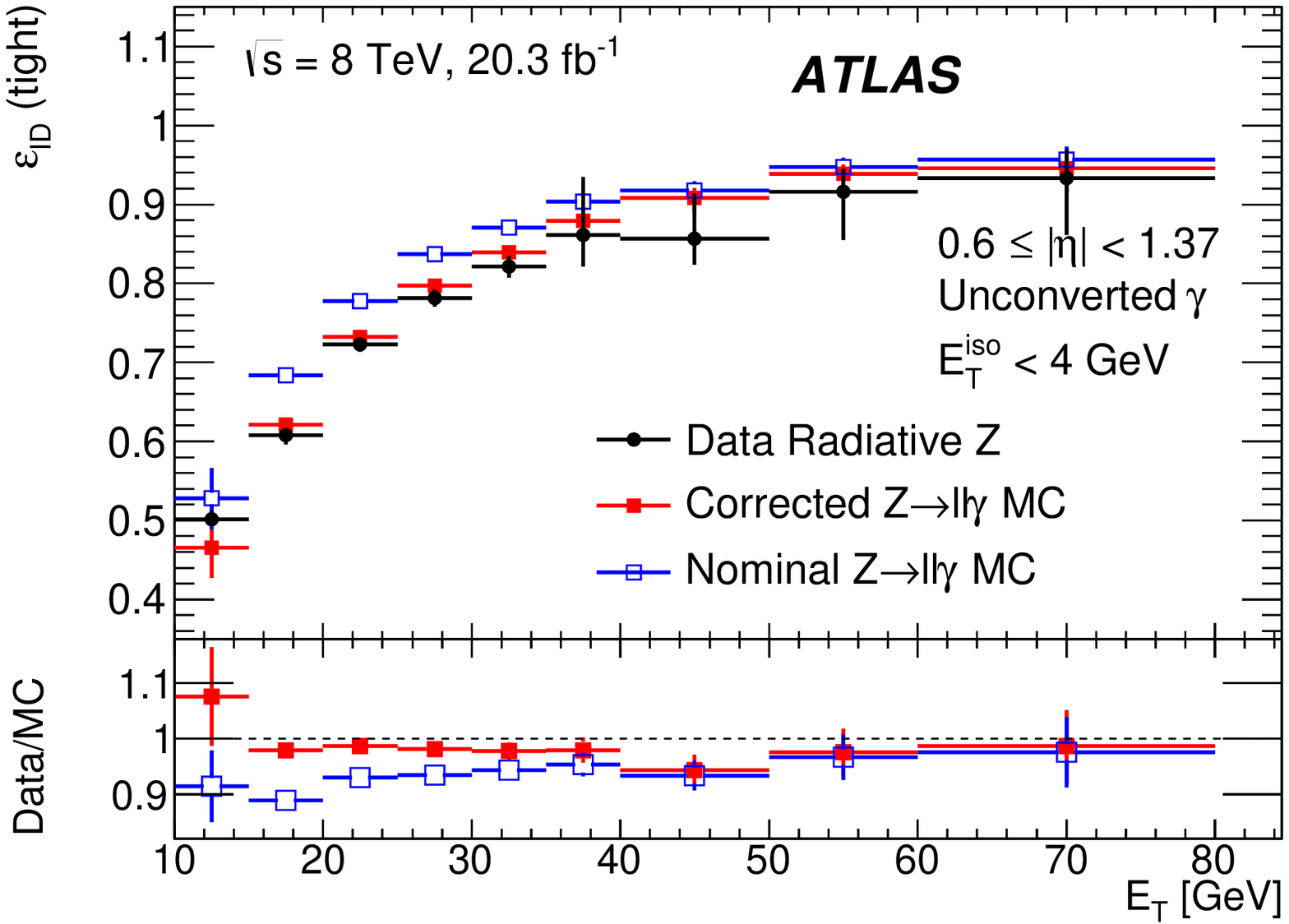}}\\
   \subfloat[]{\label{fig:radz_eff_unconv_endcap_c}\includegraphics[width=.5\textwidth]{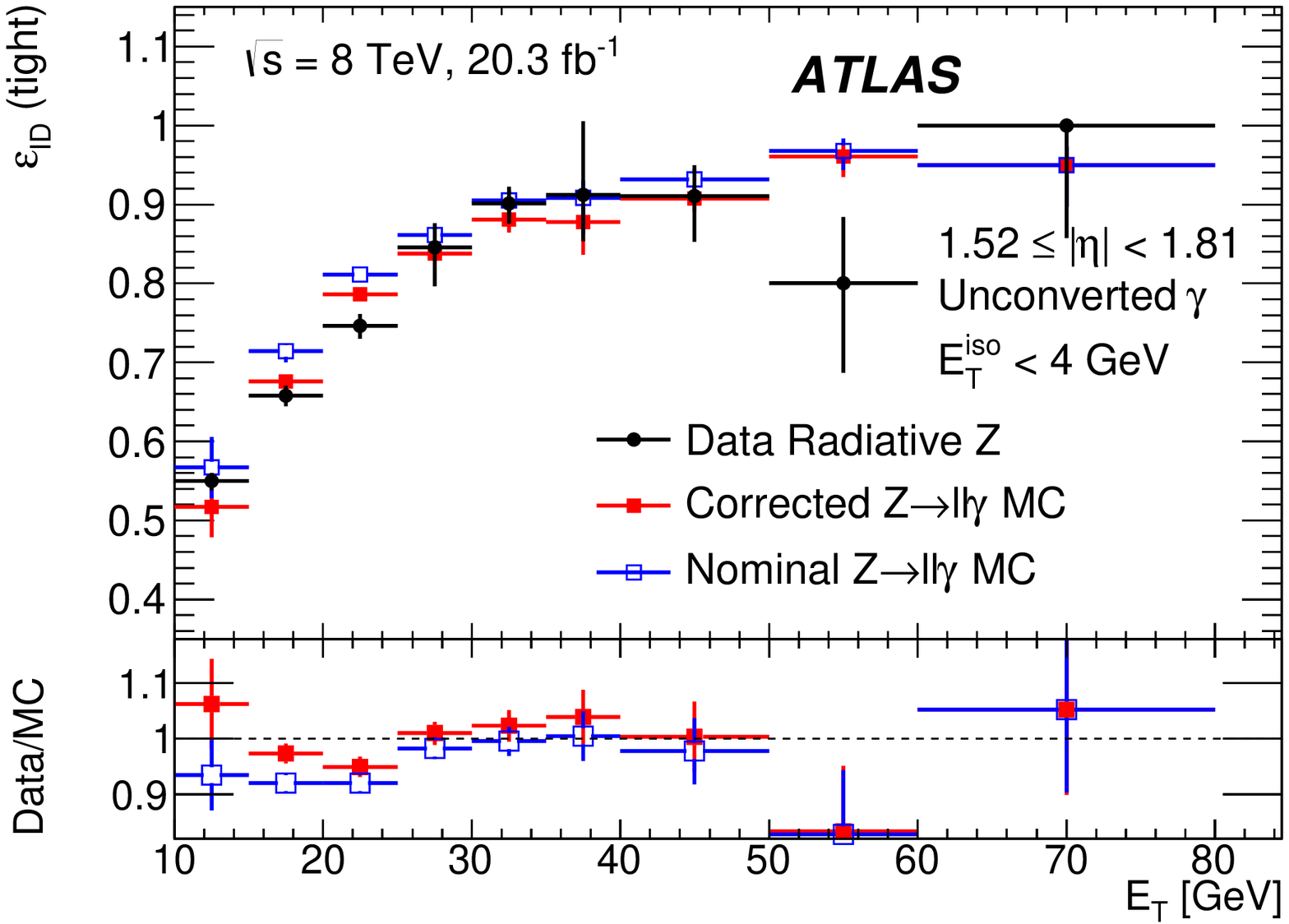}}
   \subfloat[]{\label{fig:radz_eff_unconv_endcap_d}\includegraphics[width=.5\textwidth]{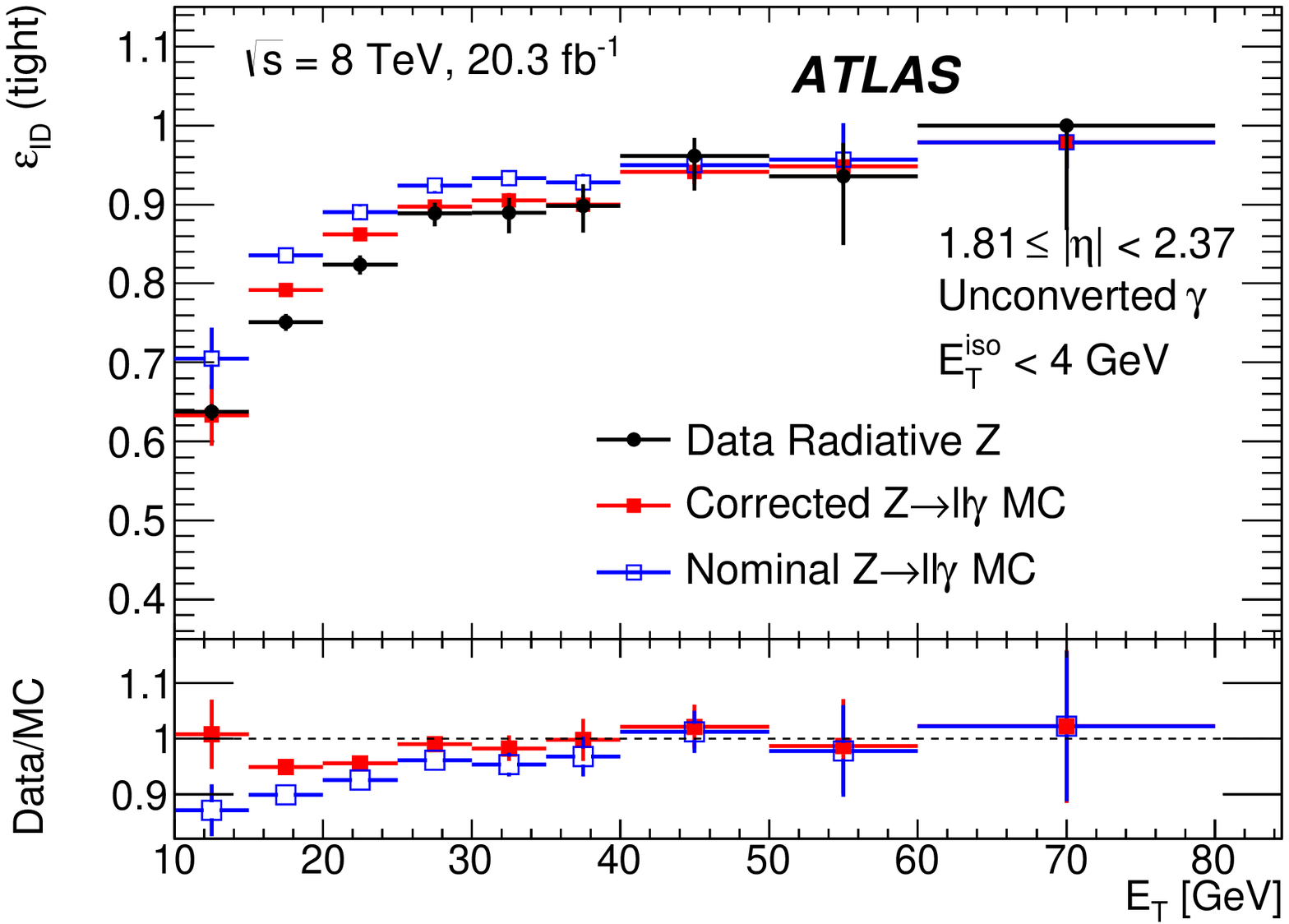}}  
   \caption{Comparison of the radiative \Zboson\ boson data-driven
     efficiency measurements of unconverted photons to the nominal and
     corrected $Z\to\ell\ell\gamma$ MC predictions as a function of $\ET$
     in the region $10~\GeV < \ET < 80~\GeV$, for the four
     pseudorapidity intervals (a) $|\eta|<0.6$, (b) $0.6\leq|\eta|<1.37$,
     (c) $1.52\leq|\eta|<1.81$, and (d) $1.81\leq|\eta|<2.37$.
     The bottom panels show the ratio of the data-driven results to
     the MC predictions (also called scale factors in the text).
     The error bars on the data points represent the quadratic sum
     of the statistical and systematic uncertainties.
     The error bars on the MC predictions correspond to the statistical
     uncertainty from the number of simulated events.
   }   
   \label{fig:radz_eff_unconv_vs_mc_barrel_and_endcap} 
\end{figure}
\begin{figure}[!htbp]
   \centering
   \subfloat[]{\label{fig:radz_eff_conv_barrel_a}\includegraphics[width=.5\textwidth]{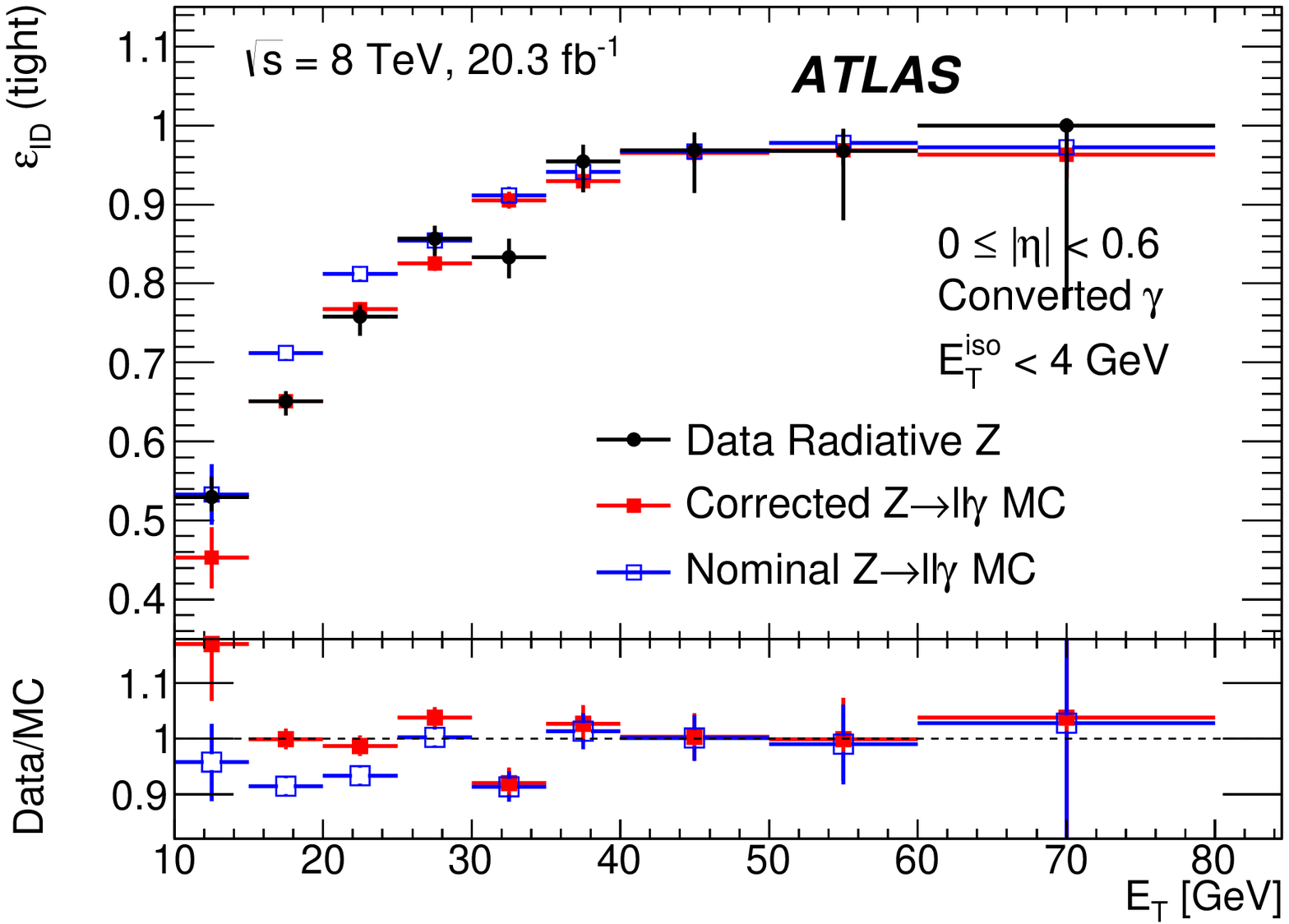}}
   \subfloat[]{\label{fig:radz_eff_conv_barrel_b}\includegraphics[width=.5\textwidth]{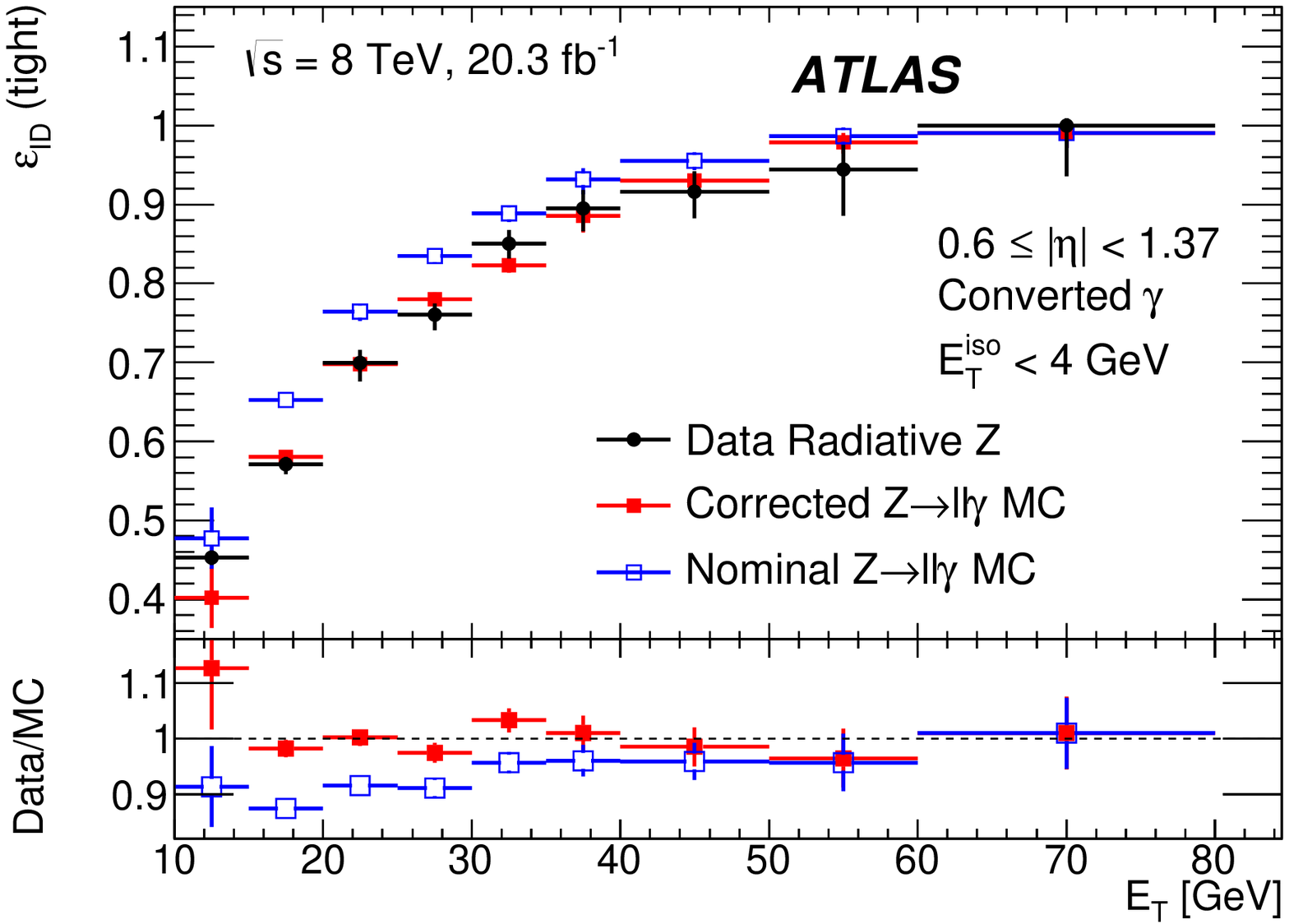}}\\
   \subfloat[]{\label{fig:radz_eff_conv_endcap_c}\includegraphics[width=.5\textwidth]{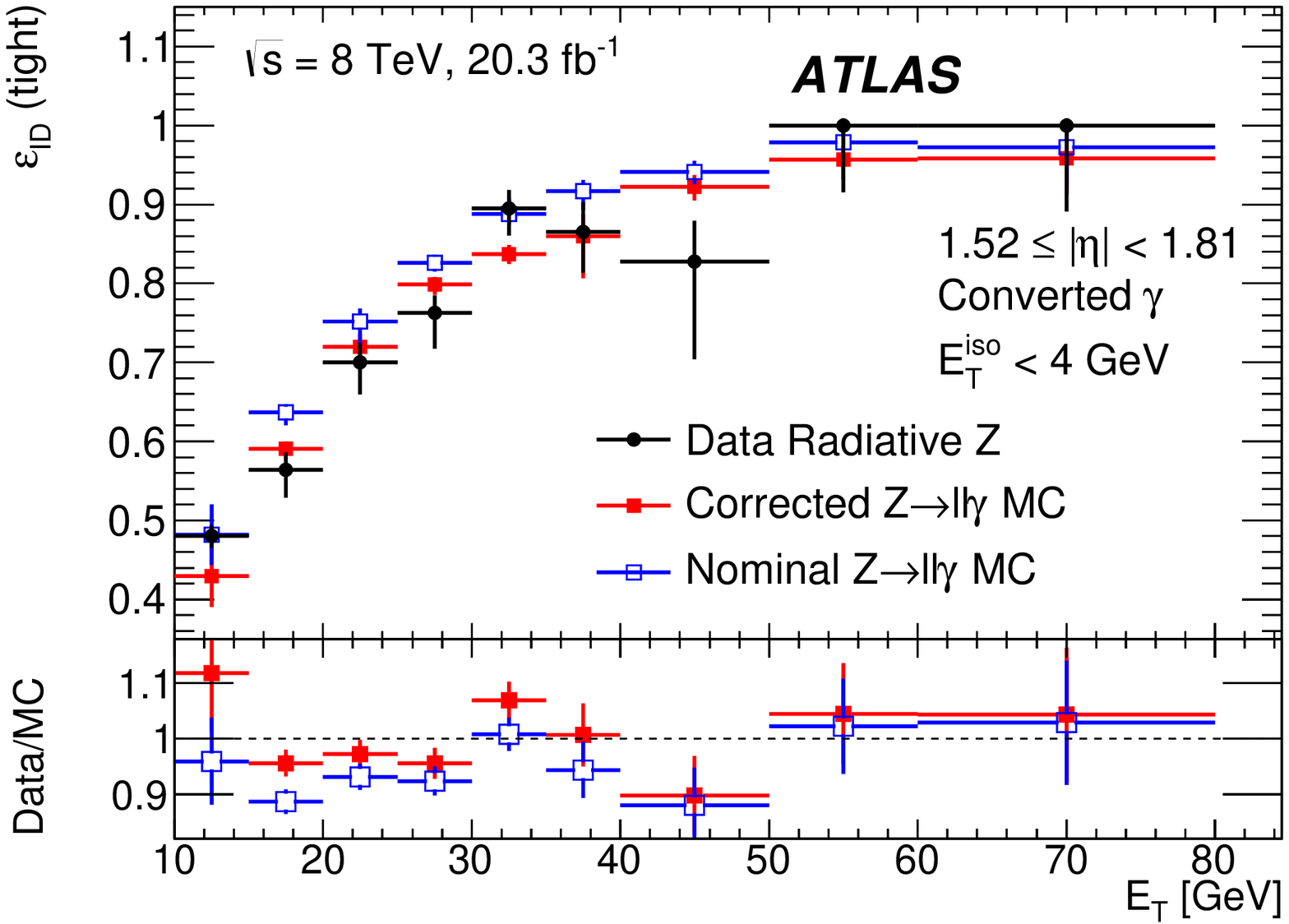}}
   \subfloat[]{\label{fig:radz_eff_conv_endcap_d}\includegraphics[width=.5\textwidth]{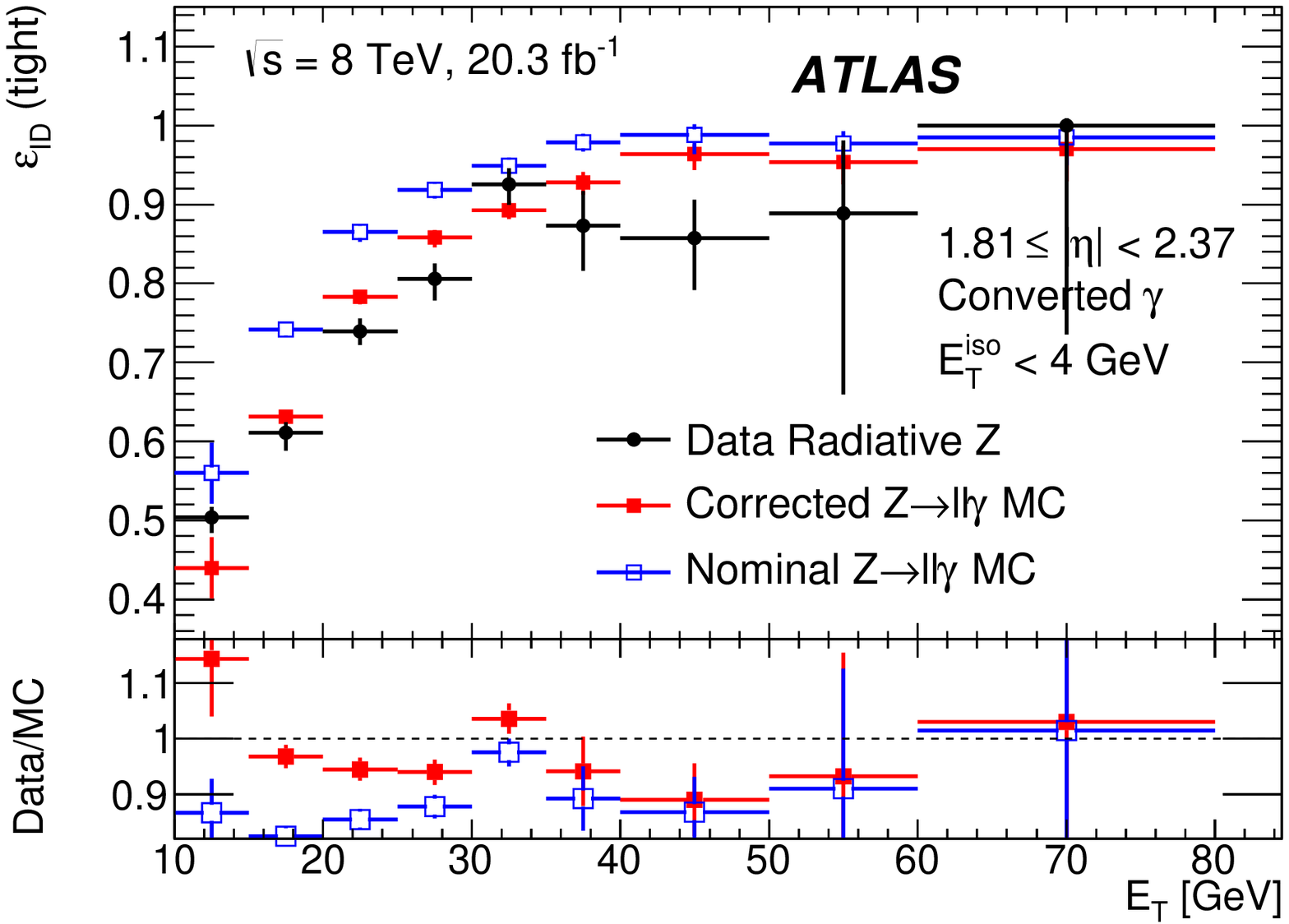}}
   \caption{Comparison of the radiative \Zboson\ boson data-driven
     efficiency measurements of converted photons to the nominal and
     corrected $Z\to\ell\ell\gamma$ MC predictions as a function of $\ET$
     in the region $10~\GeV < \ET < 80~\GeV$, for the four
     pseudorapidity intervals (a) $|\eta|<0.6$, (b) $0.6\leq|\eta|<1.37$,
     (c) $1.52\leq|\eta|<1.81$, and (d) $1.81\leq|\eta|<2.37$.
     The bottom panels show the ratio of the data-driven results to
     the MC predictions (also called scale factors in the text).
     The error bars on the data points represent the quadratic sum
     of the statistical and systematic uncertainties.
     The error bars on the MC predictions correspond to the statistical
     uncertainty from the number of simulated events.
   } 
   \label{fig:radz_eff_conv_vs_mc_barrel_and_endcap} 
\end{figure}
%
\begin{figure}[!htbp]
   \centering 
   \subfloat[]{\label{fig:eemm_eff_unconv_barrel_a}\includegraphics[width=.5\textwidth]{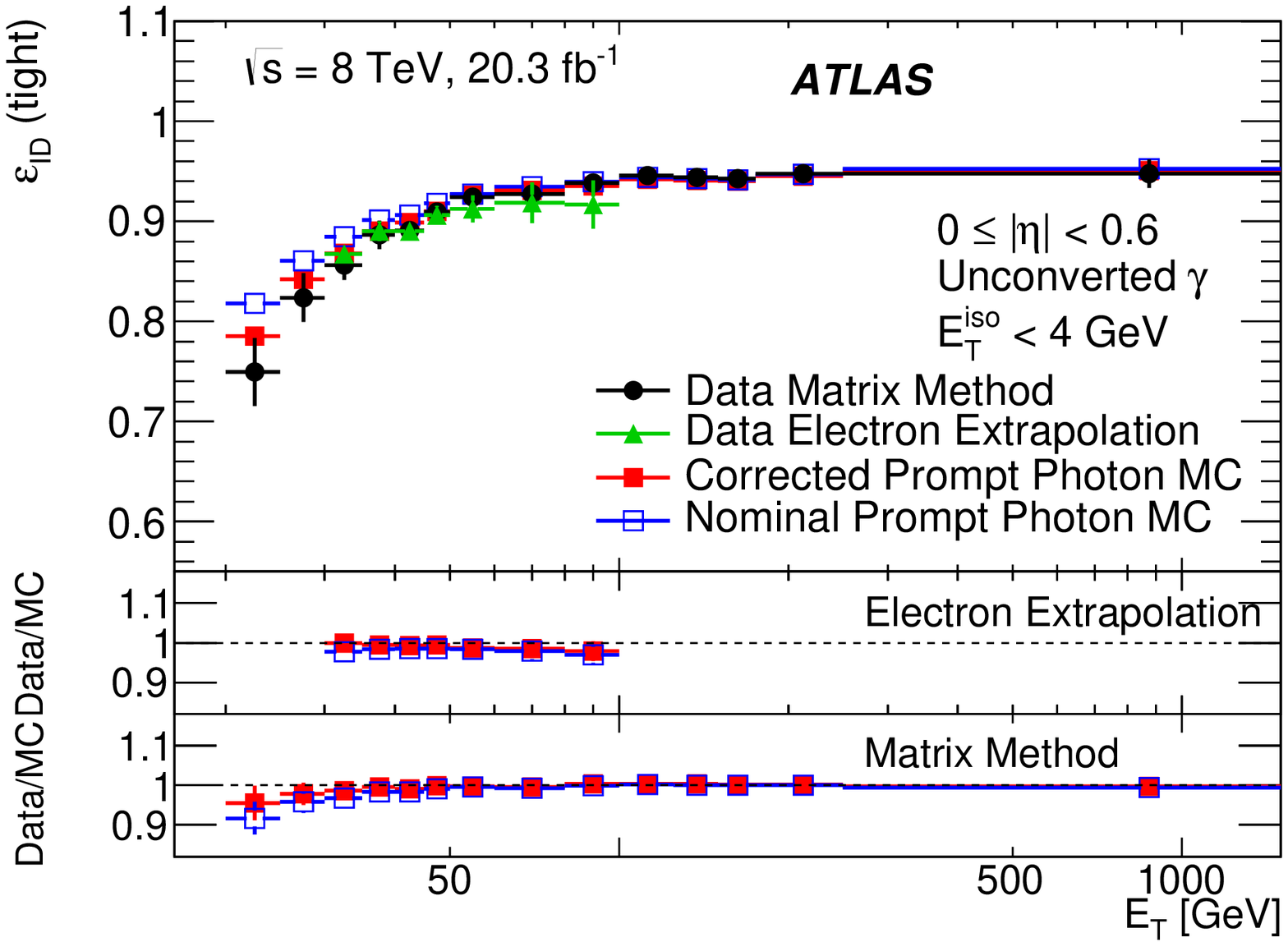}} 
   \subfloat[]{\label{fig:eemm_eff_unconv_barrel_b}\includegraphics[width=.5\textwidth]{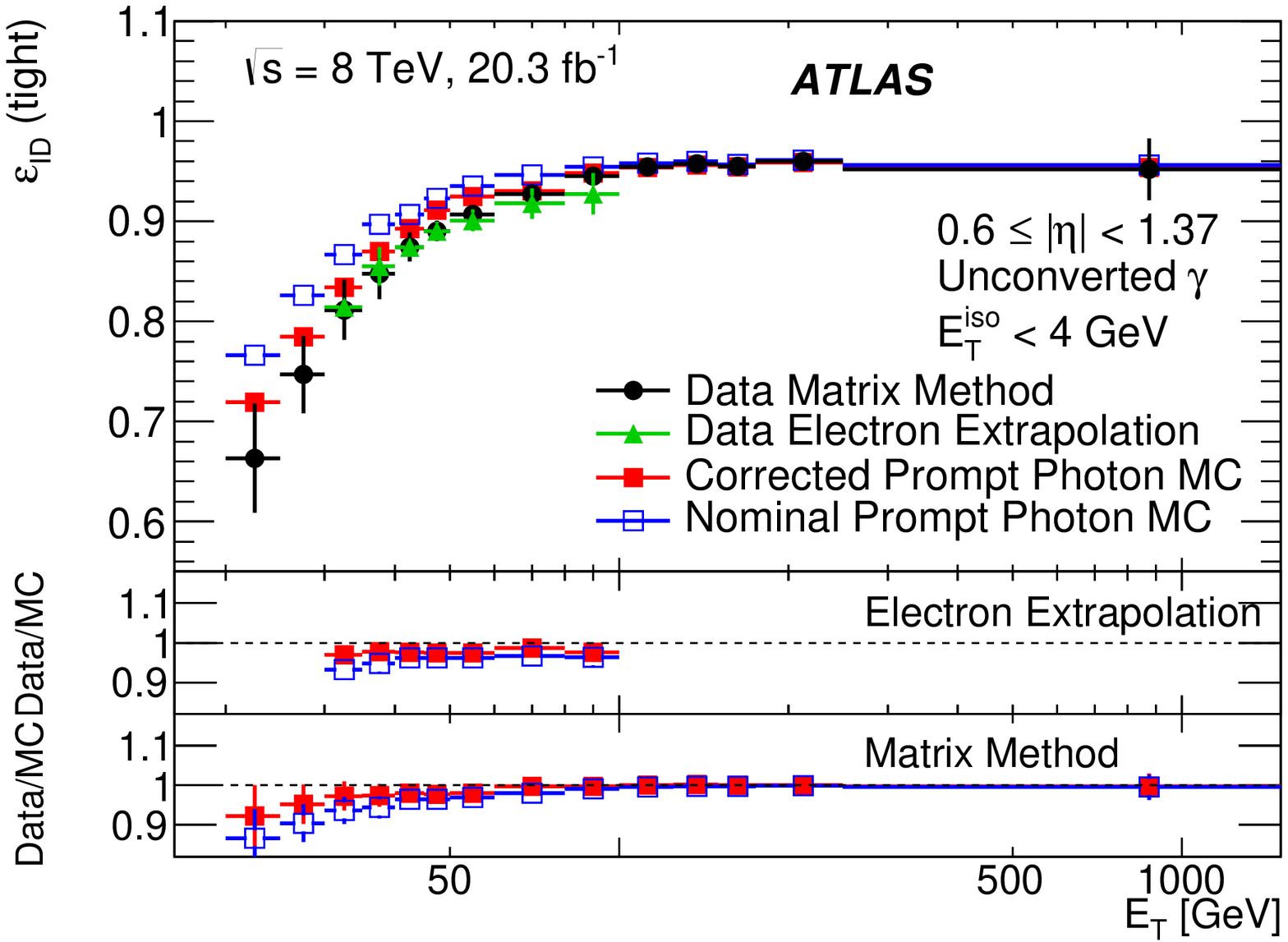}}\\
   \subfloat[]{\label{fig:eemm_eff_unconv_endcap_c}\includegraphics[width=.5\textwidth]{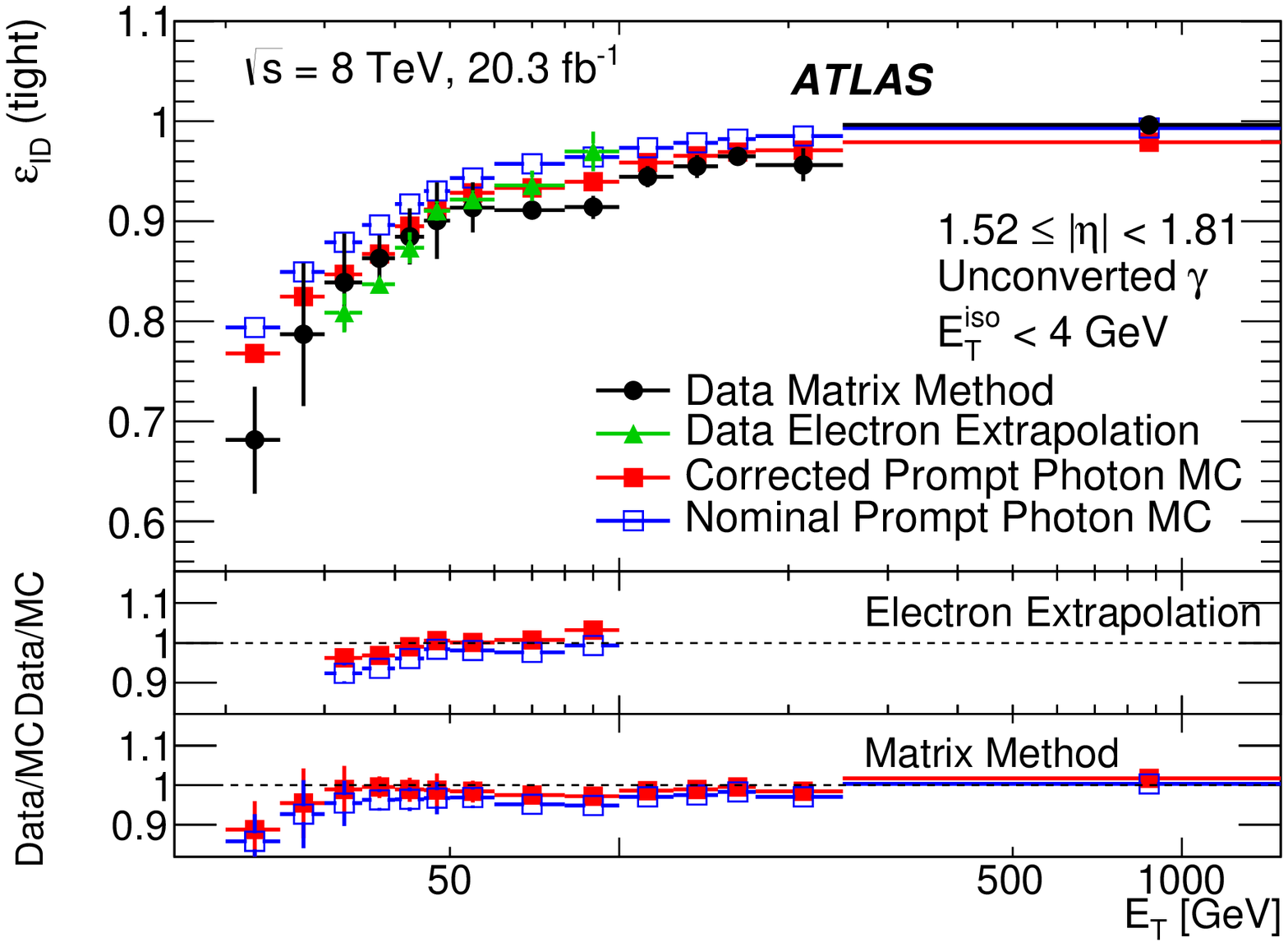}}
   \subfloat[]{\label{fig:eemm_eff_unconv_endcap_d}\includegraphics[width=.5\textwidth]{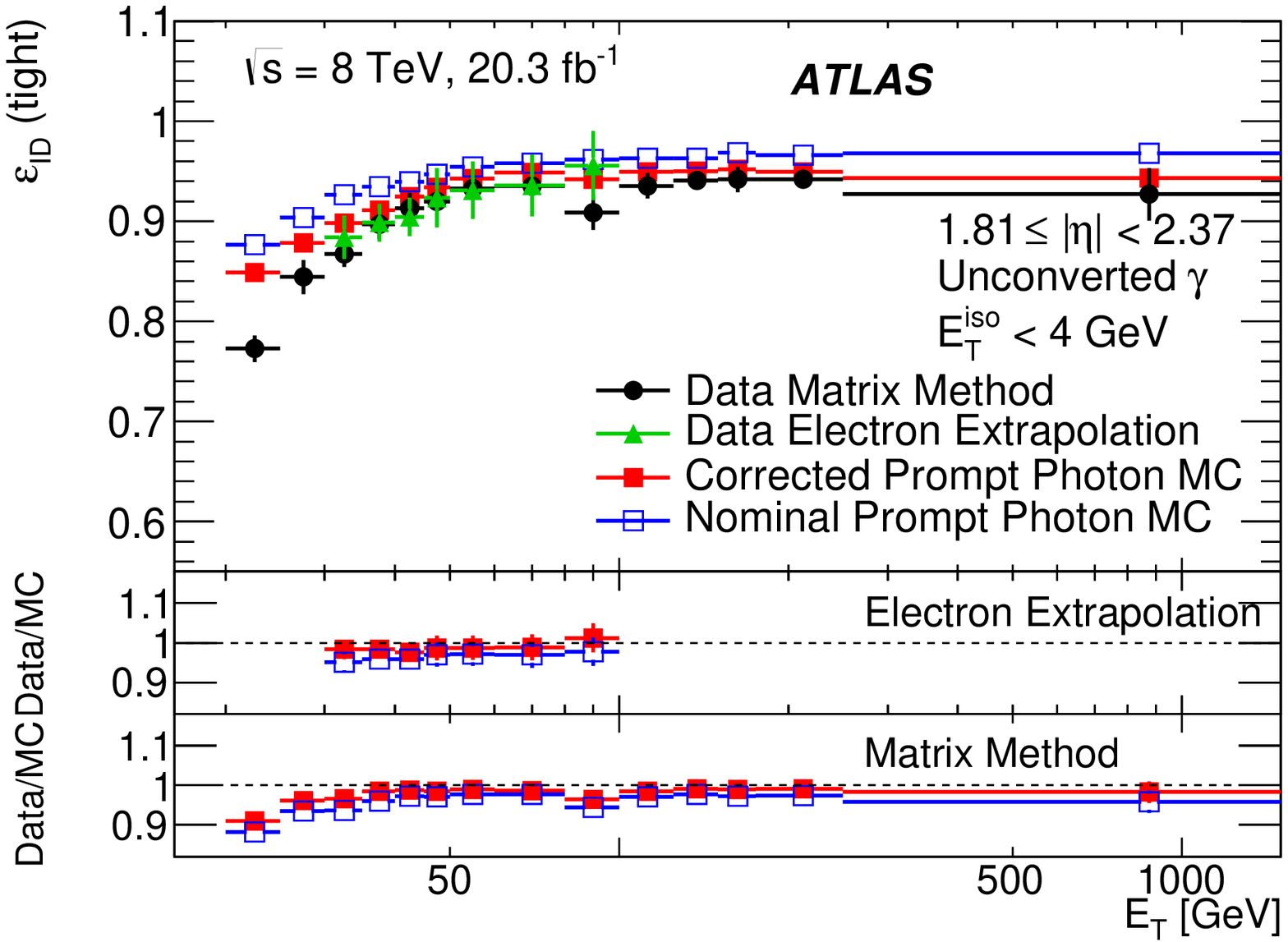}}
   \caption{Comparison of the electron extrapolation and matrix method data-driven efficiency measurements of unconverted photons to the
     nominal and corrected prompt-photon+jet MC predictions as a function
     of $\ET$ in the region $20~\GeV < \ET < 1500~\GeV$, for the four
     pseudorapidity intervals (a) $|\eta|<0.6$, (b) $0.6\leq|\eta|<1.37$,
     (c) $1.52\leq|\eta|<1.81$, and (d) $1.81\leq|\eta|<2.37$.
     The bottom panels show the ratio of the data-driven results to
     the MC predictions (also called scale factors in the text).
     The error bars on the data points represent the quadratic sum
     of the statistical and systematic uncertainties.
     The error bars on the MC predictions correspond to the statistical
     uncertainty from the number of simulated events.
   } 
   \label{fig:eemm_eff_unconv_vs_mc_barrel_and_endcap} 
\end{figure}
\begin{figure}[!htbp] 
   \centering
   \subfloat[]{\label{fig:eemm_eff_conv_barrel_a}\includegraphics[width=.5\textwidth]{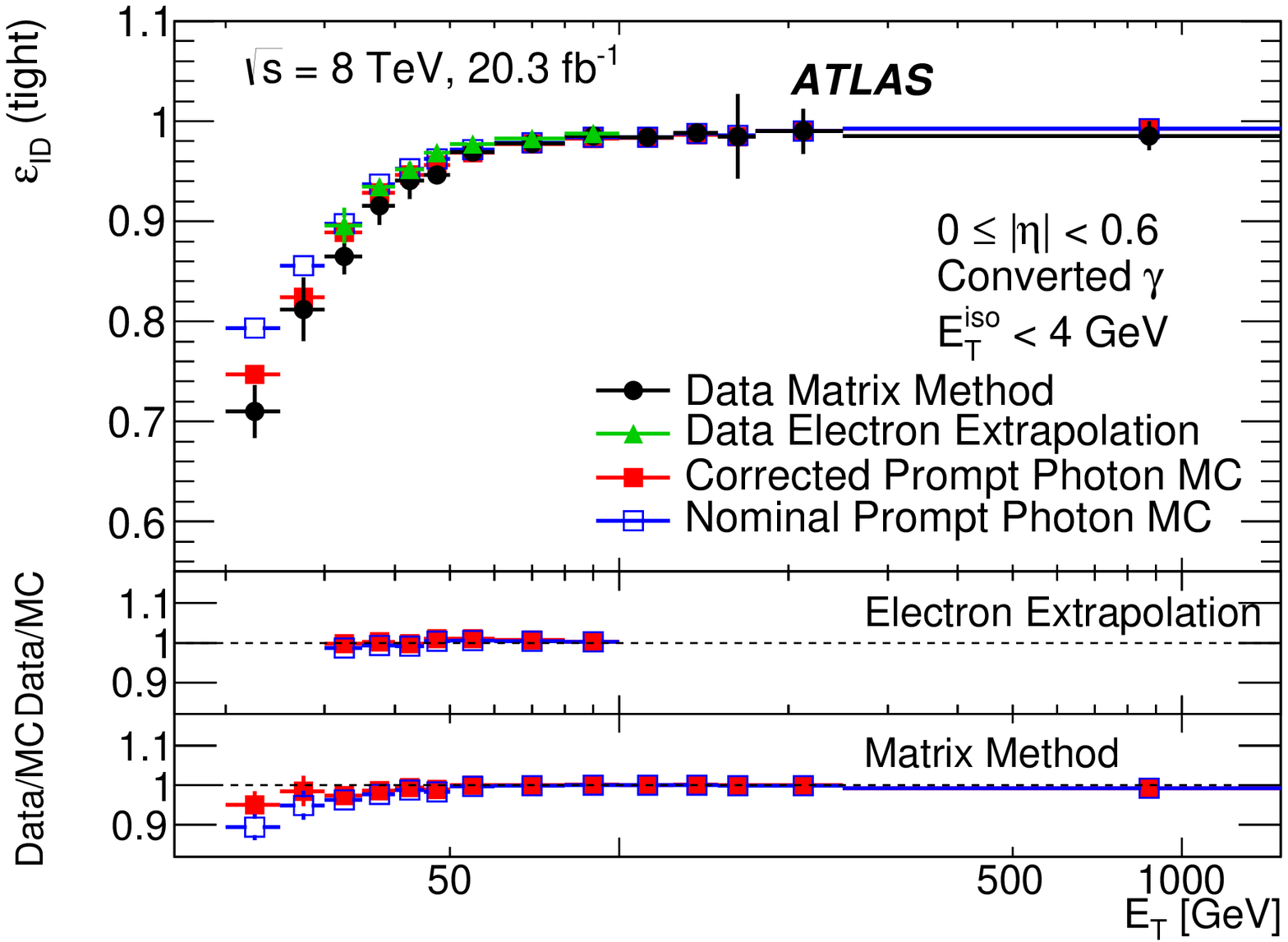}}
   \subfloat[]{\label{fig:eemm_eff_conv_barrel_b}\includegraphics[width=.5\textwidth]{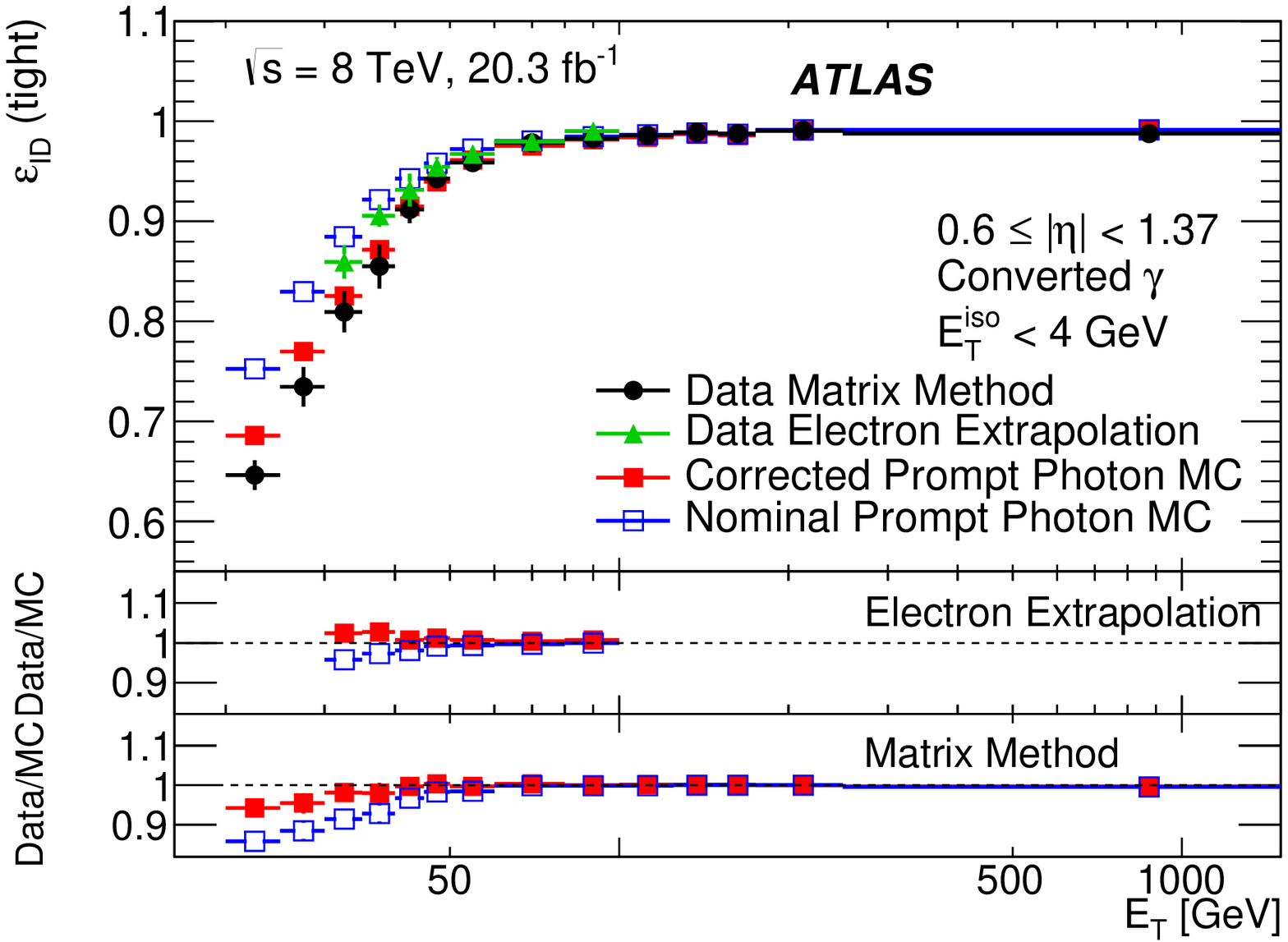}}\\
   \subfloat[]{\label{fig:eemm_eff_conv_endcap_c}\includegraphics[width=.5\textwidth]{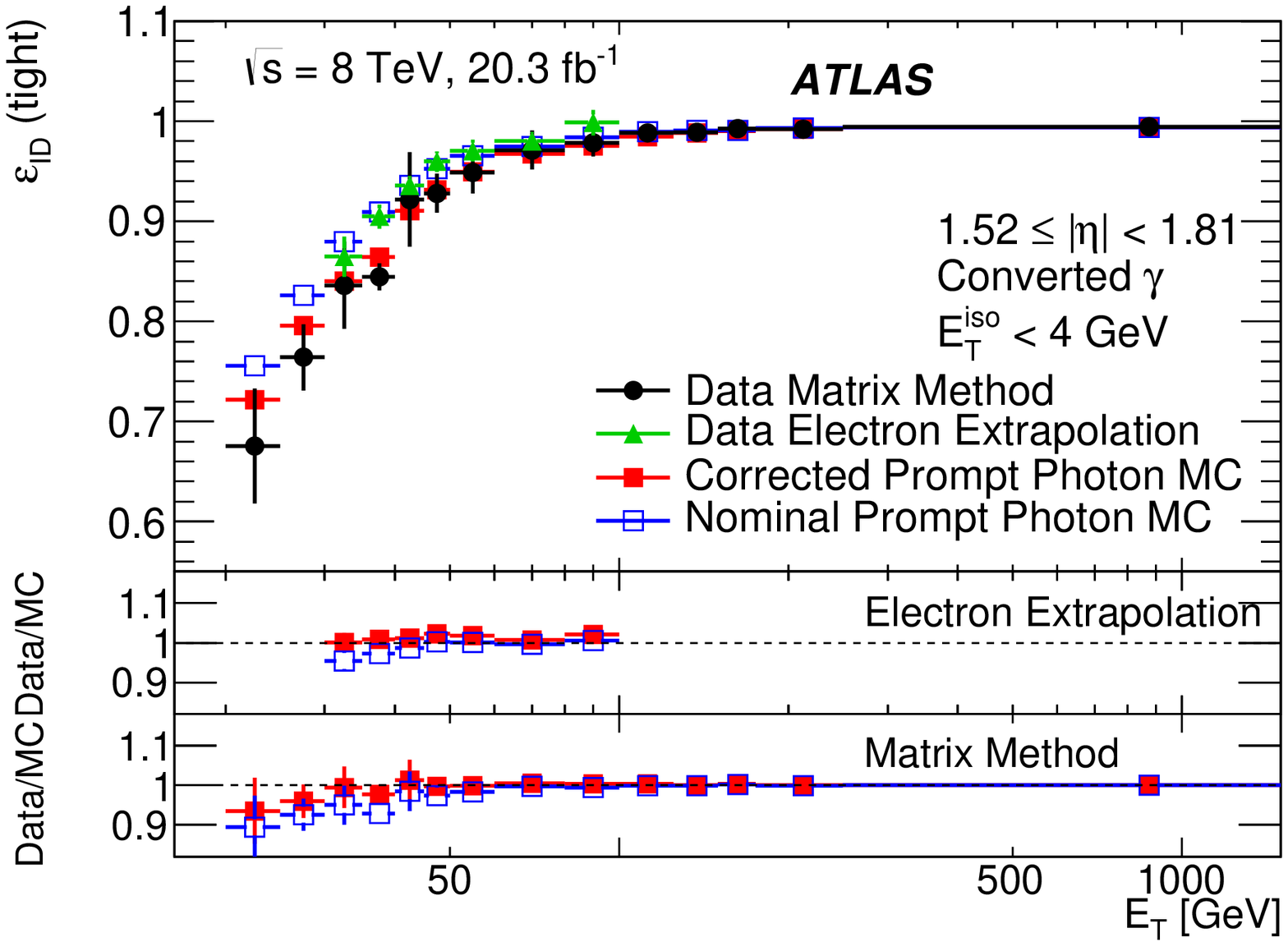}}
   \subfloat[]{\label{fig:eemm_eff_conv_endcap_d}\includegraphics[width=.5\textwidth]{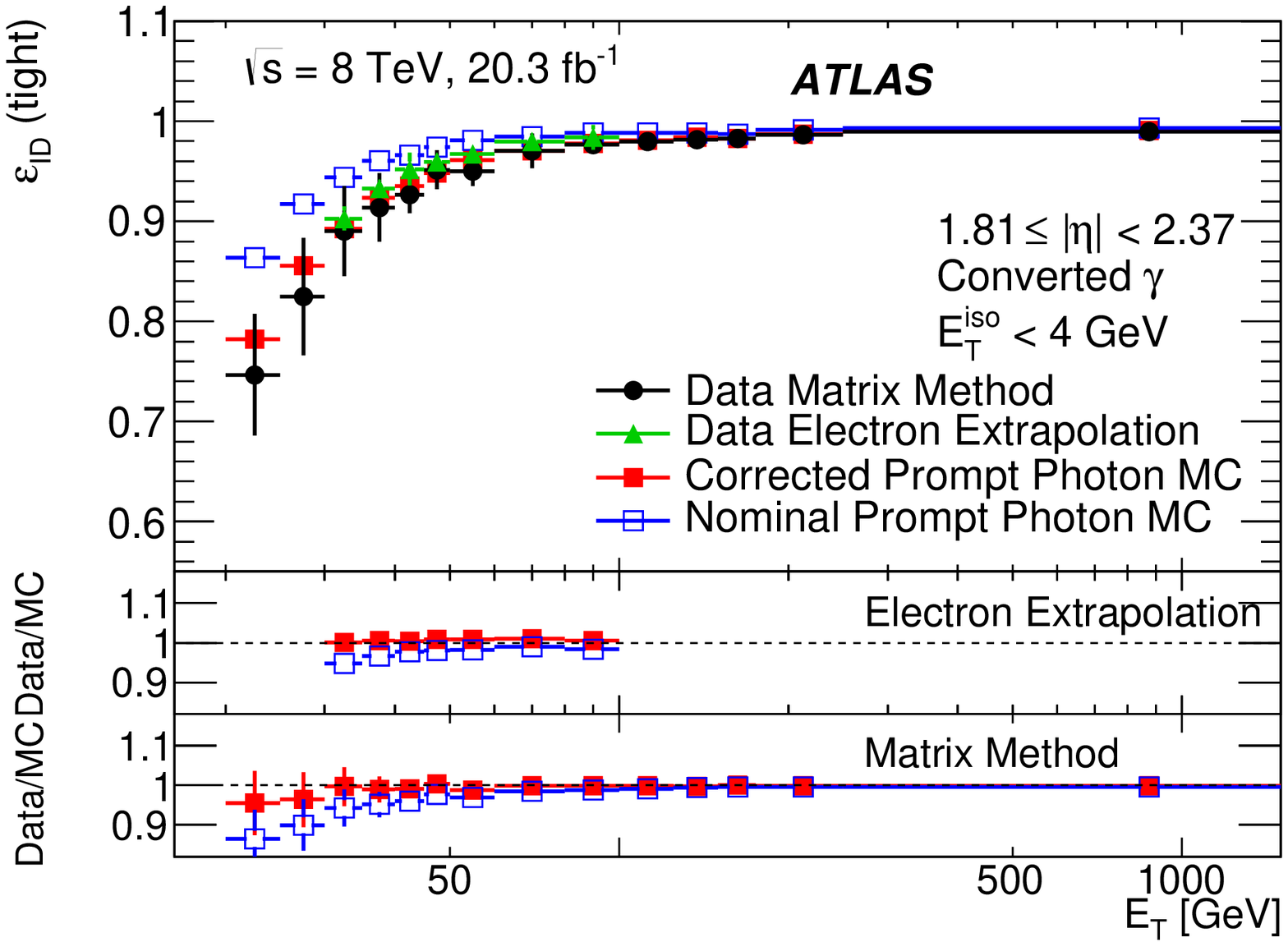}}
   \caption{Comparison of the electron extrapolation and matrix method data-driven efficiency measurements of converted photons to the
     nominal and corrected prompt-photon+jet MC predictions as a function
     of $\ET$ in the region $20~\GeV < \ET < 1500~\GeV$, for the four
     pseudorapidity intervals (a) $|\eta|<0.6$, (b) $0.6\leq|\eta|<1.37$,
     (c) $1.52\leq|\eta|<1.81$, and (d) $1.81\leq|\eta|<2.37$.
     The bottom panels show the ratio of the data-driven values to the
     MC predictions (also called scale factors in the text).
     The error bars on the data points represent the quadratic sum
     of the statistical and systematic uncertainties.
     The error bars on the MC predictions correspond to the statistical
     uncertainty from the number of simulated events.
   } 
   \label{fig:eemm_eff_conv_vs_mc_barrel_and_endcap} 
\end{figure}

The level of agreement among the different \effID\ values improves
with increasing \ET: no significant difference is observed between
the data-driven
measurements and the nominal or corrected simulation for 
$\ET> 60$~\GeV. At lower transverse momenta, the nominal simulation 
tends to overestimate the efficiency by up to 10--15\%, as the
electromagnetic showers from photons are typically narrower
in the simulation than in data.
In the same transverse momentum range, the corrected 
simulation agrees with the data-driven measurements within a few percent.

The remaining difference between the corrected simulation and the
data-driven measurements is taken into account by computing data-to-MC
efficiency ratios, also referred to as {\normalfont \itshape scale factors} (SF). 
The data-to-MC efficiency ratios are computed separately for each
method and then combined. The efficiencies from the 
$\Zboson\to\ell\ell\gamma$ data control sample are divided by the 
prediction of the simulation of radiative photons from 
\Zboson\ boson decays, while the results from the other two methods 
are divided by the predictions of the photon+jet simulation. 
The data-to-MC efficiency ratios are shown in the bottom plots of 
Figs.~\ref{fig:radz_eff_unconv_vs_mc_barrel_and_endcap}--\ref{fig:eemm_eff_conv_vs_mc_barrel_and_endcap} 
and are used to correct the
predictions in the analyses using photons.

Because of their good agreement and the mostly independent data samples
used, the data-to-MC efficiency ratios as a function of photon
\ET\ are combined into a single, more precise result in the
overlapping regions. 
The combination is performed independently in the different
pseudorapidity and transverse energy bins, using the Best 
Linear Unbiased Estimate (BLUE) method~\cite{Lyons, Valassi}.   
The combined data-to-MC efficiency ratio SF is calculated as a linear
combination of the input measurements, $\mathrm{SF}_i$,
with coefficients $w_{i}$ that
minimise the total uncertainty in the combined result.  
In the algorithm, both the statistical and systematic uncertainties, as
well as the correlations of systematic sources between input
measurements, are taken into account assuming that all
uncertainties have Gaussian distributions.
In practice, the quantity that is minimised is a $\chi^2$ built from
the various results and their statistical and systematic 
covariance matrices.
Since the three measurements use different data samples and
independent MC simulations, their systematic and statistical 
uncertainties are largely uncorrelated.
The background-induced uncertainties in the $Z\to ee\gamma$ and
$Z\to\mu\mu\gamma$ results, originating from the same background
process ($Z$+jet events with a jet misreconstructed as a photon)
and evaluated with the same method, are considered to be 100\% correlated.
The uncertainties in the results of the matrix method and the
electron extrapolation method
due to limited knowledge of the detector material in the
simulation are also partially correlated, both being determined with
alternative MC samples based on the same detector simulation with a
conservative estimate of additional material in front of the
calorimeter. The exact value of this correlation is difficult to
estimate. However, it was checked by varying the amount of
correlation that its effect on the final result is negligible. 

After the combination, for each averaged scale factor SF,
the $\chi^2=\sum_{i=1}^{N} w_i(\mathrm{SF}-\mathrm{SF}_i)^2$
is computed and compared to $N-1$, where $N$ is the number of
measurements included in the combined result for that point, and $N-1$
is the expectation value of $\chi^2$ from a Gaussian
distribution. Only a few bins among all photon $\eta$ and $\ET$ bins
for unconverted and converted photons are found to have $\chi^2/(N-1)
>1$. These $\chi^2$ values are smaller than 2.0, confirming that
the different measurements are consistent. For the
points with $\chi^2/(N-1) > 1$, the error in the combined value,
$\delta \mathrm{SF}$, is increased by a factor
$S=\sqrt{\chi^2/(N-1)}$, following the prescription in
Ref.~\cite{PDG}. 
The combined data-to-MC efficiency ratios differ from one by as much as 10\% at
\ET\ = 10~\GeV\ and by only a few percent above \ET = 40~\GeV.  

A systematic uncertainty in the data-to-MC efficiency ratios
is associated with the uncertainty in  photon+jet simulation's modelling
of the fraction of photons emitted in the fragmentation of partons.
In order to estimate the effect on the data-to-MC efficiency ratio,
the number of fragmentation photons in the photon+jet MC sample is
varied by $\pm 50\%$, and the maximum variation of the data-to-MC
efficiency ratio is taken as an additional systematic 
uncertainty. This uncertainty decreases with increasing transverse momentum and
is always below 0.5\% and 0.7\% for unconverted and converted
photons, respectively.
This uncertainty is also larger than the efficiency differences
observed in the simulation between different event generators, which
are thus not considered as a separate systematic uncertainty in the
data-to-MC efficiency ratios.

The effect of the isolation requirement on the data-to-MC efficiency
ratios is checked by varying it between 3~\GeV\ and 7~\GeV\ and recomputing
the data-to-MC efficiency ratios using \Zboson\ boson radiative
decays.
The study is performed in different regions of pseudorapidity and
integrated over \ET\ to reduce statistical fluctuations.
The deviation of the alternative data-to-MC efficiency ratios from the 
nominal value is typically 0.5\% and always lower than 1.2\%, almost 
independent of pseudorapidity.
This deviation is thus considered as an additional uncertainty and
added in quadrature in ATLAS measurements with final-state photons
to which an isolation requirement different from $\Etiso<4$~\GeV\ is
applied.

The combined data-to-MC efficiency ratios with their total uncertainties
are shown as a function of \ET\ in Figs.~\ref{fig:combSF_unconv} and
Figs.~\ref{fig:combSF_conv}.
In the low transverse energy region these ratios decrease from values higher than one to values smaller than one because the data and MC efficiency curves cross between 10~\GeV\ and 20~\GeV, as can be seen in Figure~\ref{fig:radz_eff_unconv_vs_mc_barrel_and_endcap} and \ref{fig:radz_eff_conv_vs_mc_barrel_and_endcap}.
The change of shape at \ET\ = 30~\GeV\ can be explained by the fact that the electron extrapolation method starts entering the combination, changing the central values but also decreasing the uncertainties.

\begin{figure}[!htbp]
   \centering
   \includegraphics[width=.9\textwidth]{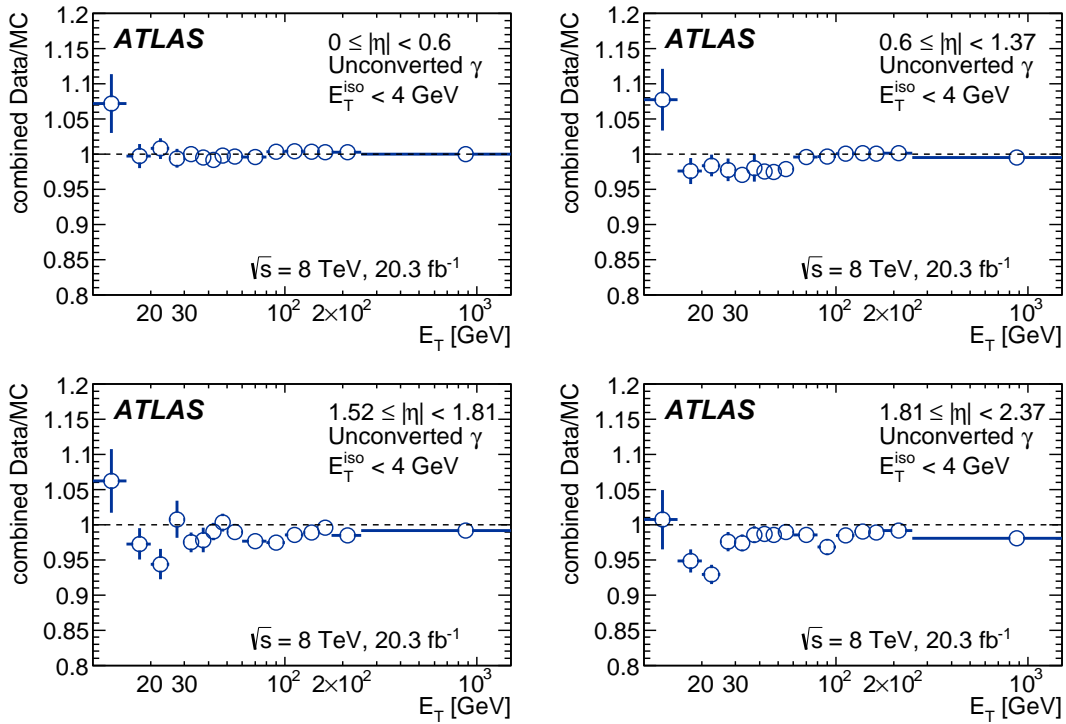}
   \caption{Combined data-to-MC efficiency ratios (SF) of unconverted
     photons in the region $10~\GeV < \ET < 1500~\GeV$.}
   \label{fig:combSF_unconv}
\end{figure}

\begin{figure}[!htbp]
   \centering
   \includegraphics[width=.9\textwidth]{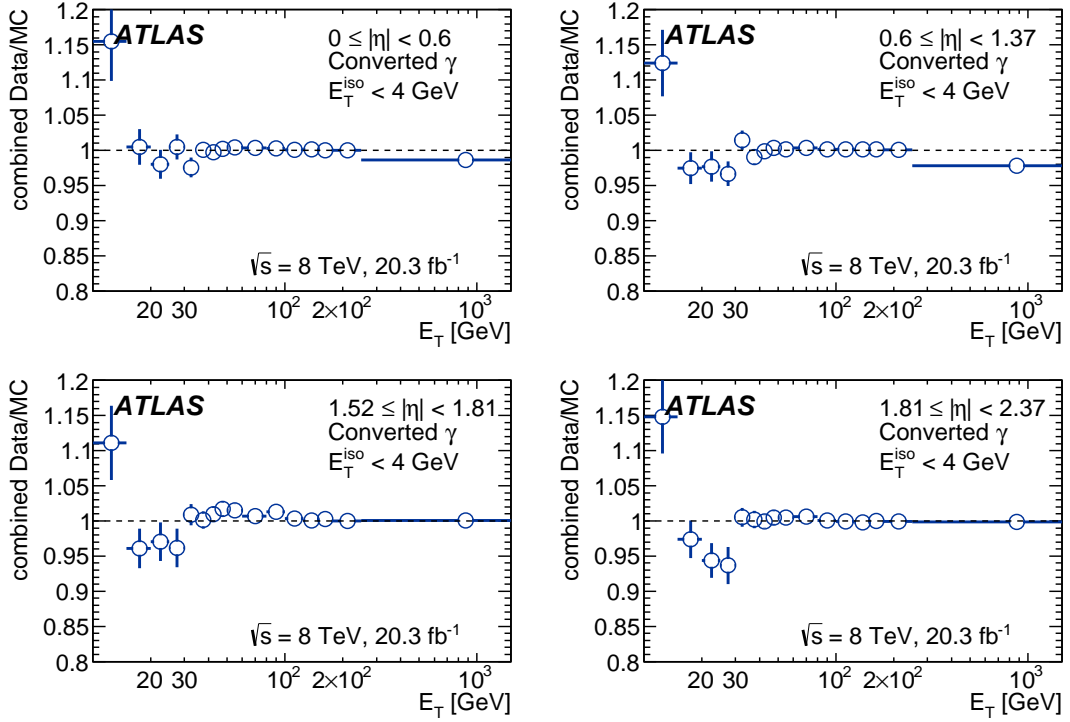}
   \caption{Combined data-to-MC efficiency ratios (SF) of converted
     photons in the region $10~\GeV < \ET < 1500~\GeV$.}
   \label{fig:combSF_conv} 
\end{figure}
The total uncertainty in the data-to-MC efficiency ratio is
1.4--4.5\% (1.7--5.6\%) for unconverted (converted) photons for
$10~\GeV<\ET < 30~\GeV$, it decreases to 0.2--2.0\% (0.2--1.5\%) for
$30~\GeV<\ET < 100~\GeV$, and it further decreases to 0.2--0.8\%
(0.2--0.5\%) for higher transverse momenta.
The $\approx 5\%$ uncertainty at low transverse momenta is due
to the systematic uncertainty affecting the measurement with
radiative $Z$ boson decays for $10~\GeV<\ET<15$~\GeV. Above 15~\GeV\
the total uncertainty is below 2.5\% (3.0\%)
for unconverted (converted) photons.
A summary of the contributions to the final uncertainty on the
data-to-MC efficiency ratios of the different sources of uncertainties
described in Sect.~\ref{sec:datadrivenefficiencymeasurements} is
given in Table~\ref{tab:uncertainties}.
The background systematic
uncertainties correspond to the background subtraction done in the three
methods. The material uncertainty comes from limited knowledge of
the material upstream of the calorimeter which affects the
shower-shape description for the electron extrapolation method
(Sect.~\ref{sec:electronextrapolation}) and the track isolation
efficiency for the matrix method (Sect.~\ref{sec:matrixmethod}). The
non-closure test uncertainty of the Smirnov transform appears only in
the electron extrapolation method
(Sect.~\ref{sec:electronextrapolation}). 

\begin{table}[!htbp]
  \setlength{\tabcolsep}{3pt}
  \begin{center}
    \footnotesize
    \begin{tabular}{l l c c c }
      \hline
      \hline
      &           & 10--30~\GeV\ & 30--100~\GeV\ & 100--1500~\GeV\  \\ 
      \hline
      \multirow{ 6}{*}{Unconverted $\gamma$} & Total uncertainty            & 1.4\%--4.5\% & 0.2\%--2.0\% & 0.2\%--0.8\% \\
                                             & Statistical uncertainty      & 0.5\%--2.0\% & 0.1\%--0.7\% & 0.1\%--0.4\% \\
                                             & Total systematic uncertainty & 1.0\%--4.1\% & 0.1\%--1.2\% & 0.1\%--0.8\% \\
                                             & Background uncertainty       & 0.6\%--1.3\% & 0.0\%--0.8\% & 0.0\%--0.7\% \\
                                             & Material uncertainty         & 0.0\%--0.8\% & 0.0\%--1.1\% & 0.0\%--0.8\% \\
                                             & Non closure                  & 0.0\%        & 0.0\%--0.9\% & 0.0\%        \\
      \hline
      \multirow{ 6}{*}{Converted $\gamma$}   & Total uncertainty            & 1.7\%--5.6\% & 0.2\%--1.5\% & 0.2\%--0.5\% \\
                                             & Statistical uncertainty      & 0.9\%--3.2\% & 0.1\%--0.6\% & 0.1\%--0.4\% \\
                                             & Total systematic uncertainty & 1.4\%--4.3\% & 0.2\%--1.4\% & 0.1\%--0.5\% \\
                                             & Background uncertainty       & 0.7\%--1.7\% & 0.0\%--0.6\% & 0.0\%--0.4\% \\
                                             & Material uncertainty         & 0.0\%--1.3\% & 0.0\%--1.0\% & 0.0\%--0.5\% \\
                                             & Non closure                  & 0.0\%        & 0.0\%--0.9\% & 0.0\%        \\
      \hline
      \hline
    \end{tabular}
    \caption{Ranges of total uncertainty on the data-to-MC photon identification efficiency ratios and breakdown of the different sources of uncertainty for unconverted and converted photons, in three bins of transverse energy, giving the minimum and maximum values in the four pseudorapidity regions.}
    \label{tab:uncertainties}
  \end{center}
\end{table}

In multi-photon processes, such as Higgs boson decays to two photons, a
per-event efficiency correction to the simulated events
is computed by applying scale factors
to each of the photons in the event.
The associated uncertainty depends on the correlation between SF
uncertainties in different regions of photon $|\eta|$ and $\ET$, and
for  converted and unconverted photons. Among the systematic
uncertainties considered in the analysis, the impact of correlations
is found to be negligible in all cases but one, that of the
uncertainty in the background level in the matrix method determination
(see Sect.~\ref{sec:matrixmethod}).
Its contribution to the SF uncertainty is conservatively assumed to be
fully correlated across all regions of $|\eta|$ and $\ET$ and between
converted and unconverted photons, while the rest of the SF
uncertainty is assumed to be uncorrelated. The correlated and
uncorrelated components of the uncertainty in each region are then
propagated to the per-event uncertainty using a toy-experiment
technique.

\section{Photon identification efficiency at $\sqrt{s}=7$~\TeV}
\label{sec:seventev}
As described in Sect.~\ref{ssec:identification}, photon
identification in the analysis of 7~\TeV\ data relies on the same
cut-based algorithms used for the 8~\TeV\ data, with different
thresholds. Such thresholds were first determined using simulated
samples prior to the 2010
data-taking and then loosened in order to reduce the observed
inefficiency and the systematic uncertainties arising from the
differences found between the distributions of the discriminating
variables in data and in the simulation.  

The efficiency of the identification algorithms used for the analysis
of the 7~\TeV\ data is measured with the same techniques described in
Sect.~\ref{sec:datadrivenefficiencymeasurements}. 
Small differences between the 7 and 8~\TeV\ measurements concern the simulated
samples that were used, and the criteria used to
select the data control samples. The 7~\TeV\ simulations are based on 
a different detector material model, as described in
Sect.~\ref{sec:dataset}; the number of simulated pile-up interactions and
the correction factors for the lepton efficiency and momentum scale and 
resolution also differ from those of the 8~\TeV\ study, as do
the lepton triggers and the algorithms used to identify the leptons
in data.
Due to the smaller number of events, the 7~\TeV\ measurements cover a
narrower transverse momentum range, $20~\GeV<\ET<250$~\GeV.
The nominal efficiency is measured with respect to photons having a
calorimeter isolation transverse energy lower than 5~\GeV, 
a typical requirement used in 7~\TeV\ ATLAS measurements.
The isolation energy is computed using all the
calorimeter cells in a cone of $\Delta R=0.4$ around the photon and
corrected for pile-up and the photon energy. 

The number of selected candidates is 12000 in the 
$\Zboson\to\ell\ell\gamma$ study, $1.8\times 10^6$ in the $\Zboson
\rightarrow ee$ one, and $1.5\times 10^7$ in the measurement 
with the matrix method. 
All data-driven measurements are combined using the same
procedure described in Sect.~\ref{sec:comparisonwithsimulation} for the scale factors,
and then compared to a simulation of prompt-photon+jet events. 
In the combination, the differences between the efficiencies of photons
from radiative $Z$ boson decays and of photons from $\gamma+$jet events
mentioned in Sect.~\ref{sec:comparisonwithsimulation} are neglected.
Such differences after the photon isolation requirement are estimated
to be much smaller than the uncertainties of the measurements performed
with the $\sqrt{s}=7$~TeV data.
The combined efficiency measurements for the cut-based identification
algorithms at $\sqrt{s}=7$~\TeV\ are shown in 
Figs.~\ref{fig:Comb_unconv_2011_barrel_and_endcap}
and~\ref{fig:Comb_conv_2011_barrel_and_endcap}.
The identification efficiency increases from 60--70\%
for \ET\ = 20~\GeV\ to 87--95\% (90--99\%) for $\ET\ > 100~\GeV$ for
unconverted (converted) photons. The uncertainty in the efficiency and
on the data-to-MC efficiency ratios decreases from 3--10\% at low
\ET\ to about 0.5--5\% for $\ET\ > 100~\GeV$, being typically larger
at higher $|\eta|$.

In the search of the Higgs boson decays to diphoton final states
with 7~\TeV\ data~\cite{ATLAS_Higgs}, an alternative photon
identification algorithm based on an artificial neural network (NN)
was used. 
The neural network uses as input the same discriminating variables
exploited by the cut-based selection. Multi-layer perceptrons are
implemented with the Toolkit for Multivariate Data
Analysis~\cite{tmva}, using 13 nodes in a single hidden
layer. Separate networks are optimised along bins of photon
pseudorapidity and transverse momentum. Different networks are created
for photons that are reconstructed as unconverted, single-track
converted and double-track converted, due to their different 
distributions of the discrimiminating variables.
The final identification is performed by requiring the output
discriminant to be larger than a certain threshold, tuned to reproduce
the background photon rejection of the cut-based algorithm.
For the training of the NN, simulated signal events 
and jet-enriched data are used. In the simulation, the discriminating
variables are
corrected for the average differences observed with respect to the data.
For the NN-based photon identification algorithm, the efficiency
increases from 85--90\% for \ET\ = 20~\GeV\ to about 97\% (99\%)
for $\ET>100~\GeV$ for unconverted (converted) photon candidates,
with uncertainties varying between 4\% and 7\%.

\begin{figure}[!htbp]
  \centering
   \subfloat[]{\label{fig:Comb_unconv_2011_barrel_a}\includegraphics[width=.45\textwidth]{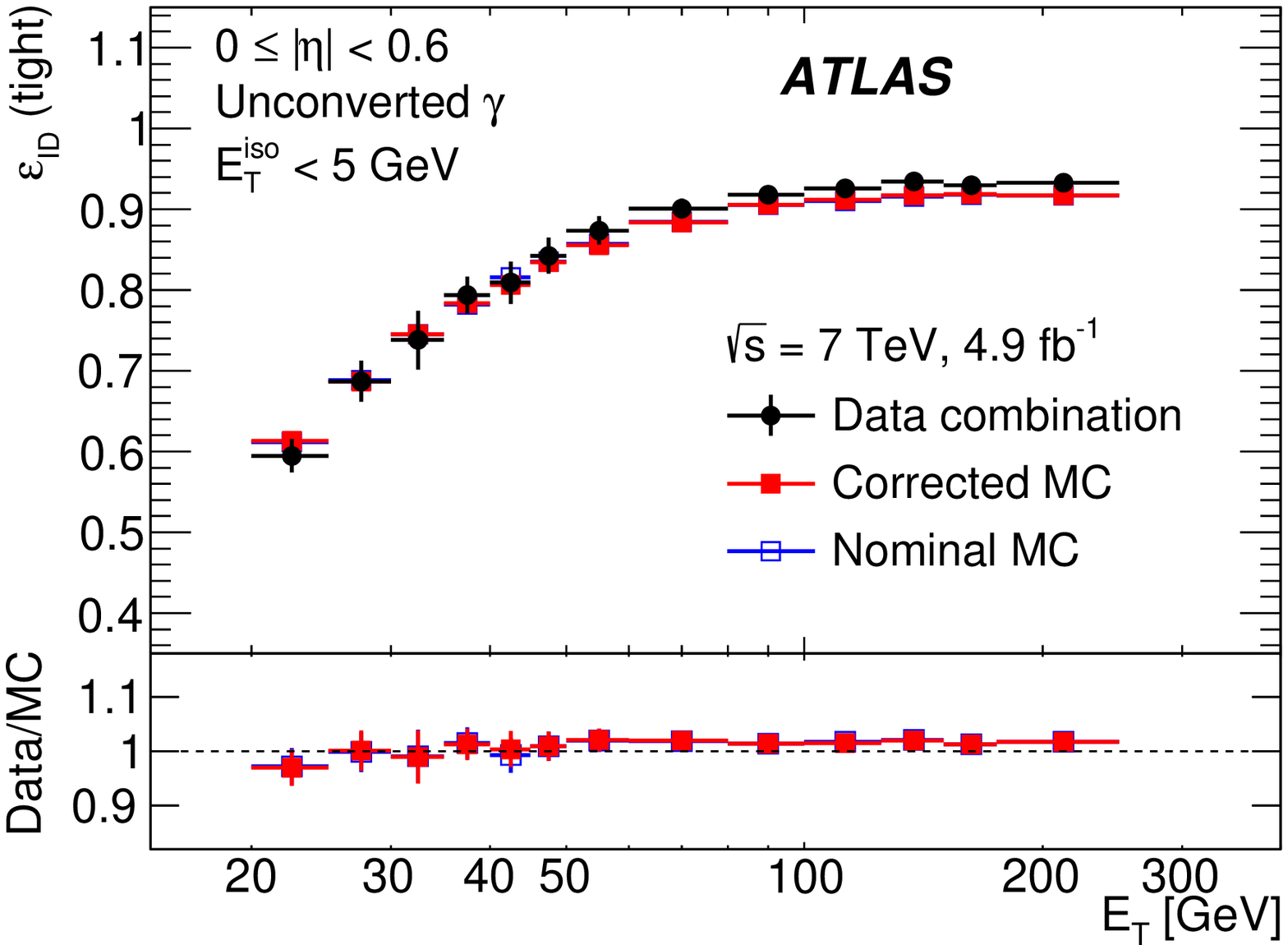}}
   \subfloat[]{\label{fig:Comb_unconv_2011_barrel_b}\includegraphics[width=.45\textwidth]{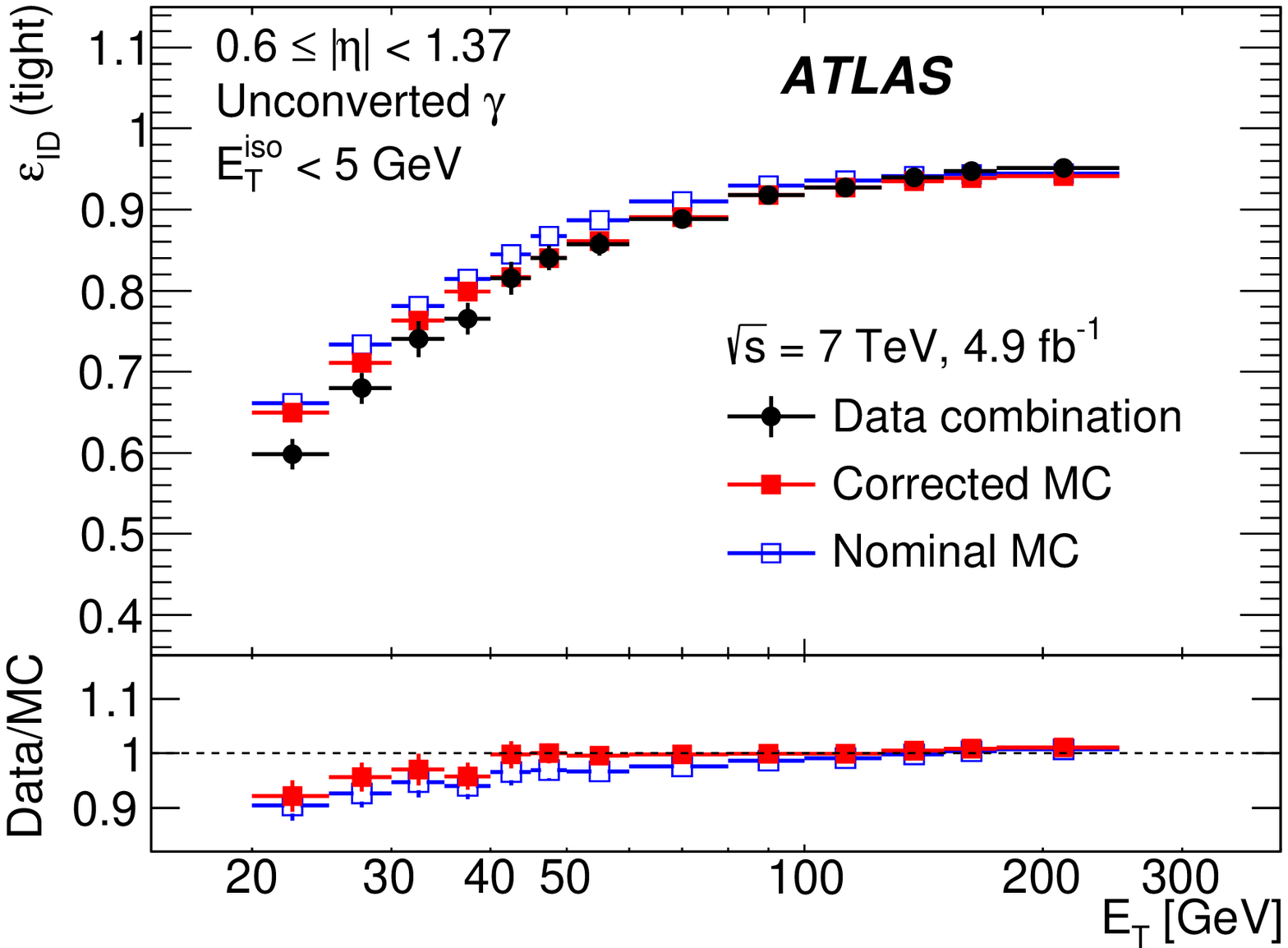}}\\
   \subfloat[]{\label{fig:Comb_unconv_2011_endcap_a}\includegraphics[width=.45\textwidth]{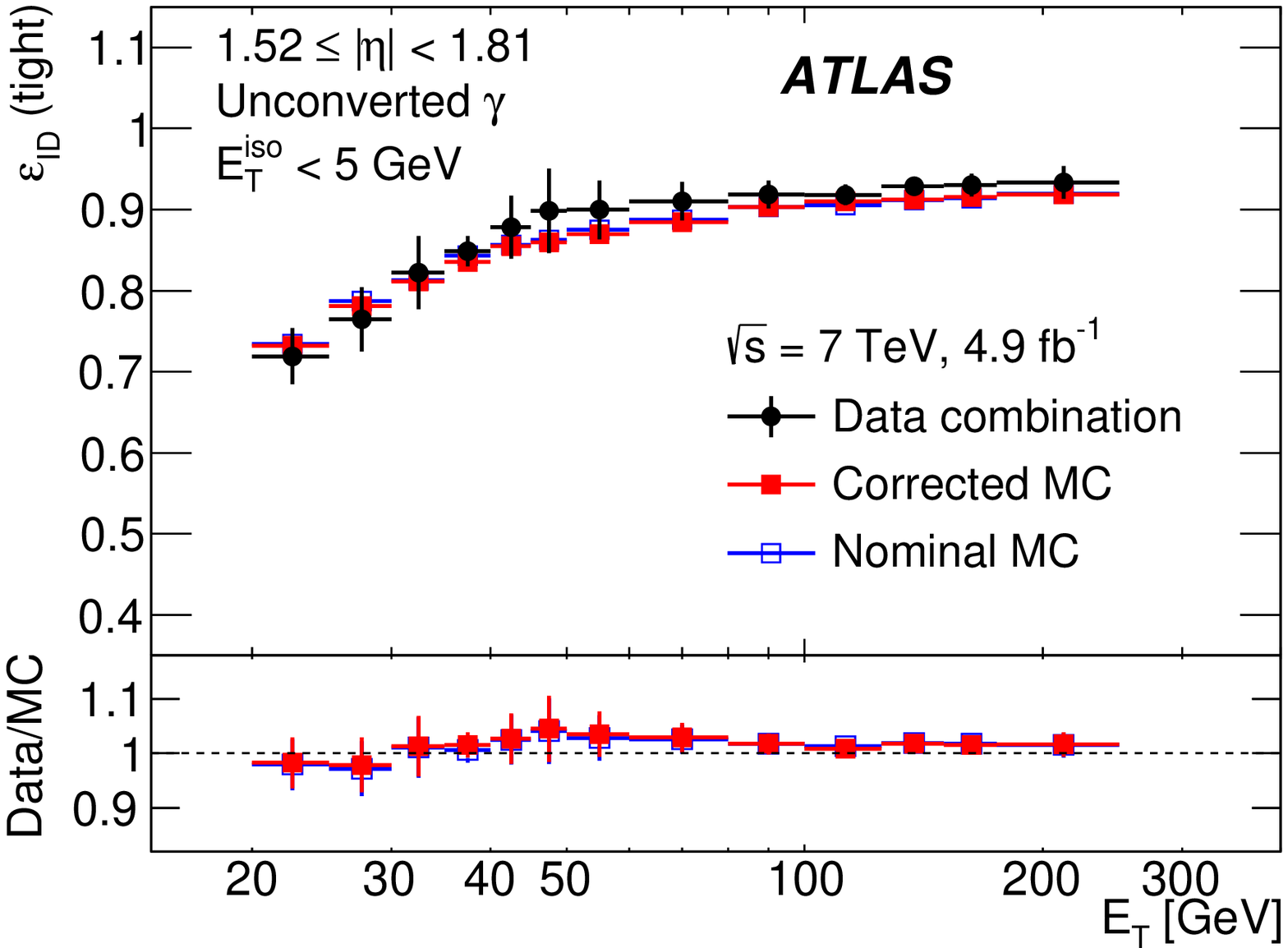}}
   \subfloat[]{\label{fig:Comb_unconv_2011_endcap_b}\includegraphics[width=.45\textwidth]{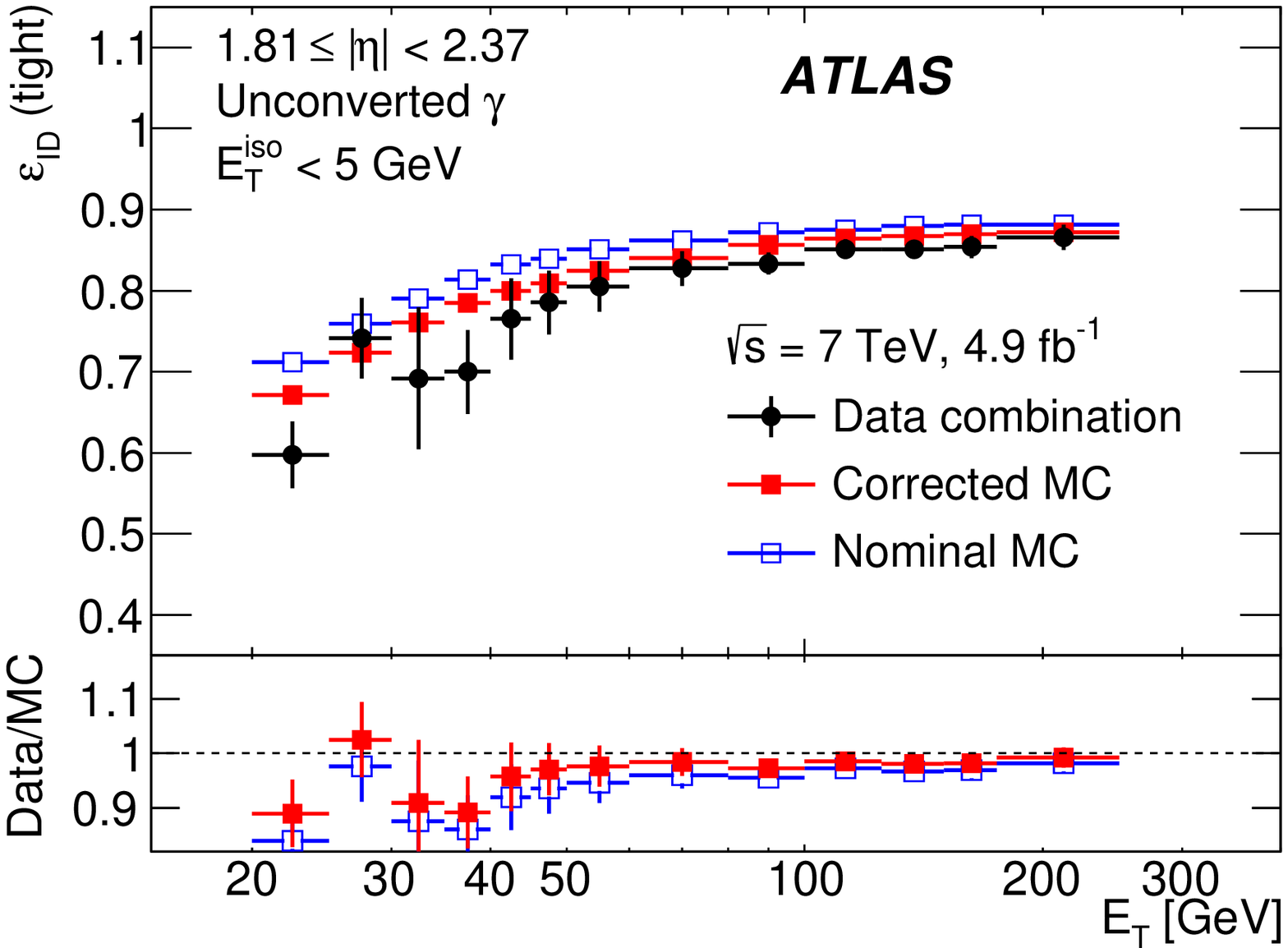}}
  \caption{Comparison between the identification efficiency $\effID$
    of unconverted photon candidates in $\sqrt{s} = 7~\TeV$ data and in the nominal and
    corrected MC predictions in the region $20~\GeV < \ET < 250~\GeV$,
    for the four pseudorapidity intervals (a) $|\eta|<0.6$, (b) $0.6\leq|\eta|<1.37$,
     (c) $1.52\leq|\eta|<1.81$, and (d) $1.81\leq|\eta|<2.37$.
     The black error bars correspond to the
    sum in quadrature of the statistical and systematic
    uncertainties estimated for the combination of the data-driven
    methods. Only the statistical uncertainties are shown for the MC
    predictions. The bottom panels show the ratio of the
    data-driven results to the nominal and corrected MC predictions.} 
 \label{fig:Comb_unconv_2011_barrel_and_endcap}
\end{figure}

\begin{figure}[!htbp]
  \centering
   \subfloat[]{\label{fig:Comb_conv_2011_barrel_a}\includegraphics[width=.45\textwidth]{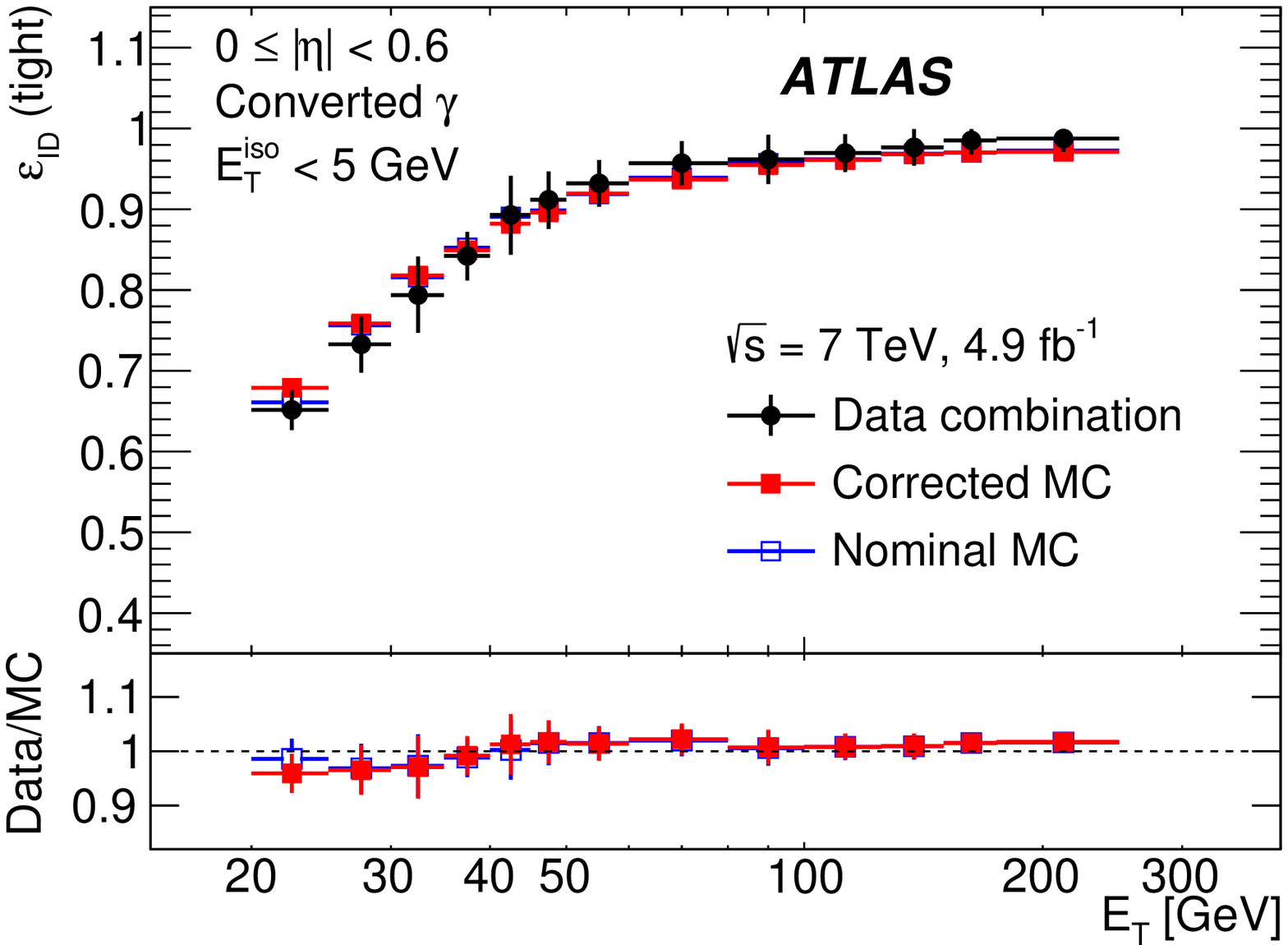}}
   \subfloat[]{\label{fig:Comb_conv_2011_barrel_b}\includegraphics[width=.45\textwidth]{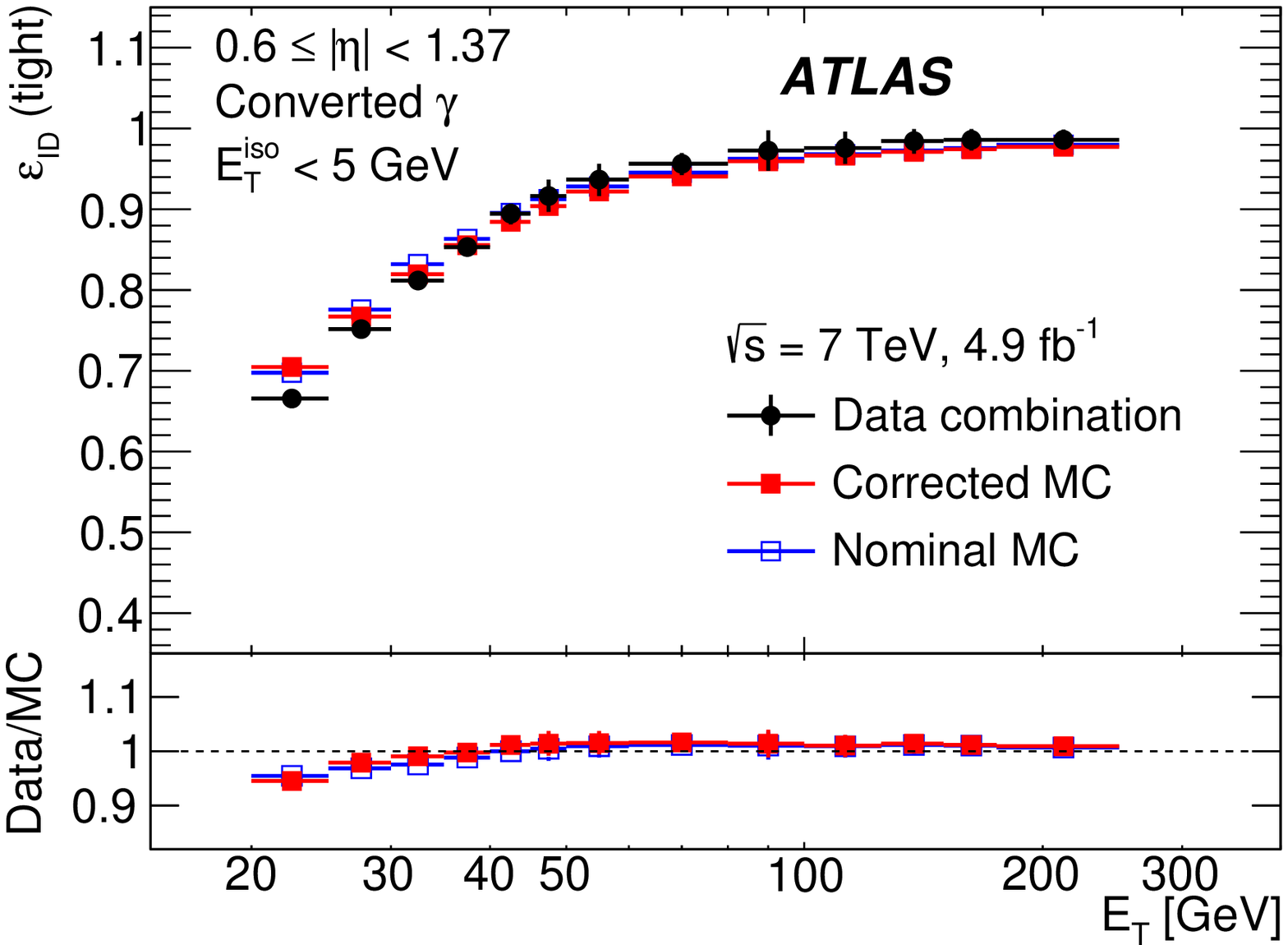}}\\
   \subfloat[]{\label{fig:Comb_conv_2011_endcap_a}\includegraphics[width=.45\textwidth]{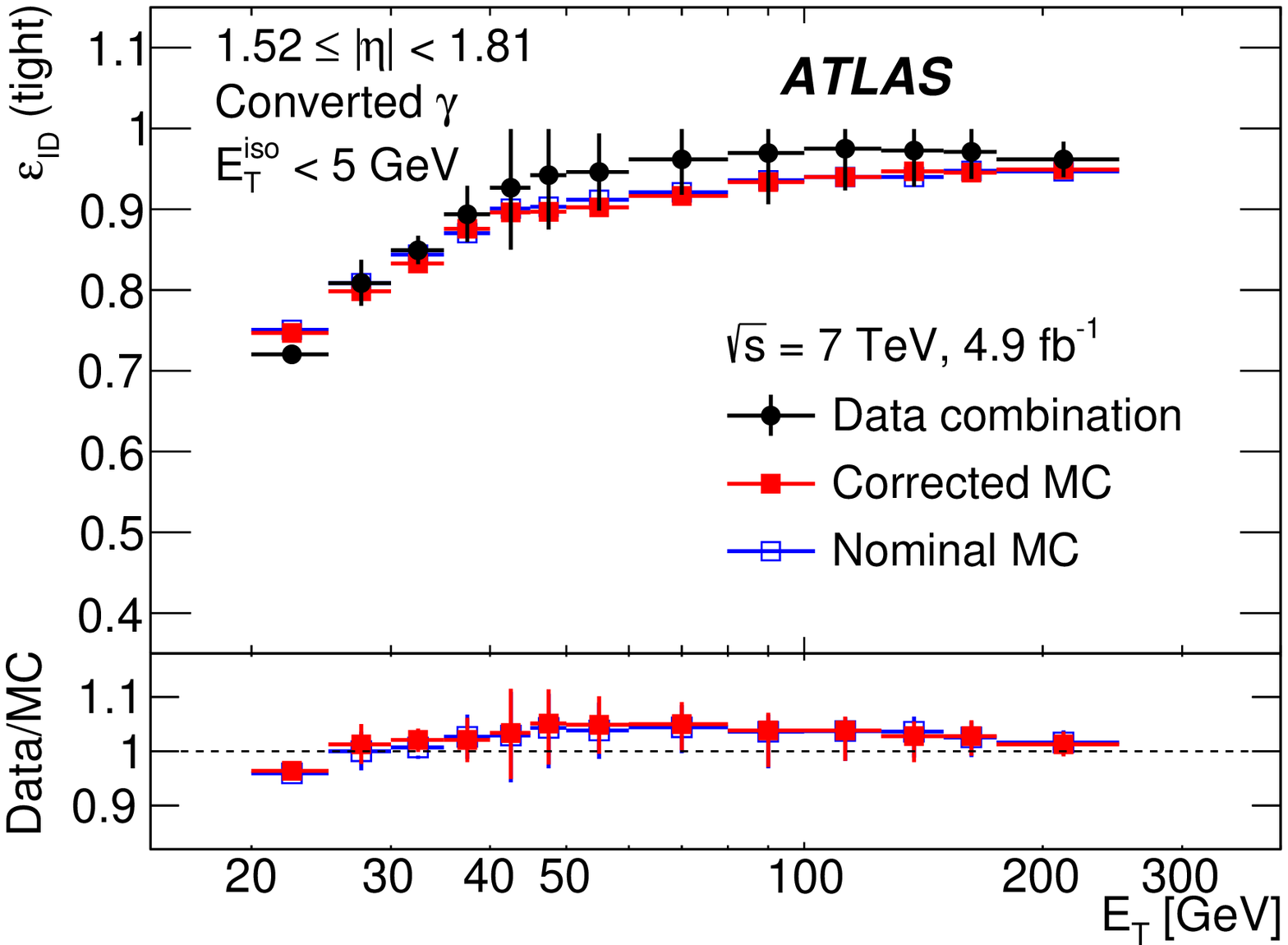}}
   \subfloat[]{\label{fig:Comb_conv_2011_endcap_b}\includegraphics[width=.45\textwidth]{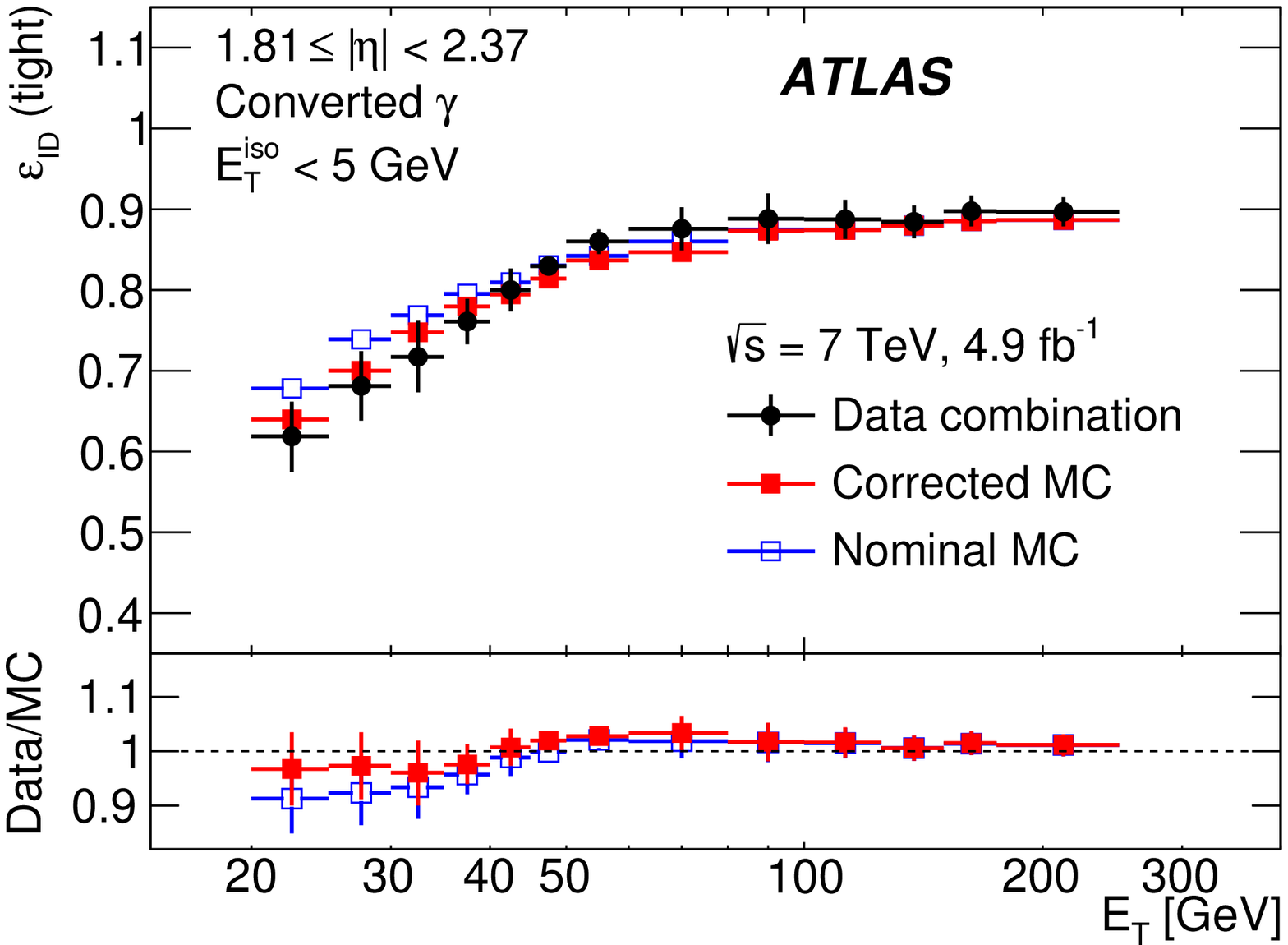}}
  \caption{Comparison between the identification efficiency $\effID$
    of converted photon candidates in $\sqrt{s}=7~\TeV$ data and in the nominal and
    corrected MC predictions in the region $20~\GeV < \ET <
    250~\GeV$, for the four
    pseudorapidity intervals (a) $|\eta|<0.6$, (b) $0.6\leq|\eta|<1.37$,
    (c) $1.52\leq|\eta|<1.81$, and (d) $1.81\leq|\eta|<2.37$.
    The black errors bars correspond to the sum in
    quadrature of the statistical and systematic uncertainties
    estimated for the combination of the data-driven methods. Only the
    statistical uncertainties are shown for the MC predictions. The
    bottom panels show the ratio of the data-driven results to
     MC predictions (also called scale factors in the text).} 
 \label{fig:Comb_conv_2011_barrel_and_endcap}
\end{figure}

\FloatBarrier
\section{Dependence of the photon identification efficiency on pile-up}
\label{sec:pileupdependence}

The dependence of the identification efficiency and of the
data/MC efficiency scale factors on pile-up was investigated
with both 7~\TeV\ and 8~\TeV\ data. The efficiencies are measured as a
function of the number of reconstructed primary vertex candidates
with at least three associated tracks, $N_\mathrm{PV}$, a quantity
which is highly correlated to $\mu$, the expected number of interactions per
bunch crossing.

In 2012 $pp$ collisions, $\mu$ was typically between 1 and 40, with an
average value of 21. 
In the range $10~\GeV<\ET<30$~\GeV\ the pile-up dependence of
the $\sqrt{s}=8$~\TeV\ identification efficiency is measured
using \Zboson\ boson radiative decays, integrating over the
photon pseudorapidity distribution because of the small size of the sample.
For higher transverse momenta the dependence is measured using the
results obtained with the electron extrapolation method, in four
$|\eta|$ bins.

In $\sqrt{s}=7$~\TeV\ $pp$ collisions, the pile-up dependence is
measured using the results obtained the matrix method,
in four $|\eta|$ bins, integrated over the $\ET>20$~\GeV\ range.

The results of the data measurements are shown in
Figs.~\ref{fig:eff_pileup_radz}--\ref{fig:eff_pileup_conv}.
The efficiency variation with $N_\mathrm{PV}$ in $\sqrt{s}=8$~\TeV\
data for $\ET<30$~\GeV\ is shown in Fig.~\ref{fig:eff_pileup_radz}.
The variation is rather large, up to 15\% in the range $0<N_\mathrm{PV}\leq 20$
(corresponding to about $0<\mu\leq 30$).
The efficiency variation with $N_\mathrm{PV}$ in $\sqrt{s}=8$~(7)~\TeV\
data for $\ET>30$~(20)~\GeV\ is shown in Figs.~\ref{fig:eff_pileup_unconv}
and~\ref{fig:eff_pileup_conv}.
In the 8~\TeV\ data the efficiency dependence on pile-up
for $\ET>30$~\GeV\ is similar in the pseudorapidity
intervals that have been studied, with a decrease of about 3--4\%
when $N_\mathrm{PV}$ increases from 1 to 20.
The pile-up dependence of the photon identification efficiency is
smaller in 8~\TeV\ data than in 7~\TeV\ data, since the 
photon identification criteria were specifically re-optimised
to be less sensitive to pile-up before the start of the 8~\TeV\ data taking.

To further study the pile-up dependence of the efficiency at high
photon transverse momenta, the $\sqrt{s}=8$~\TeV\ measurements
with the electron extrapolation have been repeated using only
electron probes with $\ET>45$~\GeV.
The efficiency for $\ET > 45~\GeV$ photons decreases by only 1--3\% when
$N_\mathrm{PV}$ increases from 1 to 20.

\begin{figure}[!htbp]
   \centering
   \subfloat[]{\label{fig:eff_pileup_radz_a}\includegraphics[width=.49\textwidth]{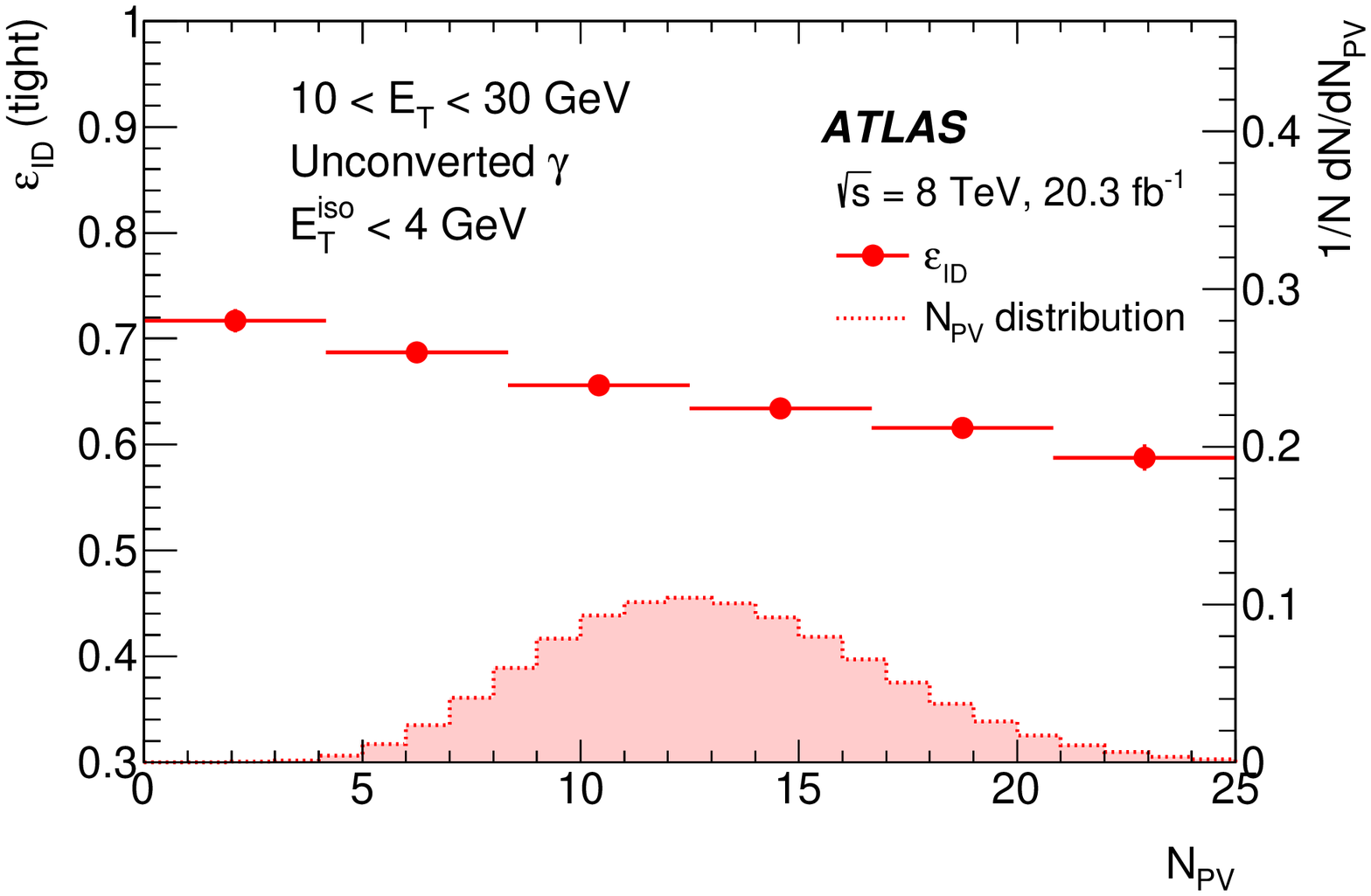}}
   \subfloat[]{\label{fig:eff_pileup_radz_b}\includegraphics[width=.49\textwidth]{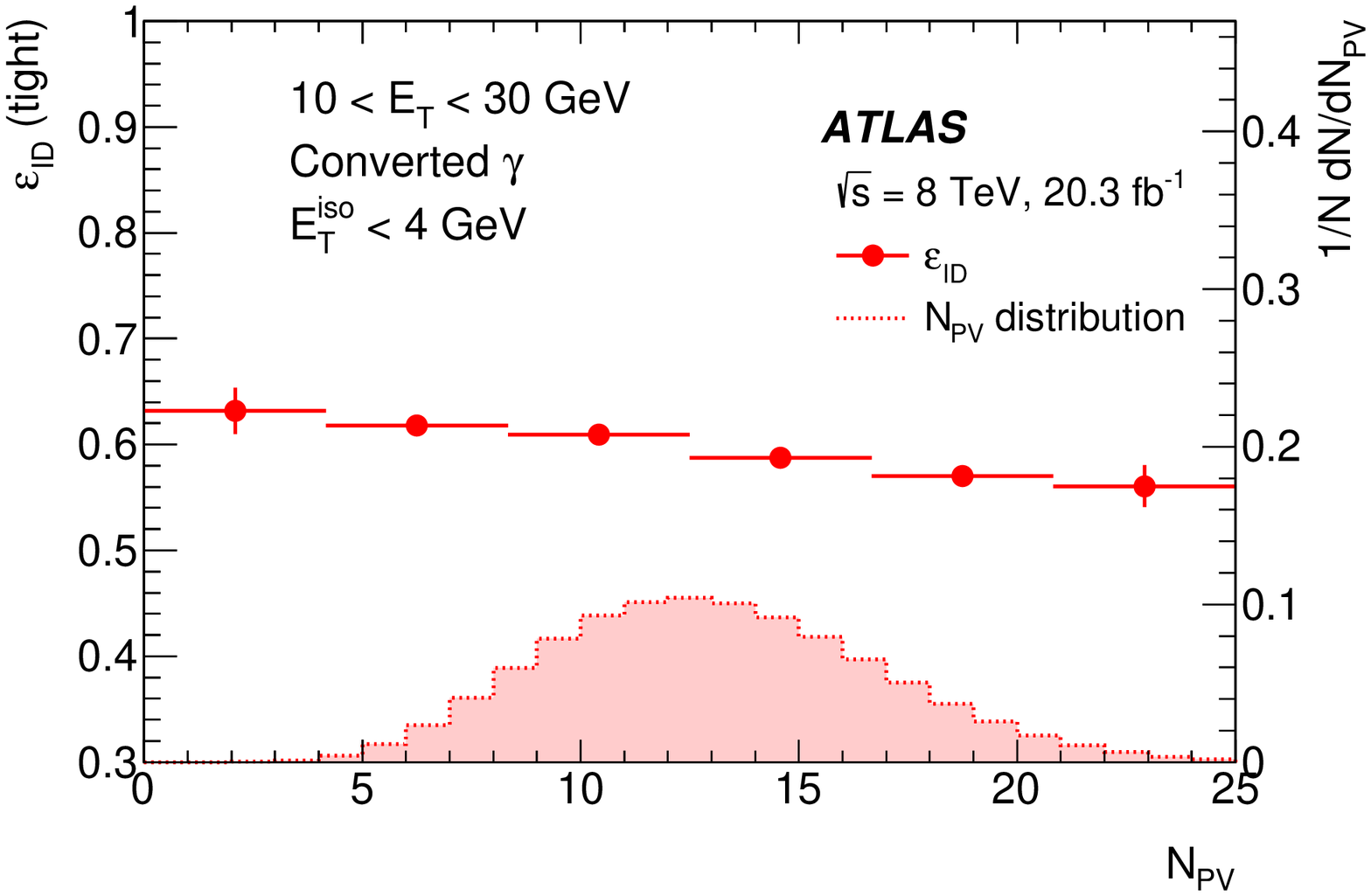}}
   \caption{Efficiency (red dots) of (a) unconverted and (b) converted
     photons candidates as a function of the number $N_\mathrm{PV}$
     of reconstructed
     primary vertices, measured in 2012 data from radiative 
     $Z$ boson decays.
     The measurements are integrated in pseudorapidity and in the 
     transverse momentum range $10~\GeV < \ET\ <30~\GeV$.
     The red histograms indicate the $N_\mathrm{PV}$ distribution
     in 2012 data.}
   \label{fig:eff_pileup_radz} 
\end{figure} 

\begin{figure}[!htbp]
   \centering 
   \subfloat[]{\label{fig:eff_pileup_unconv_a}\includegraphics[width=.5\textwidth]{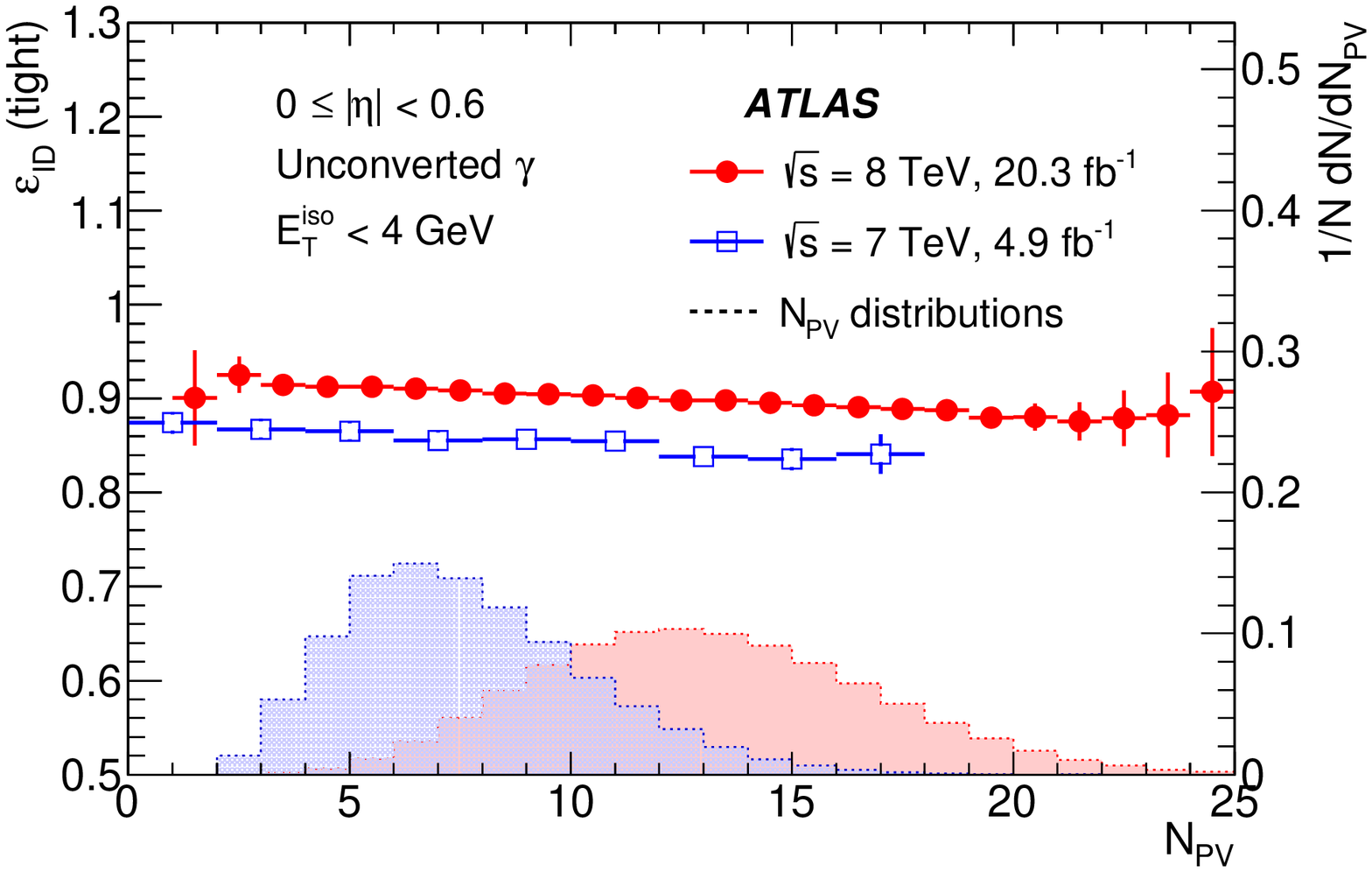}}
   \subfloat[]{\label{fig:eff_pileup_unconv_b}\includegraphics[width=.5\textwidth]{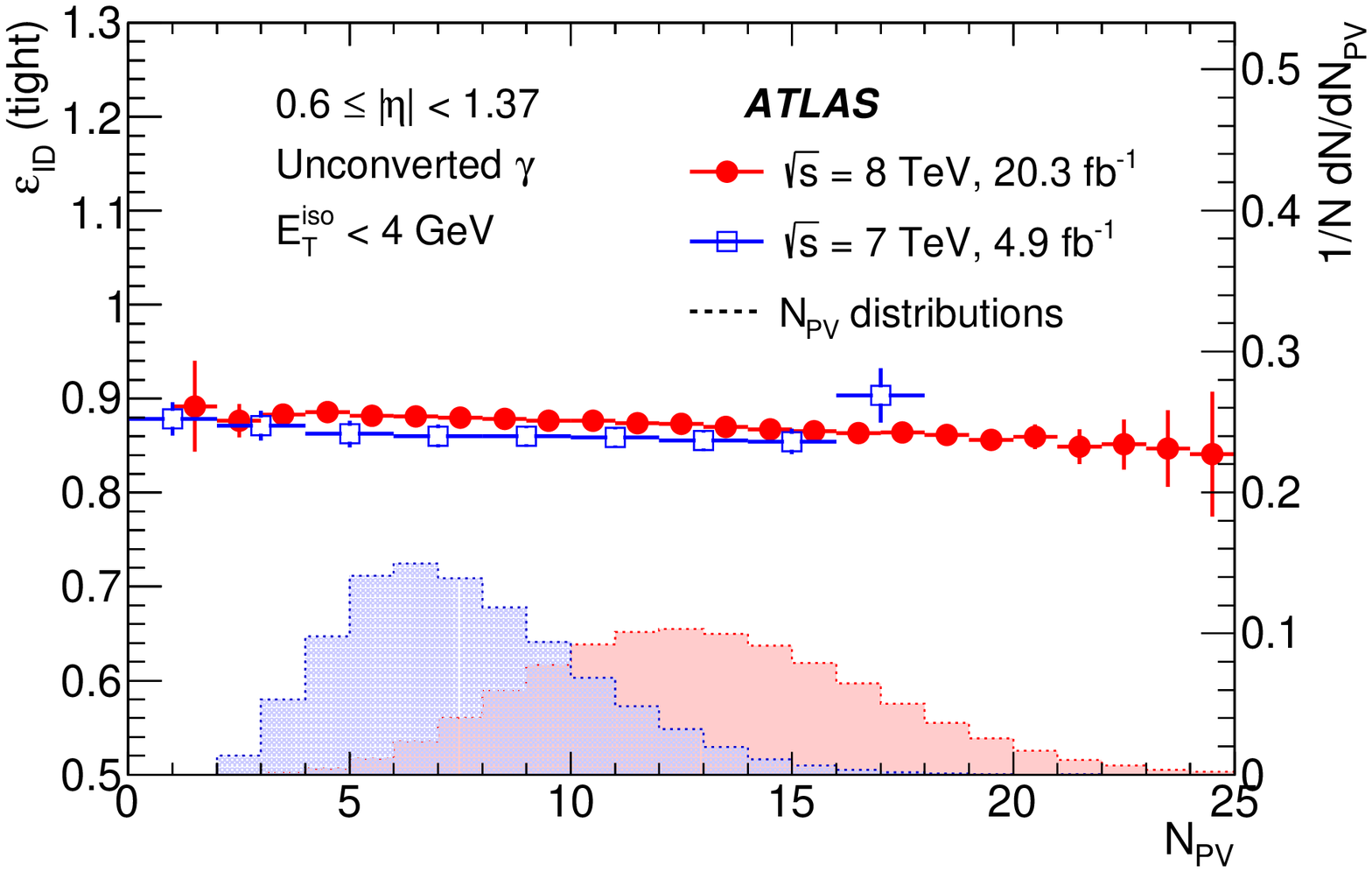}}\\
   \subfloat[]{\label{fig:eff_pileup_unconv_c}\includegraphics[width=.5\textwidth]{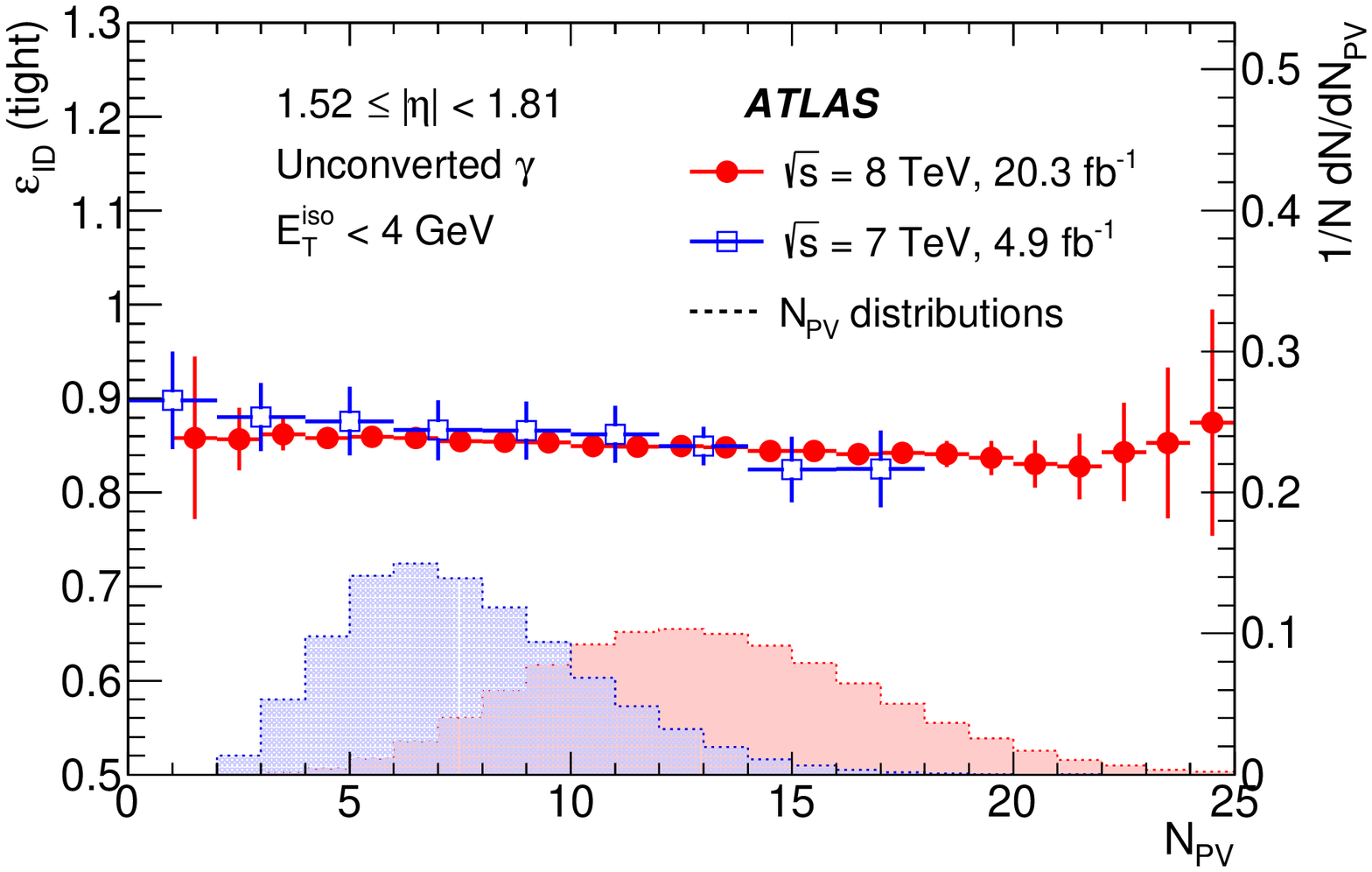}}
   \subfloat[]{\label{fig:eff_pileup_unconv_d}\includegraphics[width=.5\textwidth]{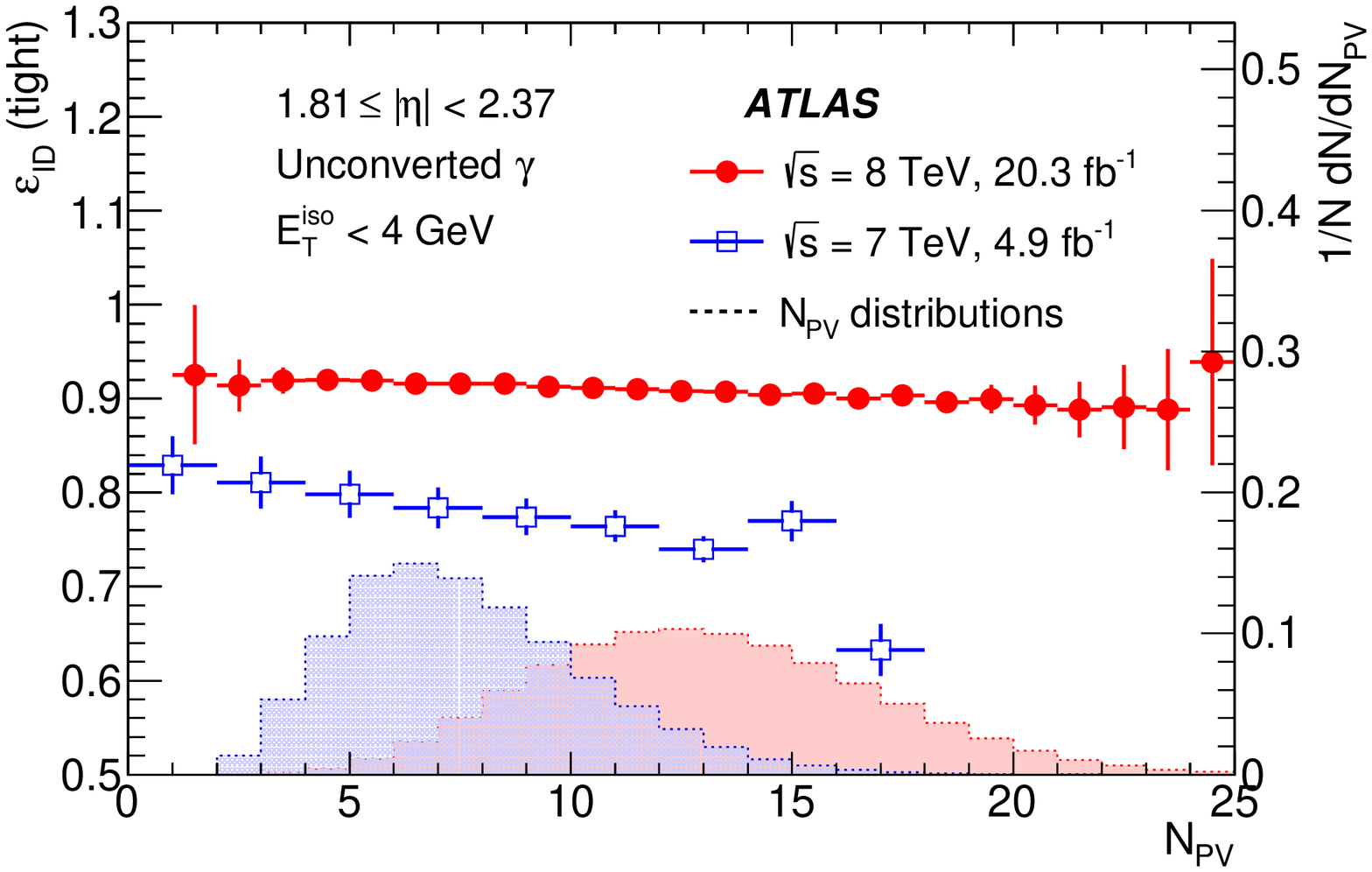}} 
   \caption{Comparison of data-driven efficiency measurements for
     unconverted photons performed with the 2011 (blue squares) and 2012
     (red circles) datasets as a function of the number $N_\mathrm{PV}$ of reconstructed
     primary vertex candidates, for for the four
     pseudorapidity intervals (a) $|\eta|<0.6$, (b) $0.6\leq|\eta|<1.37$,
     (c) $1.52\leq|\eta|<1.81$, and (d) $1.81\leq|\eta|<2.37$.
     The 2011 measurements are performed with the matrix method
     for photons with $\ET\ > 20~\GeV$ and the 2012
     measurements with the electron extrapolation method for photons
     with $\ET\ > 30~\GeV$. The two (blue/red) histograms indicate the
     $N_\mathrm{PV}$ distribution in 2011/2012 data.}  
   \label{fig:eff_pileup_unconv} 
\end{figure}
\begin{figure}[!htbp]
   \centering
   \subfloat[]{\label{fig:eff_pileup_conv_a}\includegraphics[width=.5\textwidth]{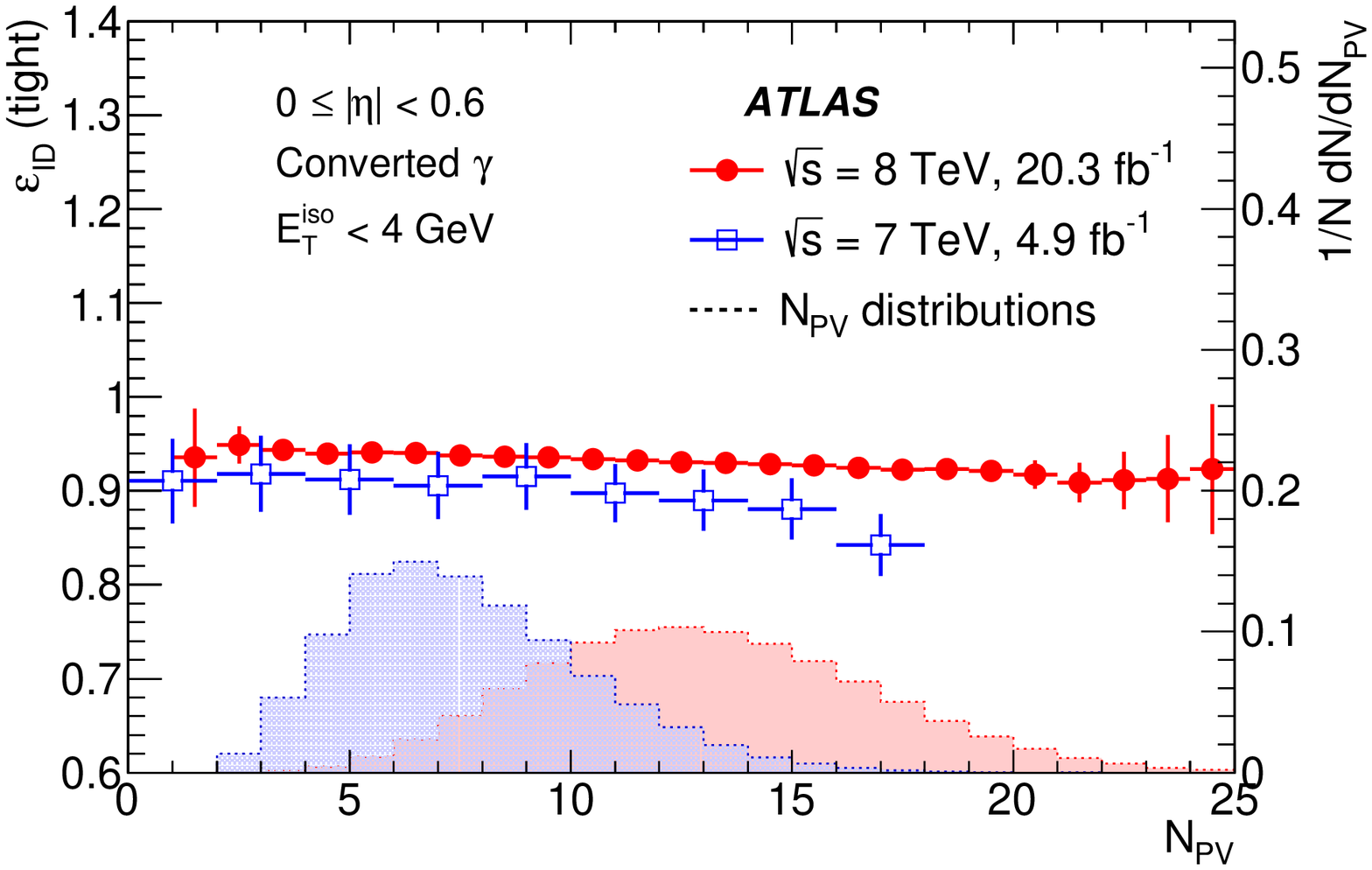}}
   \subfloat[]{\label{fig:eff_pileup_conv_b}\includegraphics[width=.5\textwidth]{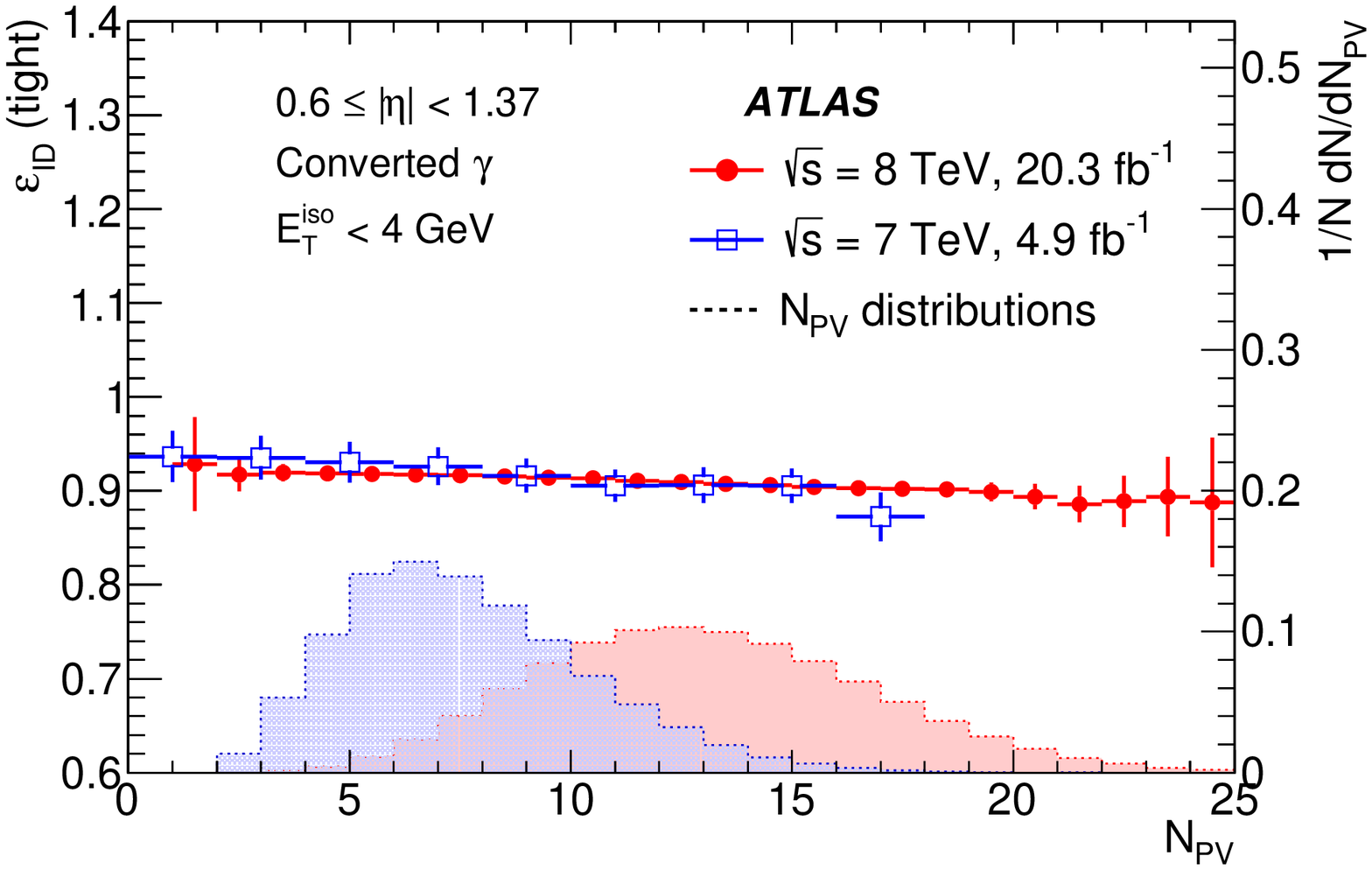}}\\
   \subfloat[]{\label{fig:eff_pileup_conv_c}\includegraphics[width=.5\textwidth]{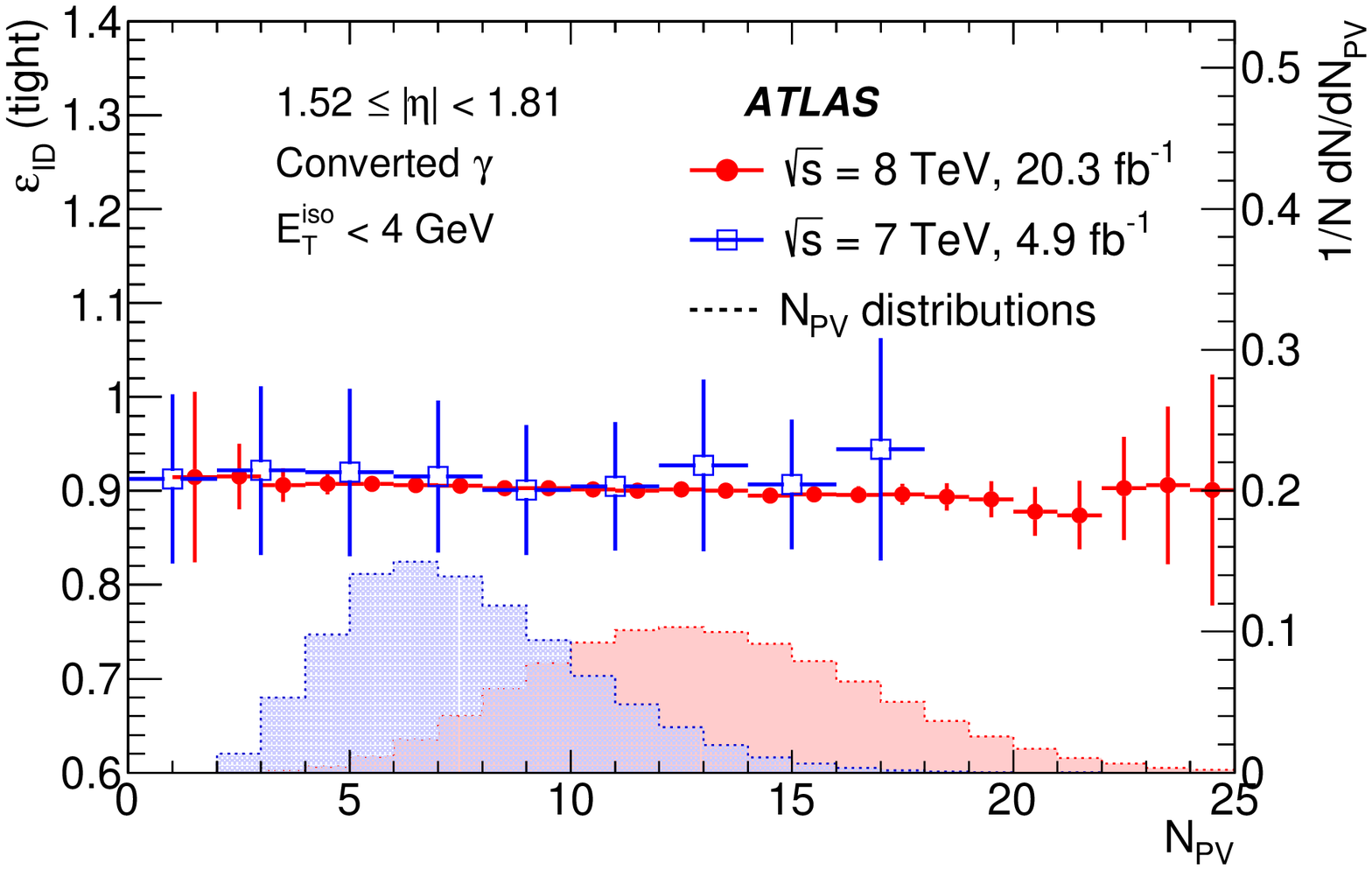}}
   \subfloat[]{\label{fig:eff_pileup_conv_d}\includegraphics[width=.5\textwidth]{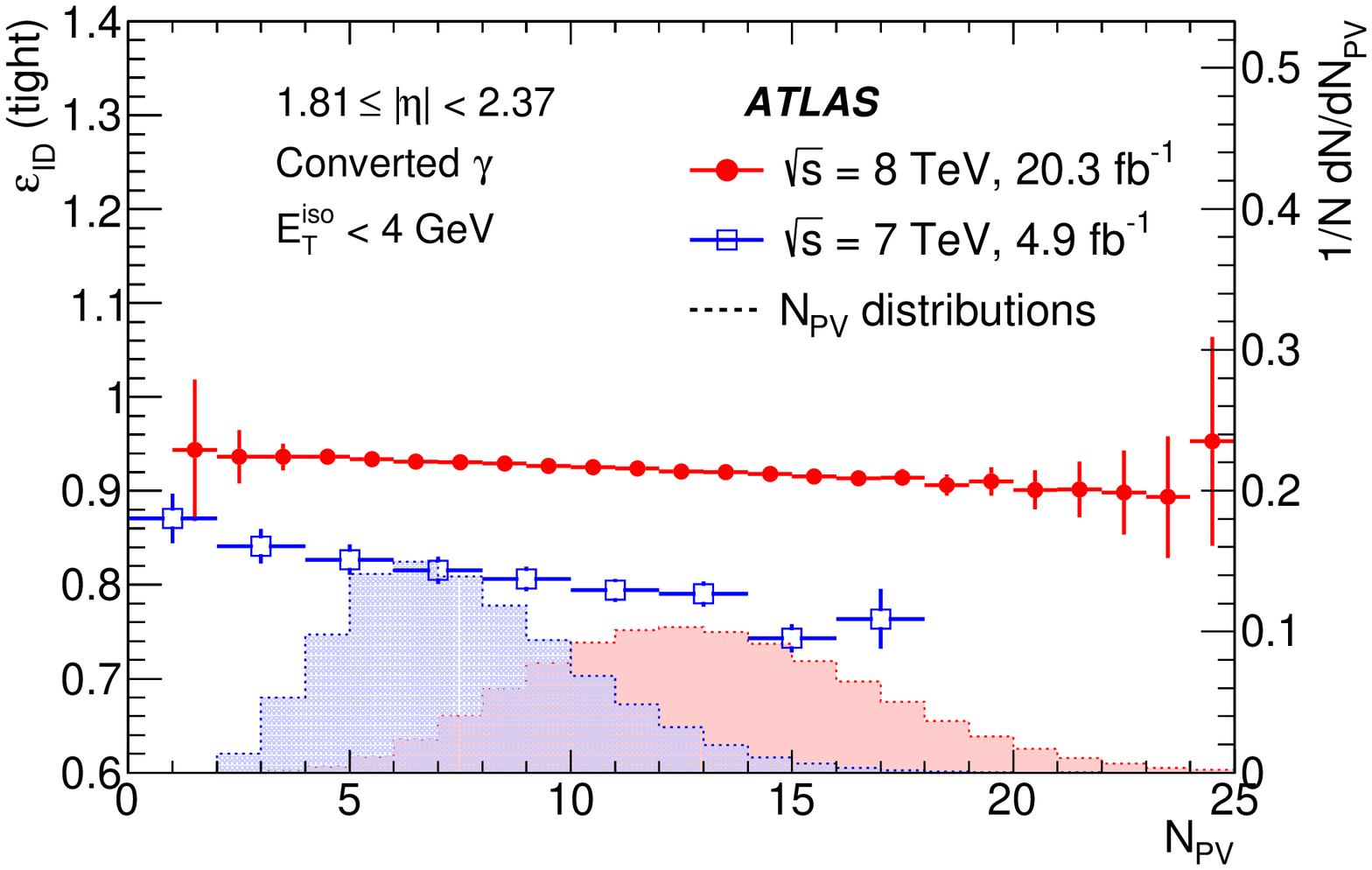}} 
   \caption{Comparison of data-driven efficiency measurements for
     converted photons performed with the 2011 (blue squares) and 2012
     (red circles) datasets as a function of the number $N_\mathrm{PV}$ of reconstructed
     primary vertex candidates, for the four
     pseudorapidity intervals (a) $|\eta|<0.6$, (b) $0.6\leq|\eta|<1.37$,
     (c) $1.52\leq|\eta|<1.81$, and (d) $1.81\leq|\eta|<2.37$.
     The 2011 measurements are performed with the matrix
     method for photons with $\ET\ > 20~\GeV$ and the 2012
     measurements with the electron extrapolation method for photons
     with $\ET\ > 30~\GeV$. The two (blue/red) histograms indicate the
     $N_\mathrm{PV}$ distribution in 2011/2012 data.}
   \label{fig:eff_pileup_conv} 
\end{figure}

The pile-up dependence of the efficiency in data is compared to the prediction
of the simulation by calculating the data-to-MC efficiency ratios as a
function of the number of reconstructed primary vertex candidates
$N_\mathrm{PV}$.
The pile-up dependence of the data-to-MC efficiency ratios is assessed
through a linear fit of the efficiency ratios as a function
of $N_\mathrm{PV}$. 
The slopes of these fits are always consistent with zero within the
uncertainties, which are of the order of 0.2\%.
Therefore, while the efficiency itself varies significantly as a function of
$N_\mathrm{PV}$, the dependence of the
data-to-MC efficiency ratios on $N_\mathrm{PV}$ in the range
$0<N_\mathrm{PV}\leq 26$ (corresponding to about $0<\mu\leq 40$) is compatible with zero.
This observation suggests that the simulation correctly models the
effect of pile-up on the distributions of the discriminating variables.

\FloatBarrier
\section{Conclusion}
\label{sec:conclusion}

The efficiency $\effID$ of the algorithms used by ATLAS to identify
photons during the LHC Run 1 has been measured from
$pp$ collision data using three independent methods in different
photon $\ET$ ranges. 
The three measurements agree within their uncertainties in the
overlapping $\ET$ ranges, and are combined.

For the data taken in 2011, 4.9 fb$^{-1}$ at $\sqrt{s}=7$~\TeV,
the efficiency of the
cut-based identification algorithm increases from 60--70\%
at $\ET=20$~\GeV\ up to 87--95\% (90--99\%) at $\ET>100$~\GeV\ for
unconverted (converted) photons.
With an optimised neural network this efficiency increases from
85--90\% at $\ET=20$~\GeV\ to about 97\% (99\%) at $\ET>100$~\GeV\ for
unconverted (converted) photon candidates for a similar background
rejection. 
For the data taken in 2012, 20.3 fb$^{-1}$ at $\sqrt{s}=8$~\TeV,
the efficiency of 
a re-optimised cut-based photon identification algorithm increases from 50--65\%
(45--55\%) for unconverted (converted) photons at $\ET=10$~\GeV\ to 
95--100\% at $\ET> 100$~\GeV, being larger than $\approx 90\%$ for $\ET>40$~\GeV. 

The nominal MC simulation of prompt photons in ATLAS predicts
significantly higher identification efficiency values than those
measured in some regions of the phase space, particularly at low
$\ET$. A simulation with shower shapes corrected for the average
shifts observed with respect to the data describes the values of $\effID$ better in the entire $\ET$ and
$\eta$ range accessible by the data-driven methods.
The residual difference between the efficiencies in data and
in the corrected simulation are taken into account by computing 
data-to-MC efficiency scale factors.
These factors differ from one by up to 10\% at $\ET=10$~\GeV\ and
by only a few percents above $\ET = 40$~\GeV, with an uncertainty
decreasing from 1.4--4.5\% (1.7--5.6\%) at $\ET=10$~\GeV\ for unconverted
(converted) photons to 0.2--0.8\% (0.2--0.5\%) at high
$\ET$ for $\sqrt{s}=8$~\TeV. 
The uncertainties are slightly larger for $\sqrt{s}=7$~\TeV\ data due 
to the smaller size of the control samples.

\section*{Acknowledgements}

We thank CERN for the very successful operation of the LHC, as well as the
support staff from our institutions without whom ATLAS could not be
operated efficiently.

We acknowledge the support of ANPCyT, Argentina; YerPhI, Armenia; ARC, Australia; BMWFW and FWF, Austria; ANAS, Azerbaijan; SSTC, Belarus; CNPq and FAPESP, Brazil; NSERC, NRC and CFI, Canada; CERN; CONICYT, Chile; CAS, MOST and NSFC, China; COLCIENCIAS, Colombia; MSMT CR, MPO CR and VSC CR, Czech Republic; DNRF and DNSRC, Denmark; IN2P3-CNRS, CEA-DSM/IRFU, France; GNSF, Georgia; BMBF, HGF, and MPG, Germany; GSRT, Greece; RGC, Hong Kong SAR, China; ISF, I-CORE and Benoziyo Center, Israel; INFN, Italy; MEXT and JSPS, Japan; CNRST, Morocco; FOM and NWO, Netherlands; RCN, Norway; MNiSW and NCN, Poland; FCT, Portugal; MNE/IFA, Romania; MES of Russia and NRC KI, Russian Federation; JINR; MESTD, Serbia; MSSR, Slovakia; ARRS and MIZ\v{S}, Slovenia; DST/NRF, South Africa; MINECO, Spain; SRC and Wallenberg Foundation, Sweden; SERI, SNSF and Cantons of Bern and Geneva, Switzerland; MOST, Taiwan; TAEK, Turkey; STFC, United Kingdom; DOE and NSF, United States of America. In addition, individual groups and members have received support from BCKDF, the Canada Council, CANARIE, CRC, Compute Canada, FQRNT, and the Ontario Innovation Trust, Canada; EPLANET, ERC, FP7, Horizon 2020 and Marie Sk{\l}odowska-Curie Actions, European Union; Investissements d'Avenir Labex and Idex, ANR, R{\'e}gion Auvergne and Fondation Partager le Savoir, France; DFG and AvH Foundation, Germany; Herakleitos, Thales and Aristeia programmes co-financed by EU-ESF and the Greek NSRF; BSF, GIF and Minerva, Israel; BRF, Norway; Generalitat de Catalunya, Generalitat Valenciana, Spain; the Royal Society and Leverhulme Trust, United Kingdom.

The crucial computing support from all WLCG partners is acknowledged gratefully, in particular from CERN, the ATLAS Tier-1 facilities at TRIUMF (Canada), NDGF (Denmark, Norway, Sweden), CC-IN2P3 (France), KIT/GridKA (Germany), INFN-CNAF (Italy), NL-T1 (Netherlands), PIC (Spain), ASGC (Taiwan), RAL (UK) and BNL (USA), the Tier-2 facilities worldwide and large non-WLCG resource providers. Major contributors of computing resources are listed in Ref.~\cite{ATL-GEN-PUB-2016-002}.


\clearpage
\appendix
\part*{Appendix}
\addcontentsline{toc}{part}{Appendix}
\section{Definition of the photon identification discriminating variables}
\label{app:IsEM-DV}

In this Appendix, the quantities used in the selection of photon
candidates, based on the reconstructed energy deposits in the ATLAS
calorimeters, are summarised.

\begin{itemize}

\item {\normalfont \bfseries Leakage in the hadronic calorimeter}

  The following discriminating variables are defined, based on the
  energy deposited in the hadronic calorimeter:

\begin{itemize}

\item {\em Normalised hadronic leakage} 
\begin{equation}
\Rhad = \frac{\ET^\mathrm{had}}{\ET}
\end{equation}
is the transverse energy $\ET^\mathrm{had}$ deposited in cells of the hadronic
calorimeter whose centre is in a window $\Delta\eta \times \Delta\phi$
= 0.24 $\times$ 0.24 behind the photon cluster, normalised to the
total transverse energy \ET\ of the photon candidate.

\item {\em Normalised hadronic leakage in first layer} 
\begin{equation}
\Rhadone = \frac{\ET^\mathrm{had,1}}{\ET}
\end{equation}
is the transverse energy $\ET^\mathrm{had,1}$ deposited in cells of
the first layer of the hadronic calorimeter whose centre is in a
window $\Delta\eta \times \Delta\phi$ = 0.24 $\times$ 0.24 behind the
photon cluster, normalised to the total transverse energy \ET\ of the
photon candidate.

\end{itemize}
The $\Rhad$ variable is used in the selection of photon candidates
with pseudorapidity $|\eta|$ between 0.8 and 1.37 while the $\Rhadone$ 
variable is used otherwise.

\item{\normalfont \bfseries Variables using the second (``middle'') layer of the
  electromagnetic calorimeter}

The discriminating variables based on the energy deposited in the
second layer of the electromagnetic calorimeter are the following:

\begin{itemize}

\item{\em Middle $\eta$ energy ratio}
\begin{equation}
\Reta = \frac{E_{3\times7}^{S2}}{E_{7\times7}^{S2}}
\end{equation}
is the ratio of the sum $E_{3\times7}^{S2}$ of the energies of
the second-layer cells of the electromagnetic calorimeter contained in
a 3$\times$7 rectangle in $\eta\times\phi$ measured in cell units
($0.025\times 0.0245$),
to the sum $E_{7\times7}^{S2}$ of the energies in a 7$\times$7
rectangle, both centred around the cluster seed.

\item{\em Middle $\phi$ energy ratio}
\begin{equation}
\Rphi = \frac{E_{3\times3}^{S2}}{E_{3\times7}^{S2}}
\end{equation}
is defined similarly to $R_{\eta}$. $R_{\phi}$ behaves very
differently for unconverted and converted photons, since the electrons
and positrons generated by the latter bend in different directions in
$\phi$ because of the solenoid's magnetic field, producing
larger showers in the $\phi$~direction than the unconverted photons.

\item{\em Middle lateral width}
\begin{equation}
\wetatwo = \sqrt{
\frac{\sum E_i \eta_i^2}{\sum E_i} - \left(\frac{\sum E_i \eta_i}{\sum E_i} \right)^2}
\end{equation}
where $E_{i}$ is the energy deposit in each cell, and $\eta_i$ is the actual $\eta$ position of the cell,
measures the shower's lateral width in the second layer of the
electromagnetic calorimeter, using all cells in a window
$\eta\times\phi = 3 \times 5$ measured in cell units.

\end{itemize}

\item {\normalfont \bfseries Variables using the first (``front'') layer of the
  electromagnetic calorimeter}

The discriminating variables based on the energy deposited in the
first layer of the electromagnetic calorimeter are the following:

\begin{itemize}

\item {\em Front side energy ratio}
\begin{equation}
\Fside = \frac{E(\pm 3) - E(\pm 1)}{E(\pm 1)}
\end{equation}
measures the lateral containment of the shower, along the $\eta$
direction.
$E(\pm n)$ is the energy in the $\pm n$ strip cells around the one
with the largest energy. 

\item {\em Front lateral width (3 strips)}
\begin{equation}
\wthree = \sqrt{\frac{\sum E_i (i-i_\mathrm{max})^2}{\sum E_i}}
\end{equation}
measures the shower width along $\eta$ in the first layer of the
electromagnetic calorimeter, using a total of three strip cells
centred on the largest energy deposit.
The index $i$ is the strip identification number, $i_\mathrm{max}$
identifies the strip cells with the greatest energy, and $E_i$ is the
energy deposit in each strip cell.

\item {\em Front lateral width (total)} \\
$\wtot$ measures the shower width along $\eta$ in the first layer of
  the electromagnetic calorimeter using all cells in a window
  $\Delta\eta\times\Delta\phi = 0.0625 \times 0.196$, corresponding 
  approximately to $20\times 2$ strip cells in $\eta\times \phi$, and is
  computed as $\wthree$. 

\item {\em Front second maximum energy difference}
\begin{equation}
\DeltaE = \left[ E_{2^\mathrm{nd}\mathrm{max}}^{S1} - E_\mathrm{min}^{S1} \right] 
\end{equation}
is the difference between the energy of the strip cell with the
second largest energy $E_{2^\mathrm{nd} \mathrm{max}}^{S1}$, and the
energy in the strip cell with the lowest energy found between the
largest and the second largest energy $E_\mathrm{min}^{S1}$ 
($\DeltaE=0$ when there is no second maximum).

\item {\em Front maxima relative energy ratio} 
\begin{equation}
\Eratio = \frac{E_{1^\mathrm{st}\,\mathrm{max}}^{S1}-E_{2^\mathrm{nd}\,\mathrm{max}}^{S1}}
{E_{1^\mathrm{st}\,\mathrm{max}}^{S1}+E_{2^\mathrm{nd}\,\mathrm{max}}^{S1}}
\end{equation}
measures the relative difference between the energy of the strip cell
with the largest energy $E_{1^\mathrm{st}\,\mathrm{max}}^{S1}$ and the energy
in the strip cell with second largest energy $E_{2^\mathrm{nd}\,\mathrm{max}}^{S1}$ 
($\Eratio=1$ when there is no second maximum).
\end{itemize}

\end{itemize}


\printbibliography

\newpage 
\begin{flushleft}
{\Large The ATLAS Collaboration}

\bigskip

M.~Aaboud$^{\rm 136d}$,
G.~Aad$^{\rm 87}$,
B.~Abbott$^{\rm 114}$,
J.~Abdallah$^{\rm 65}$,
O.~Abdinov$^{\rm 12}$,
B.~Abeloos$^{\rm 118}$,
R.~Aben$^{\rm 108}$,
O.S.~AbouZeid$^{\rm 138}$,
N.L.~Abraham$^{\rm 150}$,
H.~Abramowicz$^{\rm 154}$
% ...
\bigskip
\\
% ...
$^{12}$ Institute of Physics, Azerbaijan Academy of Sciences, Baku, Azerbaijan\\
% ...
$^{65}$ University of Iowa, Iowa City IA, United States of America\\
% ...
$^{87}$ CPPM, Aix-Marseille Universit{\'e} and CNRS/IN2P3, Marseille, France\\
% ...
$^{108}$ Nikhef National Institute for Subatomic Physics and University of Amsterdam, Amsterdam, Netherlands\\
% ...
$^{114}$ Homer L. Dodge Department of Physics and Astronomy, University of Oklahoma, Norman OK, United States of America\\
% ...
$^{118}$ LAL, Univ. Paris-Sud, CNRS/IN2P3, Universit{\'e} Paris-Saclay, Orsay, France\\
% ...
$^{136}$ $^{(a)}$ Facult{\'e} des Sciences Ain Chock, R{\'e}seau Universitaire de Physique des Hautes Energies - Universit{\'e} Hassan II, Casablanca; $^{(b)}$ Centre National de l'Energie des Sciences Techniques Nucleaires, Rabat; $^{(c)}$ Facult{\'e} des Sciences Semlalia, Universit{\'e} Cadi Ayyad, LPHEA-Marrakech; $^{(d)}$ Facult{\'e} des Sciences, Universit{\'e} Mohamed Premier and LPTPM, Oujda; $^{(e)}$ Facult{\'e} des sciences, Universit{\'e} Mohammed V, Rabat, Morocco\\
% ...
$^{138}$ Santa Cruz Institute for Particle Physics, University of California Santa Cruz, Santa Cruz CA, United States of America\\
% ...
$^{150}$ Department of Physics and Astronomy, University of Sussex, Brighton, United Kingdom\\
% ...
$^{154}$ Raymond and Beverly Sackler School of Physics and Astronomy, Tel Aviv University, Tel Aviv, Israel\\
% ...
$^{a}$ Also at Department of Physics, King's College London, London, United Kingdom\\
$^{b}$ Also at Institute of Physics, Azerbaijan Academy of Sciences, Baku, Azerbaijan\\
$^{c}$ Also at Novosibirsk State University, Novosibirsk, Russia\\
$^{d}$ Also at TRIUMF, Vancouver BC, Canada\\
$^{e}$ Also at Department of Physics {\&} Astronomy, University of Louisville, Louisville, KY, United States of America\\
% ...
\end{flushleft}


\end{document}